\documentclass[11pt,onecolumn,twoside]{IEEEtran}

\usepackage{amssymb}
\usepackage[pdftex]{graphicx}
\graphicspath{{./}{figures/}}

\usepackage[table]{xcolor}
\usepackage{amsmath}
\usepackage{amssymb}
\usepackage{url}
\usepackage{setspace}
\usepackage{fixltx2e}
\usepackage{floatrow}
\newfloatcommand{capbtabbox}{table}[][\FBwidth]
\usepackage{blindtext}
\def\eg{\emph{e.g. }}
\def\ie{\emph{i.e. }}
\def\etal{\emph{et al. }}
\def\lenna{\emph{Lenna }}
\def\LSW{\textit{Latin Square Whitening}}
\def\LSS{\textit{Latin Square Substitution}}
\def\LSP{\textit{Latin Square Permutation}}
\def\LSRS{\textit{Latin Square Row S-box}}
\def\LSCS{\textit{Latin Square Column S-box}}

\def\LSCP{\textit{Latin Square Column P-box}}
\def\LSWs{\textit{Latin Square Whitening }}
\def\LSSs{\textit{Latin Square Substitution }}
\def\LSPs{\textit{Latin Square Permutation }}
\def\LSRSs{\textit{Latin Square Row S-box }}

\def\astoday{as the date of 04/10/2012}
\def\subw{0.8}
\def\subh{0.15}
\def\subhf{0.14}
\usepackage{algorithm}
\usepackage{algorithmic}

\begin{document}
%
\title{A Novel Latin Square Image Cipher}
\author{Yue~Wu,~\IEEEmembership{Student Member,~IEEE,}
        Yicong~Zhou,~\IEEEmembership{Member,~IEEE,}
        Joseph~P.~Noonan,~\IEEEmembership{Life Member,~IEEE,}
        Sos~Agaian,~\IEEEmembership{Senior Member,~IEEE,}
        and {C.~L.~Philip~Chen},~\IEEEmembership{Fellow,~IEEE,}
%
%
\thanks{Y. Wu and J. P. Noonan is with the Department of Electrical and Computer Engineering, Tufts University, Medford, MA 02155, United States; e-mail: ywu03@ece.tufts.edu.}
\thanks{Y. Zhou and C. L. P. Chen are with the Department of Computer and Information Science,University of Macau, Macau, China; email: yicongzhou@umac.mo}
\thanks{S. Agaian is with the Electrical and Computer Engineering, University of Texas at San Antonio, San Antonio, TX 78249, United States.}
\thanks{Manuscript received April 19, 2012; revised January 11, 2013.}
}

%
%

\markboth{A draft submitted to IEEE Transactions on Information Forensics and Security}%
{Wu \MakeLowercase{\textit{et al.}}}
%



\maketitle

\begin{abstract}

In this paper, we introduce a symmetric-key Latin square image cipher (LSIC) for grayscale and color images. Our contributions to the image encryption community include 1) we develop new Latin square image encryption primitives including \textit{Latin Square Whitening, Latin Square S-box} and \textit{Latin Square P-box} ; 2) we provide a new way of integrating probabilistic encryption in image encryption by embedding random noise in the least significant image bit-plane; and 3) we construct LSIC with these Latin square image encryption primitives all on one keyed Latin square in a new loom-like substitution-permutation network. Consequently, the proposed LSIC achieve many desired properties of a secure cipher including a large key space, high key sensitivities, uniformly distributed ciphertext, excellent confusion and diffusion properties, semantically secure, and robustness against channel noise. Theoretical analysis show that the LSIC has good resistance to many attack models including brute-force attacks, ciphertext-only attacks, known-plaintext attacks and chosen-plaintext attacks. Experimental analysis under extensive simulation results using the complete USC-SIPI \textit{Miscellaneous} image dataset demonstrate that LSIC outperforms or reach state of the art suggested by many peer algorithms. All these analysis and results demonstrate that the LSIC is very suitable for digital image encryption. Finally, we open source the LSIC MATLAB code under webpage \url{https://sites.google.com/site/tuftsyuewu/source-code}.

\end{abstract}


%
\IEEEpeerreviewmaketitle

\section{Introduction}
With ubiquitous digital images and digital media devices all over the world, the importance of image security has been noticed and emphasized in recent years \cite{2004yang}. In the real world, digital cameras capture the real scene in the format of digital images, and are widely used in many digital devices such as smart phones, IPADs, and laptops. In the virtual world, digital images, including those taken from cameras, scanned documents or pictures, and computer-aid virtual paintings and so on, are the most common elements within a webpage besides texts on the World Wide Web. Due to extensive information within a digital image, divulged image contents sometimes cause severe problems for its owner(s). In many cases, such information leakage seriously invades personal privacy, \eg the malicious spread of photos in personal online albums or patients' medical diagnosis images, and furthermore it may cause uncountable losses for a company or a nation, \eg a secret product design for a company or a governmental classified scanned document.

Conventionally, digital data is encrypted by bit-stream ciphers and block ciphers \cite{blowFish,twoFish,DES,AES}. The two well-known block ciphers are the Digital Encryption Standard (DES) \cite{DES} and its successor Advanced Encryption Standard (AES) \cite{AES}. A digital image is a specific type of digital data and can be encrypted by these conventional ciphers.  However, they are not ideal ciphers for digital images because of
\begin{itemize}
  \item Insufficient large block size: digital images are normally of several kilobits (Kb) and megabits (Mb), while conventional bit-stream/block ciphers commonly has a block size less than 256 bits.
  \item Neglect of the nature of digital images: digital images are of two-dimensional data, while conventional bit-stream/block ciphers encrypt an image by indirectly encrypting a pixel sequence extracted from this image.
\end{itemize}
The first defect implies the low efficiency of encrypting a digital image using a bit-stream/block cipher \cite{2004yang,3DCat,3DBaker}. The second defect implies that a pixel sequence of an image is of high information redundancy with a tilted histogram and is distinctive from a common bit sequence input to a bit-stream/block cipher. Therefore, image encryption should be adaptive to image properties and natures \cite{2004yang,3DCat}.

In general, digital image encryption methods can be classified into two groups: perceptual-level and bit-level. The perceptual-level image encryption methods intend to transform an image into a unrecognized one using a fast algorithm, \eg selective encryption techniques \cite{1218805,5673978,5716133} and optical encryption techniques \cite{4939407,elashry:033002,chen:117001}. In this case, images are believed to be valuable only within a certain time period \eg a couple of hours. To some extend, therefore, an encrypted image by using perceptual-level image encryption is insecure because it maybe cracked after a sufficiently long time. In contrast, the bit-level image encryption aims to change an image into a random-like one. In this situation, images are believed to be valuable for a quite long time period \eg twenty years. Nowadays, the bit-level image encryption methods are mainly based on chaotic systems \cite{3DBaker,3DCat,Zhu2011,Liao2010,Awad2011,Kumar2011}. 
Although many existing chaotic image encryption algorithms have several good properties for cryptography security, they have defects in the following aspects regardless of the chaotic systems being used:
\begin{itemize}
  \item A chaotic system is defined on real numbers while a cryptsystem is defined on finite numbers.
  \item A chaotic system may lose its chaotic nature completely and become periodic when it is decretized.
  \item A chaotic system's parameters and initial values can be estimated by a number of existing tools and methods.
\end{itemize}
The first defect implies the difficulties of software and hardware implementation for a chaos-based image encryption method because round-off errors in real number quantizations may lead nonreversible functions for encryption and thus make the decryption process impossible \cite{Solak2008}. The second defect implies a chaotic image encryption method could be completely nonchaotic and thus vulnerable to attacks \cite{li2003security}. The third defect shows these chaotic image encryption method may be broken using existing tools and methods after a long-term observation \cite{Alvarez2007424,Alvarez2003334}. Besides chaotic image encryption methods, nonchaotic image encryption methods are also researched by using various random-like patterns \eg cellular automata \cite{ChenCA2007}, wave transmission model \cite{Liao2010}, Sudoku matrices \cite{wu:77080P},  P-Fibonacci transform \cite{Zhou2012PFibonacci}, and magic cubes \cite{MagicCube2005,MagicCube2006}. Although nonchaotic image encryption methods eliminate drawbacks of chaotic encryption methods especially those round-off errors, many of them do not have good confusion and diffusion properties \cite{ShannonCipher} as in chaotic encryption methods, due to the fact that the used random-like patterns are not truly random-like.

To address above mentioned problems in bit-level image encryption, we introduce a new Latin square image cipher in this paper. It is a nonchaotic image cipher and directly defined on finite numbers, and thus can be effectively implemented with accuracy and ease in software and hardware. LSIC uses a $256$-bit encryption key to generate keyed Latin squares for encryption/decryption, whose key space is large enough to resist today's brute force attacks and could be expended to even larger sizes if needed. LSIC contains three Latin square based encryption primitives, namely \textit{Latin Square Whitening, Latin Square Substitution} and \textit{Latin Square Permutation}, all of which are dependent on a keyed $256\times 256$ Latin square. Hence, LSIC is very sensitive to key changes. Further, these encryption primitives construct LSIC in a SPN \cite{CryptographyBook}, a structure proven to be effective and efficient to encrypt images with good confusion and diffusion properties \cite{ShannonCipher}. In plus, we integrate the probabilistic encryption \cite{springerlink:10.1007/3-540-48405-1_34,springerlink:10.1007/3-540-39799-X_12} in LSIC by embedding random noise in the least significant bit-plane of a plaintext image. Consequently, encrypting a plaintext image multiple times will generate distinctive ciphertext images even though the encryption key is unchanged. This property helps LSIC to achieve a higher level of security and it completely prevents an adversary to sense identical plaintext images by observing identical ciphertext images. We demonstrate the robustness and effectiveness of LSIC using extensive theoretical analysis, simulation results and comparisons to peer algorithms.

The rest of the paper is organized as follows: Section II gives a brief review on preliminary materials. Section III introduces the new Latin square image cipher including its key schedule, probabilistic encryption, and Latin square based encryption primitives. Section IV discusses the simulation database information, extensive simulations results and cipher encryption and decryption speed, Section V analyzes the cipher security theoretically and experimentally under various attacks, and finally Section VI concludes the paper and give discussions on open questions in LSIC.

\section{Background}
\subsection{Latin Squares}
A Latin square of order $N$ is an $N\times N$ array filled with a symbol set of $N$ distinctive elements, with each symbol appears exactly once in each row and each column. The name \textit{Latin Square} is motivated by the mathematician \textit{Leonhard Euler}, who used Latin characters as symbols.

Mathematically, we can define a Latin square $L$ of order $N$ via a tri-tuple function $f_L$ of $(r,c,i)$ as follows
\begin{equation}\label{eqn:LatinSquare}
    f_L(r,c,i) = \left\{\begin{array}{ccc} 1&,& L(r,c) = S_i\\ 0 &,& Otherwise\end{array}\right.
\end{equation}
where $r$ denotes the row index of an element in $L$ with $r\in \mathbb{N} = \{0,1,\cdots, N-1\}$; $c$ denote the column indx of an element in $L$ with $c\in \mathbb{N}$; $i$ denotes the symbol index of an element in $L$ with $I\in \mathbf{N}$; and $S_i$ is the $i$th symbol in the symbol set $\mathbb{S} = \{S_0, S_1, \cdots, S_{N-1}\}$.

Therefore, if $L$ is a Latin square of order $N$, then
\begin{itemize}
  \item for arbitrary $c,i\in \mathbb{N}$, we have
  \begin{equation}\label{eqn:LatinProp1}
  \sum\limits_{r = 0}^{N-1}{f_L(r,c,i)} = 1
  \end{equation}
  \item for arbitrary $r,i\in \mathbb{N}$, we have
  \begin{equation}\label{eqn:LatinProp2}
  \sum\limits_{c = 0}^{N-1}{f_L(r,c,i)} = 1
  \end{equation}
\end{itemize}
which implies that each symbol appears exactly once in each row and each column in $L$.

\begin{figure}[h]
\centering
\scriptsize
  \begin{minipage}[b]{.12\linewidth}
    \centerline{\includegraphics[width=.8\linewidth]{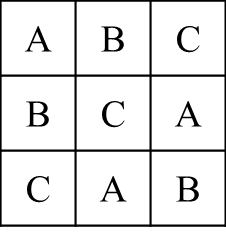}}
    \centerline{(a)  $3\times 3$}
  \end{minipage}
  \begin{minipage}[b]{.18\linewidth}
    \centerline{\includegraphics[width=.8\linewidth]{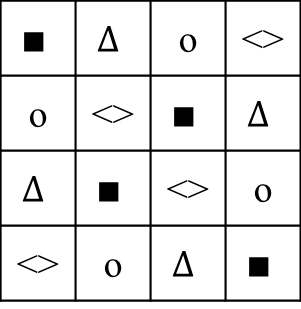}}
    \centerline{(b)  $4\times 4$}
  \end{minipage}
  \begin{minipage}[b]{.22\linewidth}
    \centerline{\includegraphics[width=.8\linewidth]{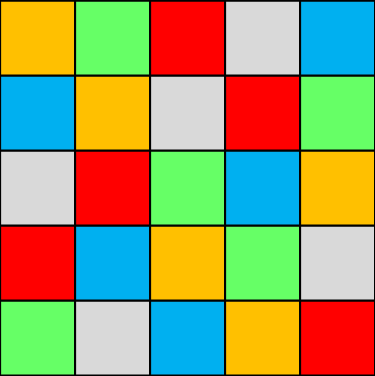}}
    \centerline{(c)  $5\times 5$}
  \end{minipage}
  \begin{minipage}[b]{.35\linewidth}
    \centerline{\includegraphics[width=.8\linewidth]{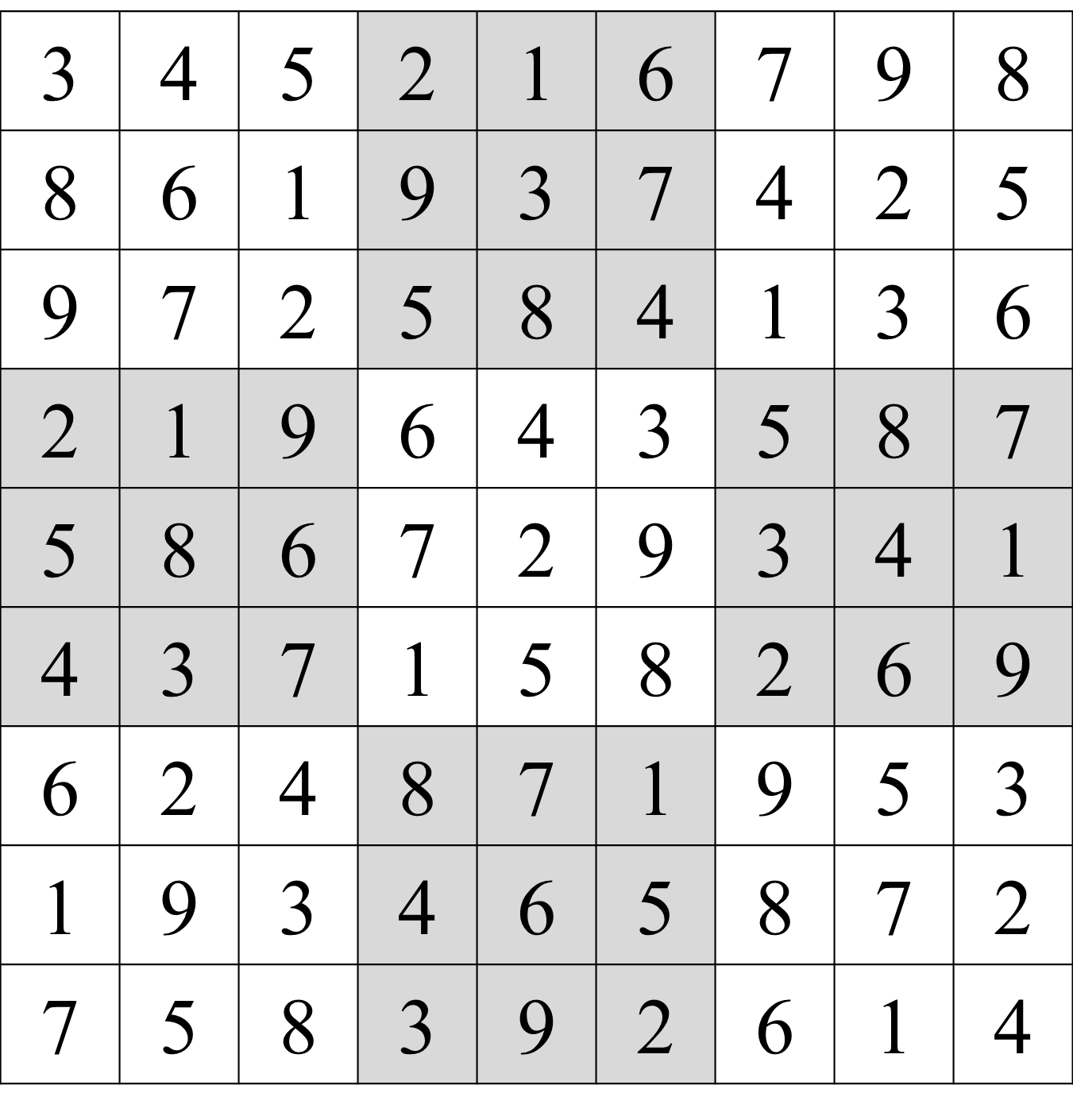}}
    \centerline{(d)  $9\times 9$}
  \end{minipage}
  \caption{ Latin square examples}\label{fig:LatinSquare}
\end{figure}

Fig. \ref{fig:LatinSquare} shows examples of Latin squares at different orders with various symbol sets. It is worthwhile to note that the popular \textit{Sudoku} puzzle \cite{SudokuSci,wu:77080P} is also a special case of Latin square with additional block constraint as shown in Fig. \ref{fig:LatinSquare}(d).

Throughout of the paper, we are interested in $N\times N$ Latin squares with the symbol set of integers from $0$ to $N-1$, \ie $\mathbb{S} = \{0,1,\cdots,N-1\}$.
\subsection{Latin Square Generator}
Although Latin squares can be generated via a variety of means, for the sake of simplicity we use Algorithm 1 described below for Latin square generation in the paper.
\begin{algorithm}
\caption{\textbf{A Latin Square Generator $L = LSG(Q_1,Q_2)$}}
\scriptsize
\begin{algorithmic}
\REQUIRE $Q_1$ and $Q_2$ are two length-$N$ sequences
\ENSURE $L$ is a Latin square of order $N$
\STATE ${Q_{seed}} = SortMap(Q_1)$
\STATE ${Q_{shift}} = SortMap(Q_2)$
\FOR{$r = 0:1:N-1$}
\STATE $L(r,:) = \textrm{RowShift}(Q_{seed},Q_{shift}(r))$
\ENDFOR
\end{algorithmic}
\end{algorithm}

In Algorithm 1,  both $Q_1$ and $Q_2$ are length-$N$ sequences from a pseudo-random number generator (PRNG), \eg Linear Congruential Generators (LCG) \cite{press2007numerical}; ${SortMap}(Q)$ is a function which finds the index mapping between a sequence $Q$ and its sorted version $Q^*$ in the ascending order; and $RowShift(Q,v)$ ring shifts the sequence $Q$ with $v$ elements towards left.

For example, if we want to generate a $4\times 4$ Latin square $L$ with
$$Q_1 = [.1,.6,.9,.7] \textrm{ and }Q_2=[.3,.9,.4,.2]$$
Then function ${SortMap}(.)$ first calculates the sorted version of the input sequence and obtains
$$Q_1^* = [.1, .6, .7, .9]\textrm{ and }Q_2^* = [.2, .3, .4, .9]$$
it then compares $Q_1$ with $Q_1^* $ and $Q_2$ with $Q_2^* $ and obtains the element mapping sequences as
$$Q_{seed} = {SortMap}(Q_1) = [0,1,3,2]$$and $$Q_{shift} = {SortMap}(Q_2)= [3,0,2,1]$$
, where the permutation sequences $Q_{seed}$ and $Q_{shift}$ indicate for $i\in \{0,1,2,3\}$
$$Q_1^* \left(Q_{seed}(i)\right) = Q_1\left(i\right)\textrm{ and } Q_2^* \left(Q_{shift}(i)\right) = Q_2\left(i\right)$$
Finally, function $\textrm{RowShift}(Q,v)$ left shifts $Q_{seed}$ with the amount $v = Q_{shift}(r)$ indicated by the $r$th element in $Q_{shift}$, and assign this row to be the $r$th row in $L$. We therefore have the $4\times 4$ Latin square $L$:
\begin{equation*}
    L = \left[
      \begin{array}{cccc}
        2 & 0 & 1 & 3 \\
        0 & 1 & 3 & 2 \\
        3 & 2 & 0 & 1 \\
        1 & 3 & 2 & 0 \\
      \end{array}
    \right]
\end{equation*}
whose $1$st row is obtained by left shifting $Q_{seed}$ for $3$ units; $2$nd row is obtained by left shifting $Q_{seed}$ for $0$ unit; $3$rd row is obtained by left shifting $Q_{seed}$ for $3$ units; and $4$th row is obtained by left shifting $Q_{seed}$ for $1$ unit. It is noticeable that this shifting amount sequence $\{3,0,2,1\}$ is indeed $Q_{shift}$.

\subsection{Substitution-Permutation Network}
In cryptography, an input message and its corresponding output message of a cryptsystem are referred to as \textit{plaintext} and \textit{ciphertext}, respectively. A substitution-permutation network is a cipher structure composed of a number of substitution and permutation ciphers with multiple iterations. This structure is widely used in many well-known block ciphers, \eg Rijndael \ie AES \cite{AES}, and ensures good confusion and diffusion properties \cite{ShannonCipher}.
\begin{figure}[h]
  \centerline{\includegraphics[width=.8\linewidth]{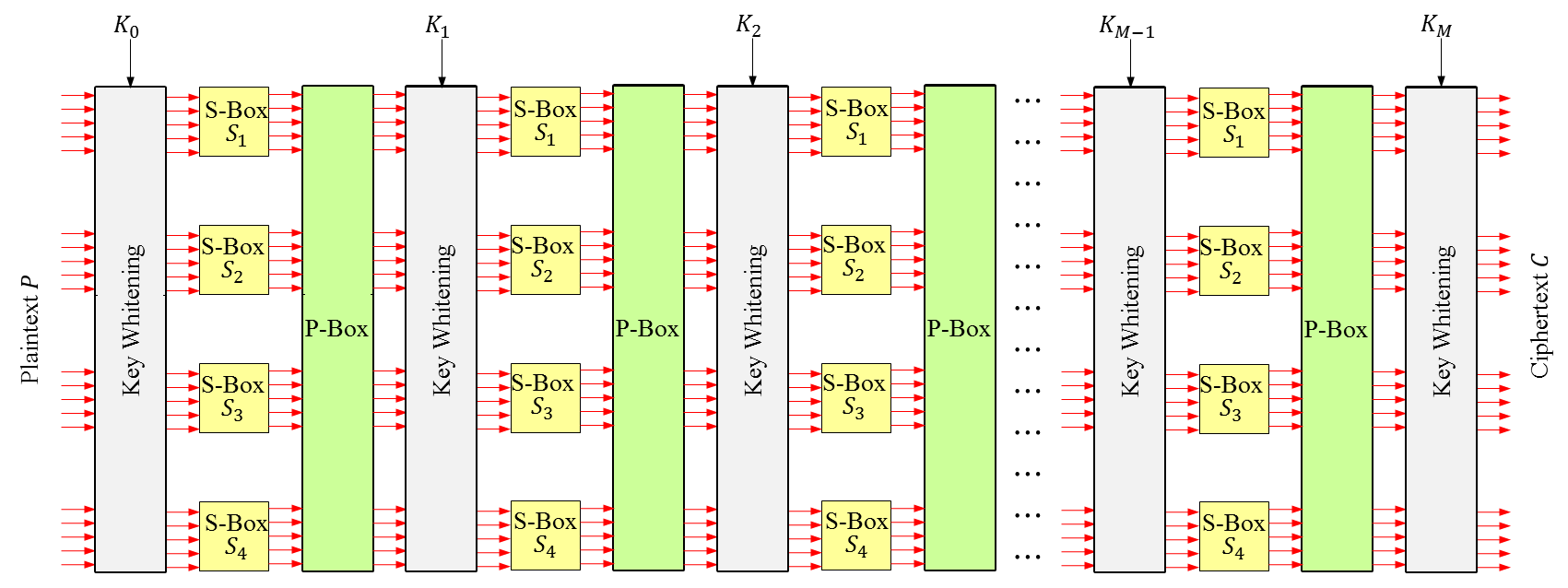}}
  \caption{ A $M$-round substitution-permutation network for block ciphers}\label{fig:SPN}
\end{figure}

A typical $M$-round SPN for block ciphers has a structure shown in Fig. \ref{fig:SPN}. Conventionally in a SPN, \textit{plaintext}, commonly in the form of a bit stream and denoted as $P$, is the original message to be encrypted; $\textit{Key Whitening}$ denotes an operation to mix the plaintext $P$ with a round key; $\textit{S-Box}$ denotes a substitution-box, which maps one input byte to another in a deterministic way; \textit{P-Box} denotes a permutation-box, which shuffles bit positions within the input bit stream in a deterministic way; and \textit{ciphertext} denotes the output bit stream $C$, which is an encrypted message by the SPN. The decryption process of a SPN cipher is simply to reverse the arrow directions of all processing and to use inverse $\textit{S-Box}$, and inverse $\textit{P-Box}$ instead.

The classic SPN ciphers are able to obtain good Shannon's confusion and diffusion properties \cite{ShannonCipher}. For the diffusion property: if one changes one bit in plaintext $P$, the corresponding ciphertext $C$ changes in many bits. This one-bit change results in a different byte after passing through a $\textit{S-Box}$, then leads more byte changes after passing through a $\textit{P-Box}$, so on and so forth in each cipher round. Finally, one-bit change leads to significant changes in ciphertext $C$. The confusion property is the similar to the diffusion property. One bit change in encryption key $K,$ will spread over all bits and result significant changes in ciphertext $C$.

\subsection{Psychovisual Redundancy in Image}
As a typical type of two-dimensional data, digital images contain many kinds of redundancies \cite{DIP}:  1) coding redundancy, which requires to express a pixel with the number of bits than the optimal number; 2) temporal
or spatial redundancy, which implies strong correlations between pixels within a neighborhood; and 3) psychovisual redundancy, which is visually negligible details in the human vision system. The first two redundancies about real visual information are commonly removed by lossless compression techniques, such as Huffman coding and lossless predictive coding. In contrast, psychovisual redundancy within an image is not essential for normal visual processing, and thus it is commonly removed by lossy compression techniques, such as quantization techniques used in JPEG \cite{JPEG}. Besides image compression, psychovisual redundancy is also commonly used for data hiding, especially for the least significant bit-plane (LSB) data hiding techniques \cite{5411758,4908965,4598830}, which encode secret information within these unintelligible redundancies for human visual perception.

\begin{figure}[h]
\scriptsize
\centering
\begin{minipage}[b]{.2\linewidth}
  \centerline{\includegraphics[width=.95\linewidth]{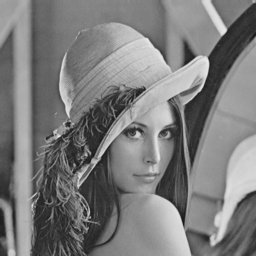}}
  \centerline{(a)}
\end{minipage}\hfill
\begin{minipage}[b]{.2\linewidth}
  \centerline{\includegraphics[width=.95\linewidth]{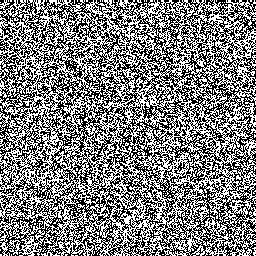}}
  \centerline{(b)}
\end{minipage}\hfill
\begin{minipage}[b]{.2\linewidth}
  \centerline{\includegraphics[width=.95\linewidth]{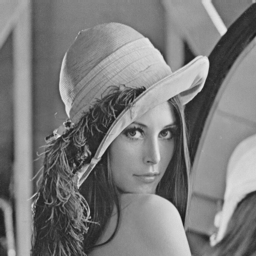}}
  \centerline{(c)}
\end{minipage}\hfill
\begin{minipage}[b]{.2\linewidth}
  \centerline{\includegraphics[width=.95\linewidth]{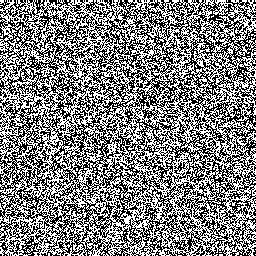}}
  \centerline{(d)}
\end{minipage}\hfill
\begin{minipage}[b]{.2\linewidth}
  \centerline{\includegraphics[width=.95\linewidth]{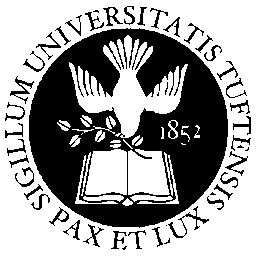}}
    \centerline{(e)}
\end{minipage}\hfill
  \caption{Psychovisual redundancy used in LSB data hiding - (a) original image \textit{Lenna} $O$; (b) $O_{LSB}$; (c) secret embedded image $E$; (d) $E_{LSB}$; and (e) secret information $I = O_{LSB}\bigoplus E_{LSB}$.}\label{fig:lsb}
\end{figure}

Fig. \ref{fig:lsb} shows an example of using the psychovisual redundancy for data hiding in the least significant image bit-plane. The secret embedded image Fig. \ref{fig:lsb}-(c) contains identical bit-planes to those in the original image -(a) , except its least significant bit-plane is modified by bitwise XORing $O_{LSB}$ and secrete binary image $I$. As can be seen from this example, the secret embedded image is visually indistinguishable from the original one, although the least significant bit-plane has been dramatically changed.

Consequently, psychovisual redundancy in an image provides us a feasible way of introducing true randomness in an image without harming the image visual quality and expanding the image size for image encryption. More details are discussed in future sections.

\section{Latin square image cipher}
\subsection{Overview}
To standardize the encryption/decryption processing, the cipher processing block is set to a $256\times 256$ grayscale block, \ie its pixel intensity is denoted as a 8-bit byte. In the rest of the paper, we use $P$ to denote a $256\times 256$ plaintext image block, $C$ to denote a corresponding ciphertext image block of $P$, $L$ to denote a keyed Latin square of order $256$, and $K$ to denote a $256$-bit encryption key.

\begin{figure*}[!h]
\scriptsize
\centering
  \begin{minipage}[c]{.9\linewidth}
  \centerline{\includegraphics[width=1\linewidth]{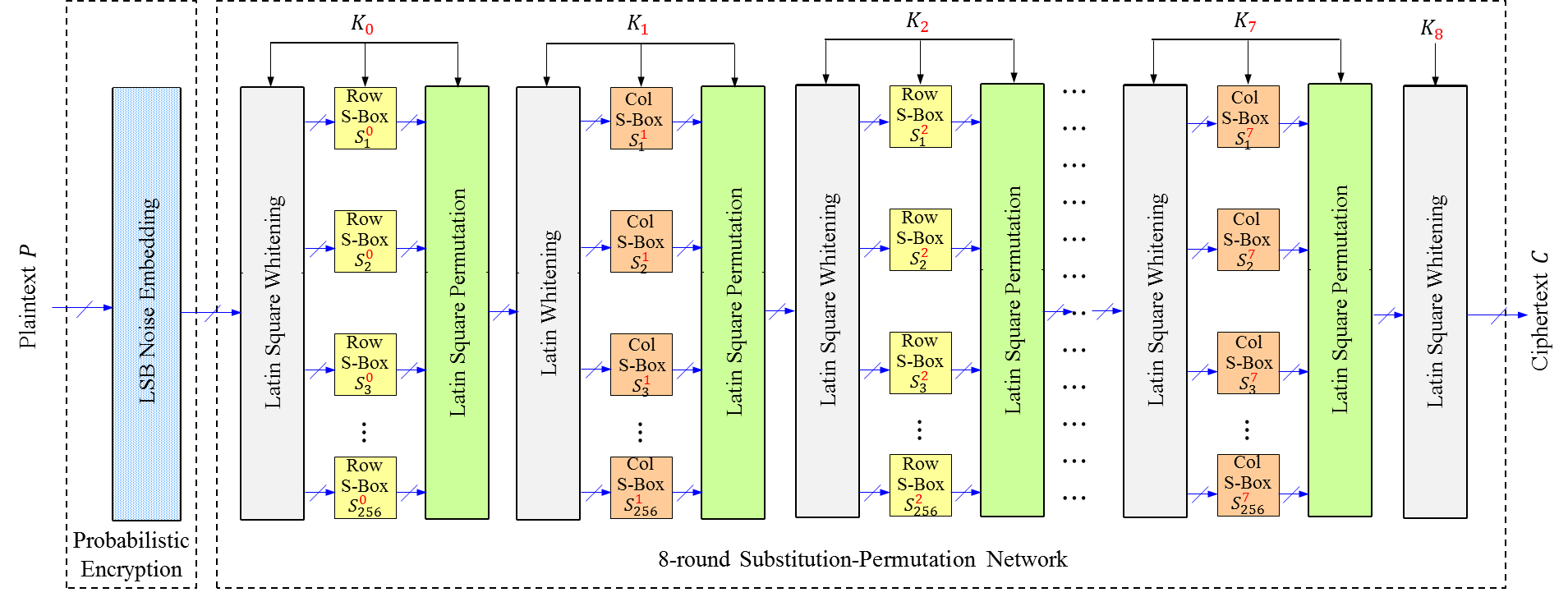}}
  \centerline{(a)}
  \end{minipage}\hfill
  \caption{Overview of the Latin square image cipher}\label{fig:SPN_Latin}
\end{figure*}

The new proposed Latin square image cipher is of a SPN structure with eight rounds as shown in Fig. \ref{fig:SPN_Latin}. It is composed of the probabilistic encryption stage \textit{noise embedding in LSB} and the SPN stage containing three encryption primitives $\LSW$, $\LSS$ and $\LSP$. It is worthwhile to note that this SPN is of a loom-like structure designed for image data, which encrypts plaintext image along rows and columns iteratively. In the rest of this section, we will discuss these stages and the detail encryption/decryption algorithms for LSIC.

\subsection{LSB Noise Embedding}
Probabilistic encryption \cite{Fujisaki:1999:SIA:646764.706343,Galil:1986:SPE:18262.25394} means to use randomness in a cipher, so that this cipher is able to encrypt one plaintext with the exact same encryption key to distinctive ciphertexts. It is well known that such randomness is crucial to achieve semantic security \cite{646128}.

In this paper, we introduce such randomness by embedding noise in the least significant bit-plane of an image. More specifically, we XOR a randomly generated $256\times 256$ bit-plane with the least significant bit-plane of the plaintext image, where the generation of this random bit-plane is completely independent of the encryption key.

\begin{figure}[h]
\scriptsize
\centering
\begin{minipage}[b]{.68\linewidth}
  \begin{minipage}[b]{.33\linewidth}
    \centerline{\includegraphics[width=.85\linewidth]{Lenna}}
        \centerline{(a)}
    \centerline{\includegraphics[width=.95\linewidth]{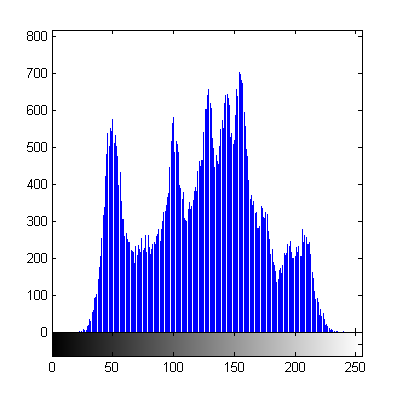}}
  \end{minipage}\hfill
  \begin{minipage}[b]{.33\linewidth}
    \centerline{\includegraphics[width=.85\linewidth]{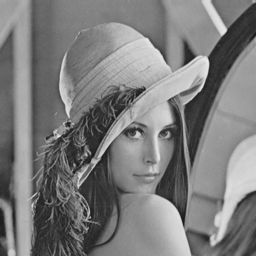}}
    \centerline{(b)}
        \centerline{\includegraphics[width=.95\linewidth]{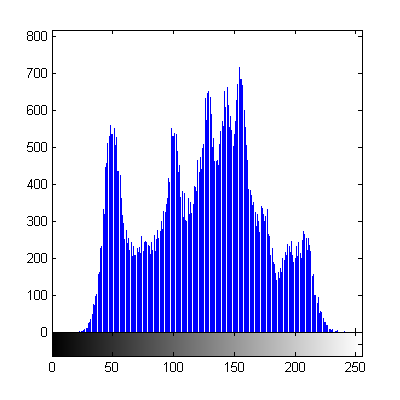}}
  \end{minipage}\hfill
  \begin{minipage}[b]{.33\linewidth}
    \centerline{\includegraphics[width=.85\linewidth]{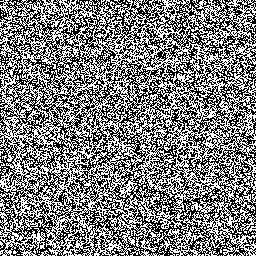}}
    \centerline{(c)}
        \centerline{\includegraphics[width=.95\linewidth]{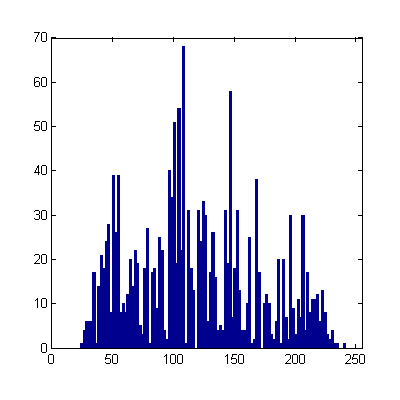}}
  \end{minipage}\hfill
\end{minipage}\hfill
    \caption{Noise embedding in LSB - (a) plaintext \textit{Lenna} $P$ with histogram (b) noise embedded plaintext $P'$ with histogram, and (c) $|P-P'|$ with the difference of histograms}\label{fig:permutation2}
\end{figure}

Fig. \ref{fig:permutation2} shows an example of LSB noise embedding. Once again, these introduced noise in LSB does not affect any image visual quality from the point view of human visual perceptibility. However, any slight change in plaintext here will lead to significant changes in ciphertext after it is encrypted by by the SPN.

\subsection{Key Translation}
In conventional block ciphers, keys are directly used without translation, \eg in key whitening process, but the proposed LSIC uses a $256$-bit encryption key $K$ with key translation to eight key-dependent Latin squares of order $256$ before using in LSIC. Specifically speaking, for a given $256$-bit encryption key $K$, we
\begin{enumerate}
  \item Divide the encryption key $K$ using function $SubKeyDiv$ into eight 32-bit subkeys, \ie
        $$K = [k_0,k_1,\cdots,k_7]$$
  \item Generate pairs of pseudo-random sequences $(Q_1^0,Q_2^0),(Q_1^1,Q_2^1),\cdots,(Q_1^8,Q_2^8)$, each pair with $2\times 256$ elements by using PRNGs by feeding these subkeys as seeds.
  \item Generate key-dependent Latin squares \ie $L_0,L_1,\cdots, L_8$ with the order of $256$ by feeding these pseudo-random sequences in Algorithm 1. Namely, $\forall n\in \{0,1,\cdots,8\}$, we have $$L_n = LSG(Q_1^n,Q_2^n)$$
\end{enumerate}
The first two steps can be realized via Algorithm 2 with encryption key $K$ and $M=8$.
\begin{algorithm}
\caption{\textbf{Key Dependent Sequence Generator $\left(\mathbf{Q_1},\mathbf{Q_2}\right) = KDSG(Key,M)$}}
\scriptsize
\begin{algorithmic}
\REQUIRE $K$ is a $256$-bit key
\REQUIRE $n$ is a nonnegative integer
\ENSURE $\mathbf{Q_1} = \{Q_1^0,\cdots,Q_1^{M}\}$ and $\mathbf{Q_2} = \{Q_2^0,\cdots,Q_2^{M}\}$ are $n$-element set of random sequences, each of a length $256$.
\STATE $K_0 = K$
\FOR{$n = 0:1:M$}
\STATE $[k_0,k_1,\cdots,k_7] = SubKeyDiv(K_n)$
\FOR{$i = 0:1:8$}
\STATE $q^i(0) = PRNG(k_i)$\footnotemark[1]
\FOR{$j = 1:1:63$}
\STATE $q^i(j) = PRNG(q^i(j-1))$
\ENDFOR
\ENDFOR
\STATE $Q^n_1=\left[q^0(0:31),q^1(0:31),\cdots,q^{7}(0:31)\right]$ \%\footnotemark[2]
\STATE $Q^n_2=\left[q^0(32:63),q^1(32:63),\cdots,q^{7}(32:63)\right]$
\STATE $K_{n+1} = \left[q^0(63),q^1(63),\cdots,q^{7}(63)\right]$
\ENDFOR
\end{algorithmic}
\end{algorithm}

\footnotetext[1]{Any PRNG can be used here by taking the key as its seed.}
\footnotetext[2]{$q^i(j_1:j_2)$ denotes a vector of pseudo-random numbers with elements $[q^i(j_1),q^i(j_1+1),\cdots,q^i(j_2-1),q^i(j_2)]$.}


All PRNGs can be used in Algorithm 2, but not all of them are secure for cryptography, for example, the linear congruential generator \cite{press2007numerical}. Using cryptographic secure PRNGs in Algorithm 2 can further enhance the cipher security, for example, PRNGs from eSTREAM project \footnotemark[3] with nonce \cite{bjrstad2008introduction}.
\footnotetext[3]{The eSTREAM project is available under \url{http://www.ecrypt.eu.org/stream/} \astoday.}

\subsection{Latin Square Whitening}
In the conventional SPN for block ciphers, the \textit{Whitening} stage normally mixes a plaintext message $P$ with a round key, \eg XOR operation in \cite{DES,AES}, such that
\begin{itemize}
  \item The statistics of the plaintext message $P$ is redistributed after mixing.
  \item The relationship between ciphertext and encryption key is very complicated and involved.
\end{itemize}

In image encryption, a plaintext message is an image block, $P$, composed of a number of pixels. Each pixel is represented by several binary bits (a byte). Therefore, XOR whitening scheme become inefficient for image data, in the sense that it requires to extend an encryption key to be a equal size to a plaintext image, and to impose bitwise XOR to byte pixels. And this type of image encryption is called a naive algorithm in \cite{Yang2004}. Since the objective of key whitening is to mix plaintext data with encryption keys, we therefore define whitening as a transposition cipher \cite{Stinson} over the finite field $GF(2^8)$ for image data, as shown in Eq. \eqref{eqn:whitening}
\begin{equation}\label{eqn:whitening}
y = [x+l]_{2^8}
\end{equation}
where $x$ is a byte in plaintext, $l$ is a corresponding byte in the keyed Latin square, $y$ is the whitening result and $[.]_{2^8}$ denotes the computations over $GF(2^8)$. Above whitening process can be easily reversed by applying Eq. \eqref{eqn:whitening2}.
\begin{equation}\label{eqn:whitening2}
x = [y+l]_{2^8}
\end{equation}

In image encryption, plaintext byte $x$ is a pixel, say it is located at the intersection of $r$th row and $c$th column \ie $x = P(r,c)$. Now let $l = L(r,c)$ be an element located at the corresponding position in the keyed Latin square $L$, and $y$ be the ciphertext byte with $y = C(r,c)$, then we have the pixel-level equation
\begin{equation}\label{eqn:LSW0}
\left\{\begin{array}{l}
   C(r,c)= \left[\textrm{SR}\left(P(r,c),[D]_{3}\right)+L(r,c)\right]_{2^8}\\
   P(r,c) = \textrm{SR}\left([C(r,c)+L(r,c)]_{2^8}, [D]_{3}\right)
    \end{array}\right.
\end{equation}
where symbol $n$ denotes the current round number ($n\in [0,7]$), $D = L(0,0)$ is the rotating parameter, and SR denotes the \textit{spatial rotating} function $(X,d)$ rotates an image $X$ according to different values of the direction $d$ as defined in Eq. (\ref{eqn:SR})
\begin{equation}\label{eqn:SR}
    Y = \textrm{SR}(X,d) = \left\{\begin{array}{r}
    X,\textrm{ if }d = 0 \\
    \textrm{Flip } X \textrm{ up}\rightarrow \textrm{down},\textrm{ if }d = 1\\
    \textrm{Flip } X \textrm{ left}\rightarrow \textrm{right},\textrm{ if }d = 2\\
    \end{array}
    \right.
\end{equation}
Notice that if $Y = $SR$(X,d)$, then the following identity always holds
\begin{equation}\label{eqn:SR2}
    X = \textrm{SR}(Y,d)
\end{equation}

Apply key whitening for all pixels using the pixel-level Eq. \eqref{eqn:LSW0}, the \LSW (LSW) in the image-level then can be denoted as
\begin{equation}\label{eqn:LSW}
LSW:\left\{\begin{array}{l}
    C =Ecr_w(L,P,D)\\
    P = Dcr_w(L,C,D)
    \end{array}\right.
\end{equation}
Therefore, we can restore the plaintext image block $P$ from the ciphertext image block $C$ using Eq. \eqref{eqn:LSW}.

\begin{figure}[h]
\scriptsize
\centering
\begin{minipage}[b]{.68\linewidth}
  \begin{minipage}[b]{.33\linewidth}
    \centerline{\includegraphics[width=.85\linewidth]{Lenna}}
    \centerline{(a)}
    \centerline{\includegraphics[width=.95\linewidth]{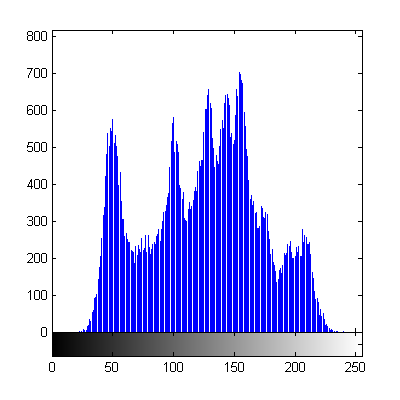}}
  \end{minipage}\hfill
  \begin{minipage}[b]{.33\linewidth}
    \centerline{\includegraphics[width=.85\linewidth]{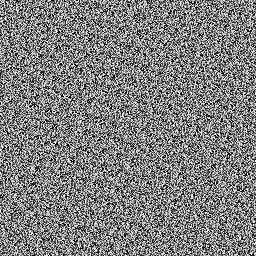}}
    \centerline{(b)}
    \centerline{\includegraphics[width=.95\linewidth]{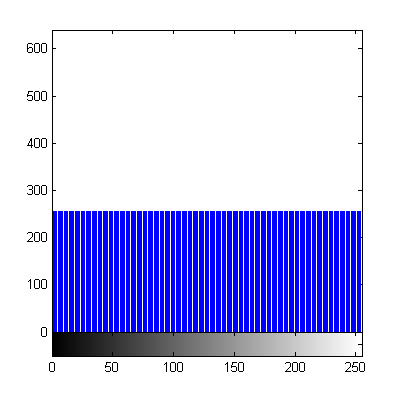}}
  \end{minipage}\hfill
  \begin{minipage}[b]{.33\linewidth}
    \centerline{\includegraphics[width=.85\linewidth]{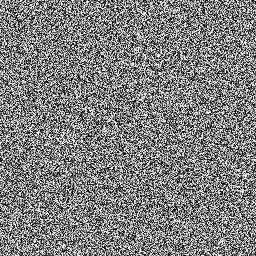}}
    \centerline{(c)}
    \centerline{\includegraphics[width=.95\linewidth]{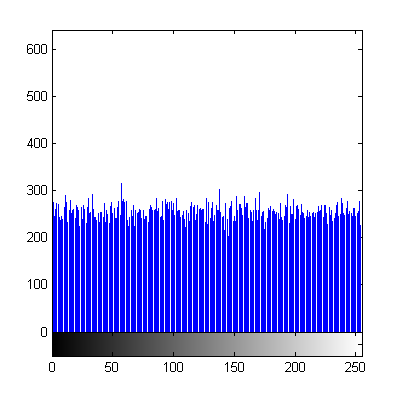}}
  \end{minipage}\hfill
\end{minipage}\hfill
    \caption{A \LSWs example - (a) plaintext \lenna $P$, (b) reference Latin square $L$, and (c) ciphertext $C = Ecr_w(L,P,0)$}\label{fig:Whitening}
\end{figure}

Fig. \ref{fig:Whitening} shows an example of \LSW, where the first row shows images and the second row shows corresponding histograms of these images. From this example, it is easy to verify that the ciphertext image after the \LSWs is unrecognizable and its pixels are redistributed to uniform-like.

\subsection{Latin Square Row and Column Bijections}
Since Eqs. (\ref{eqn:LatinProp1}) and (\ref{eqn:LatinProp2}) hold, each row and each column in a Latin square $L$ of order $N$ is a permutation of the integer number sequence $[0,1,\cdots, N-1]$, we can define bijections (one-to-one and onto mapping) by mapping this integer number sequence to either a row or a column in a Latin square, which is a permutated sequence of the integer number sequence. In other words, we are able to construct forward and inverse row mapping functions (FRM and IRM) with respect to the $r$th row in $L$ as shown in Eq. (\ref{eqn:LatinSRow}), and also forward and inverse column mapping functions (FCM and ICM) with respect to the $c$th column in $L$ as shown in Eq. (\ref{eqn:LatinSCol}), where $x$ and $y$ denote the input and output of the mapping functions, respectively.
\begin{equation}\label{eqn:LatinSRow}
    \left\{\begin{array}{l}
    y = FRM(L,r,x) = L(r,x)\\
    x = IRM(L,r,y) = \arg\max\limits_{z\in \mathbb{N}}{\left(f_L(r,z,y)\right)}
    \end{array}\right.
\end{equation}
\begin{equation}\label{eqn:LatinSCol}
    \left\{\begin{array}{l}
    y= FCM(L,x,c)=L(x,c)\\
    x= ICM(L,y,c)= \arg\max\limits_{z\in \mathbb{N}}{\left(f_L(z,c,y)\right)}
    \end{array}\right.
\end{equation}
where $f_L$ is the tri-tuple function defined in Eq. (\ref{eqn:LatinSquare}). Its maximum is equal to 1, \ie $f_L(r,x,y)=1$, only for the column number $x$ satisfying the constraint, $L(r,x) = y$. Further, we have row mapping identities hold for arbitrary $x$ and $y$ within a Latin square $L$:
\begin{equation}\label{eqn:idn1}
\left\{\begin{array}{l}
IRM(L,r,FRM(L,r,x)) = x\\
FRM(L,r,IRM(L,r,y)) = y
\end{array}\right.
\end{equation}
Similarly, we also have column mapping identities as follows:
\begin{equation}\label{eqn:idn2}
\left\{\begin{array}{l}
ICM(L,FCM(L,x,c),c) = x\\
FCM(L,ICM(L,y,c),c) = y
\end{array}\right.
\end{equation}

\begin{figure}
\centering
\scriptsize
  \includegraphics[width=.7\linewidth]{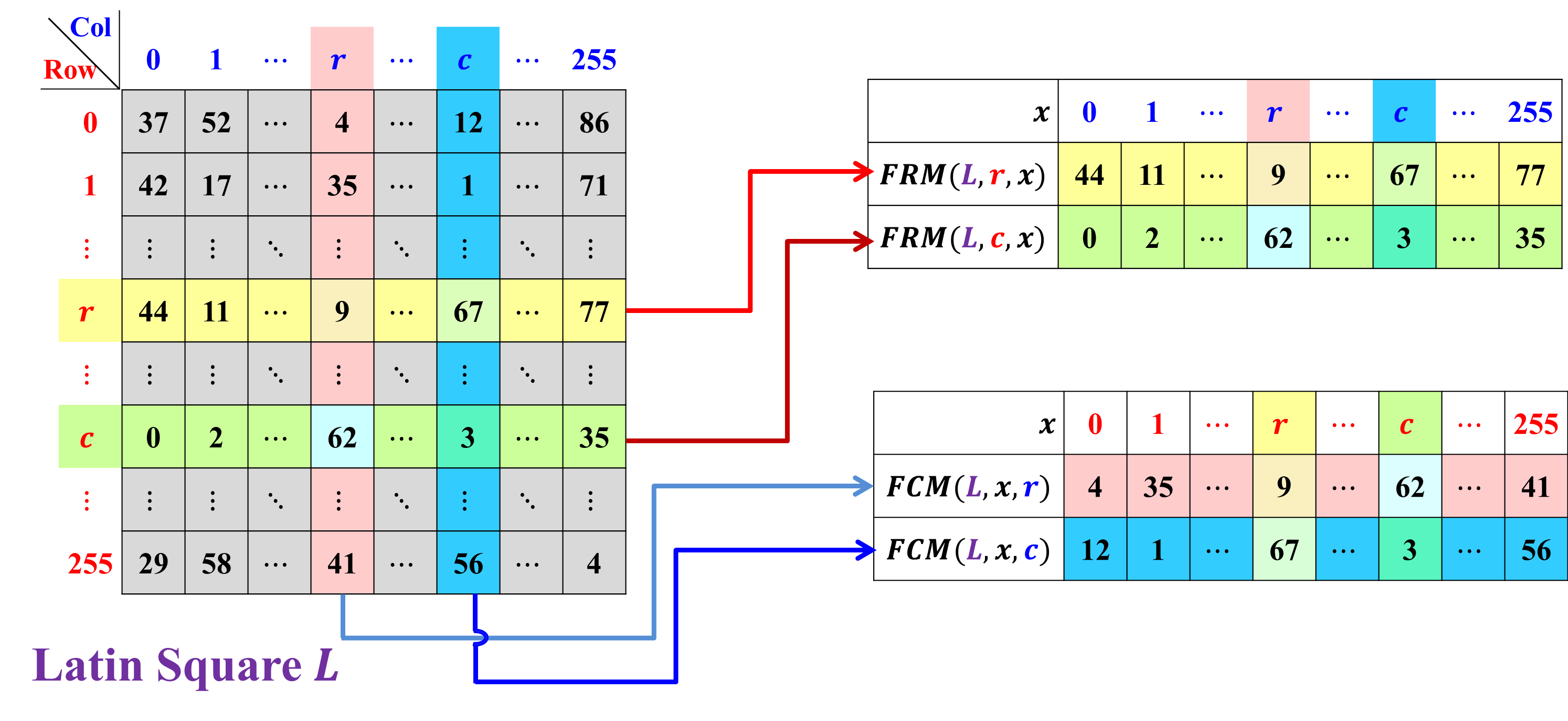}\\
  \caption{Examples of building forward row mappings and forward column mappings using a Latin square}\label{fig:LatinMapping}
\end{figure}

Fig. \ref{fig:LatinMapping} shows FRM and FCM functions defined by a Latin square. As can be seen, given a row number $r$, the effect of forward row mapping is to use this Latin square as a look-up table and to find the corresponding element in row $r$. Similarly, given a column number $c$, the effect of forward column mapping to use this Latin square as a look-up table and to find the corresponding element in column $c$. Such nice property is directly from the constructional constraint in a Latin square.

Since a $N$th order Latin square has $N$ rows and $N$ columns, there are $N$ bijections from $N$ rows and another $N$ bijections from $N$ columns in the Latin square. It is well-known that a bijection can be directly used as a P-Box \cite{CryptographyBook} and can also serve as a S-Box \cite{CryptographyBook}. We therefore are able to construct S-Boxes and P-Boxes from a Latin square.
\subsection{Latin Square Substitution}
An S-Box in cryptography is a basic component performing byte substitution. Each S-Box can be defined as a bijection, also known as a one-to-one and onto mapping. In image encryption, an image pixel is commonly represented as a byte, \ie a sequence of bits. For example, 8-bit grayscale image has 256 gray intensity scales with each intensity scale represented in an 8-bit sequence.

Because of the existence of FRM/IRM and FCM/ICM bijections in a Latin square, we are able to perform byte substitution in an image cipher using bijections from rows and columns in a Latin square. The substitution with respect to a row in a Latin square is called \LSRS (LSRS) in this paper:
\begin{eqnarray}\label{eqn:LatinRow}
    LSRS:\left\{\begin{array}{r}
    C = Ecr_s^{row}(L,P)\\
    P = Dcr_s^{row}(L,C)
    \end{array}\right.
\end{eqnarray}
In regard to pixel-level function of LSRS, each ciphertext byte is determined by the FRM function (see Eq. \eqref{eqn:LatinSRow}) using the keyed Latin square $L$ with function parameters given by plaintext bytes and ciphertext bytes as follows
\begin{equation}\label{eqn:LatinRowElement}
       Ecr_s^{row}: C(r,c) = \left\{
    \begin{array}{r}
    FRM\left(L,C(r-1,c),P(r,c)\right),\textrm{ if } r\neq 0\\
    FRM\left(L,0,P(r,c)\right),\textrm{ if } r= 0\
    \end{array}
    \right.
\end{equation}
Clearly, plaintext bytes then can be perfectly restored from ciphertext bytes, if we use IRM instead of FRM as follows
\begin{equation}\label{eqn:LatinRowElement2}
      Dcr_s^{row}:P(r,c) = \left\{
    \begin{array}{rcccc}
    IRM\left(L,C(r-1,c),C(r,c)\right),\textrm{ if } r\neq 0\\
    IRM\left(L,0,C(r,c)\right),\textrm{ if } r= 0\
    \end{array}
    \right.
\end{equation}

Similarly, we use bijections from columns in a Latin square to perform byte substitutions. And this is called \LSCS(LSCS) \ie
\begin{eqnarray}\label{eqn:LatinCol}
LSCS:\left\{\begin{array}{r}
    C = Ecr_s^{col}(L,P)\\
    P = Dcr_s^{col}(L,C)
    \end{array}\right.
\end{eqnarray}
and the corresponding LSCS encryption and decryption process then can be defined as:
\begin{equation}\label{eqn:LatinColElement}
      Ecr_s^{col}:  C(r,c) = \left\{
    \begin{array}{r}
    FCM\left(L,P(r,c),C(r,c-1)\right),\textrm{ if } c\neq 0\\
    FCM\left(L,P(r,c),0\right),\textrm{ if } c= 0\
    \end{array}
    \right.
\end{equation}
\begin{equation}\label{eqn:LatinColElement2}
     Dcr_s^{col}   P(r,c) = \left\{
    \begin{array}{rcccc}
    ICM\left(L,C(r,c),C(r,c-1)\right),\textrm{ if } c\neq 0\\
    ICM\left(L,C(r,c),0\right),\textrm{ if } c= 0\
    \end{array}
    \right.
\end{equation}

\begin{figure}[h]
\scriptsize
\centering
\begin{minipage}[b]{.9\linewidth}
  \begin{minipage}[b]{.25\linewidth}
    \centerline{\includegraphics[width=.85\linewidth]{Lenna}}
    \centerline{(a)}
    \centerline{\includegraphics[width=.95\linewidth]{histLenna}}
  \end{minipage}\hfill
  \begin{minipage}[b]{.25\linewidth}
    \centerline{\includegraphics[width=.85\linewidth]{Latin256}}
    \centerline{(b)}
    \centerline{\includegraphics[width=.95\linewidth]{histLatin256}}
  \end{minipage}\hfill
  \begin{minipage}[b]{.25\linewidth}
    \centerline{\includegraphics[width=.85\linewidth]{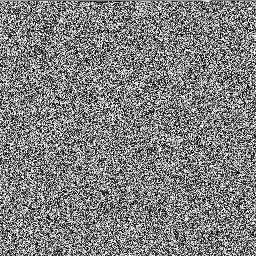}}
    \centerline{(c)}
    \centerline{\includegraphics[width=.95\linewidth]{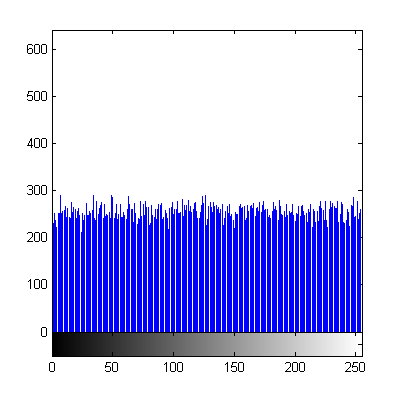}}
  \end{minipage}\hfill
  \begin{minipage}[b]{.25\linewidth}
    \centerline{\includegraphics[width=.85\linewidth]{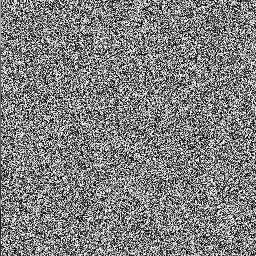}}
    \centerline{(d)}
    \centerline{\includegraphics[width=.95\linewidth]{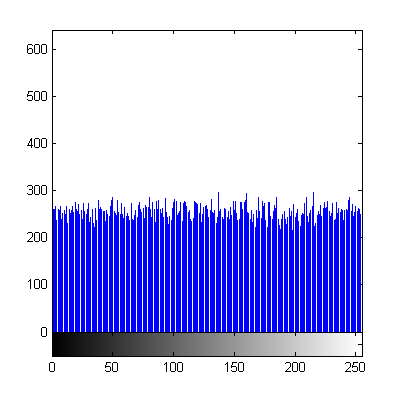}}
  \end{minipage}\hfill
\end{minipage}\hfill
    \caption{A \LSSs example - (a) plaintext $P$: \textit{Lenna Image}, (b) Latin square $L$, (c) ciphertext $C_r = Ecr_s^{row}(L,P)$, and (d) Ciphertext $C_c = Ecr_s^{col}(L,P)$}\label{fig:Substitution}
\end{figure}

Fig. \ref{fig:Substitution} shows encryption results of \LSRSs and \LSCS. As can be seen, the plaintext image block $P$ becomes unrecognizable after applying either LSRS or LSCS. Histogram analysis also shows that the statistics of the pixel intensity changes dramatically after substitution.

The Latin square row/column substitution defined above has excellent diffusion properties. One pixel change in the plaintext $P$ will diffuse to a column/row of pixels after a round of LSRS or LSCS. This diffusion quickly spreads to the entire ciphertext image in several cipher rounds.
\subsection{Latin Square Permutation}
Unlike a S-Box performing byte substitution, a P-Box performs byte shuffling or scrambling. Each P-Box can also be defined as a bijection \cite{CryptographyBook}.

If we consider both input $x$ and output $y$ in FRM and IRM as indices (see Eq. \eqref{eqn:LatinSRow}), then FRM defines a mapping $\{0,1,\cdots,255\}\rightarrow \{0,1,\cdots,255\}$ and IRM defines the corresponding inverse mapping. We therefore are able to define the \textit{Latin square row p-box} (LSRP) with respect to rows in a Latin square $L$ as follows,
\begin{equation}\label{eqn:LSRP}
LSRP:\left\{\begin{array}{l}
    C(r,c_y) = P\left(r,FRM(L,r,c_x)\right)\\
    P(r,c_x) = C\left(r,IRM(L,r,c_y)\right)
    \end{array}\right.
\end{equation}
where $c_x$ and $c_y$ denotes the column indices before and after mapping. Consequently, for any pixel in $P$ and its corresponding pixel in $C$ are in the same row $r$ after LSRP; and only column indices change before and after mapping with relationship $c_y = FRM(L,r,c_x)$ holds.

Similarly, we can also construct \LSCP (LSCP) with respect to columns in a Latin square as
\begin{equation}\label{eqn:LSRC}
LSCP:\left\{\begin{array}{l}
    C(r_y,c) = P\left(FCM(L,r_x,c),c\right)\\
    P(r_x,c) = C\left(ICM(L,r_y,c),c\right)
    \end{array}\right.
\end{equation}

In order to achieve better performance, we construct our \LSPs by cascading LSRPs s and LSCPs as follows
\begin{eqnarray}\label{eqn:LatinPBox}
    C(r,c) = C^*(FCM(L,r,c),c)\\
    C^*(r,c) = P(r,FRM(L,r,c))
\end{eqnarray}
In general, we write this \LSP function as
\begin{eqnarray}\label{eqn:LatinPerm}
    LSP:\left\{\begin{array}{rclcl}
    C =Ecr_p(L,P)\\
    P = Dcr_p(L,C)
    \end{array}\right.
\end{eqnarray}

\begin{figure}[h]
\scriptsize
\centering
\begin{minipage}[b]{1\linewidth}
  \begin{minipage}[b]{.2\linewidth}
    \centerline{\includegraphics[width=.95\linewidth]{Lenna}}
    \centerline{(a)}
  \end{minipage}\hfill
  \begin{minipage}[b]{.2\linewidth}
    \centerline{\includegraphics[width=.95\linewidth]{Latin256}}
    \centerline{(b)}
  \end{minipage}\hfill
  \begin{minipage}[b]{.2\linewidth}
    \centerline{\includegraphics[width=.95\linewidth]{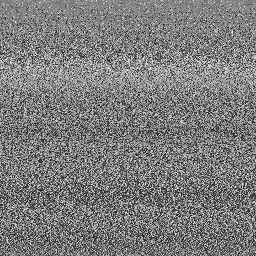}}
    \centerline{(c)}
  \end{minipage}\hfill
    \begin{minipage}[b]{.2\linewidth}
    \centerline{\includegraphics[width=.95\linewidth]{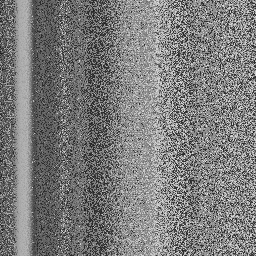}}
    \centerline{(d)}
  \end{minipage}\hfill
    \begin{minipage}[b]{.2\linewidth}
    \centerline{\includegraphics[width=.95\linewidth]{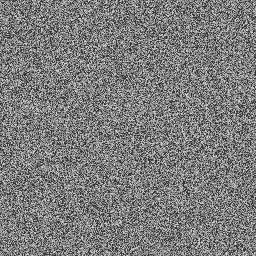}}
    \centerline{(e)}
  \end{minipage}\hfill
\end{minipage}\hfill
    \caption{A \LSPs example - (a) plaintext \lenna $P$, (b) Latin square $L$, (c) ciphertext with only LSRP, (d) ciphertext with only LSCP, and (e) ciphertext with LSP $C = Ecr_p(L,P)$}\label{fig:permutation}
\end{figure}

Fig. \ref{fig:permutation} shows permutation results of using LSRP, LSCP and LSP. It is clear that cascading LSRP and LSCP in LSP helps LSP to achieve a better pixel permutation performance in the sense that, pixels in its ciphertext image become more random-like and make the ciphertext image content unintelligible.
\subsection{Encryption/Decryption Algorithms}
After constructing Latin square based encryption primitives for image data, we now are able to construct the SPN cipher as shown in Fig. \ref{fig:SPN_Latin}. Algorithm 3 and 4 describe the encryption and decryption process of the Latin square image cipher, respectively.

\begin{algorithm}
\caption{\textbf{Latin Square Image Cipher- Encryption $C = \mathfrak{E}(P,K)$}}
\scriptsize
\begin{algorithmic}
\REQUIRE $K$ is a $256$-bit key
\REQUIRE $P$ is a $256\times 256$ 8-bit grayscale image block
\ENSURE $C$ is a $256\times 256$ 8-bit grayscale image block
\STATE $ $
\STATE $(\mathbf{Q_1},\mathbf{Q_2}) = KDSG(K,8)$
\FOR{$n = 0:1:7$}
    \IF {$n == 0$}
        \STATE{$C_{LSP} = \textrm{LSBNoiseEmbedding}(P)$}\footnotemark[4]
    \ENDIF
    \STATE{$L_n = LSG(Q_1^n,Q_2^n)$}
    \STATE{$D_n = L_n(0,0)$}
    \STATE{$C_{LSW} = Ecr_w(L_n,C_{LSP},D_n)$}
    \IF {$mod(n,2)\neq 0$}
        \STATE{$C_{LSS} = Ecr_s^{col}(L_n,C_{LSW})$}
    \ELSE
        \STATE{$C_{LSS} = Ecr_s^{row}(L_n,C_{LSW})$}
    \ENDIF
        \STATE{$C_{LSP} = Ecr_p(L_n,C_{LSS})$}
\ENDFOR
\STATE{$L_8 = LSG(Q_1^8,Q_2^8)$}
\STATE{$D_8 = L_8(0,0)$}
\STATE{$C = Ecr_w(L_8,C_{LSP},D_n)$}
\end{algorithmic}
\end{algorithm}
\footnotetext[4]{LSBNoiseEmbedding could be any function to embed random binary noise in the least significant bit-plane of a plaintext image, \eg a function randomly change the parity of a plaintext pixel on its LSB.}
\begin{algorithm}
\caption{\textbf{Latin Square Image Cipher- Decryption $P = \mathfrak{D}(C,K)$}}
\scriptsize
\begin{algorithmic}
\REQUIRE $K$ is a $256$-bit key
\REQUIRE $C$ is a $256\times 256$ 8-bit grayscale image block
\ENSURE $P$ is a $256\times 256$ 8-bit grayscale image block
\STATE $ $
\STATE $(\mathbf{Q_1},\mathbf{Q_2}) = KDSG(K,8)$
\FOR{$n = 7:-1:0$}
    \IF {$n == 7$}
        \STATE{$L_8 = LSG(Q_1^8,Q_2^8)$}
        \STATE{$D_8 = L_8(0,0)$}
        \STATE{$P_{LSW} = Dcr_w(L_8,C,D_8)$}
    \ENDIF
    \STATE{$L_n = LSG(Q_1^n,Q_2^n)$}
    \STATE{$D_n = L_n(0,0)$}
    \STATE{$P_{LSP} = Dec_p(L_n,P_{LSW})$}
    \IF{$mod(n,2)\neq 0$}
        \STATE{$P_{LSS} = Dcr_s^{col}(L_n,P_{LSP})$}
    \ELSE
        \STATE{$P_{LSS} = Dcr_s^{row}(L_n,P_{LSP})$}
    \ENDIF
    \STATE{$P_{LSW} = Dcr_w(L_n,P_{LSS},D_n)$}
\ENDFOR
\STATE{$P = P_{LSW}$}
\end{algorithmic}
\end{algorithm}

\subsection{Discussion}
\begin{figure}[h]
\scriptsize
\centering
\begin{minipage}[b]{1\linewidth}
  \begin{minipage}[b]{.95\linewidth}
    \centerline{\includegraphics[width=.95\linewidth]{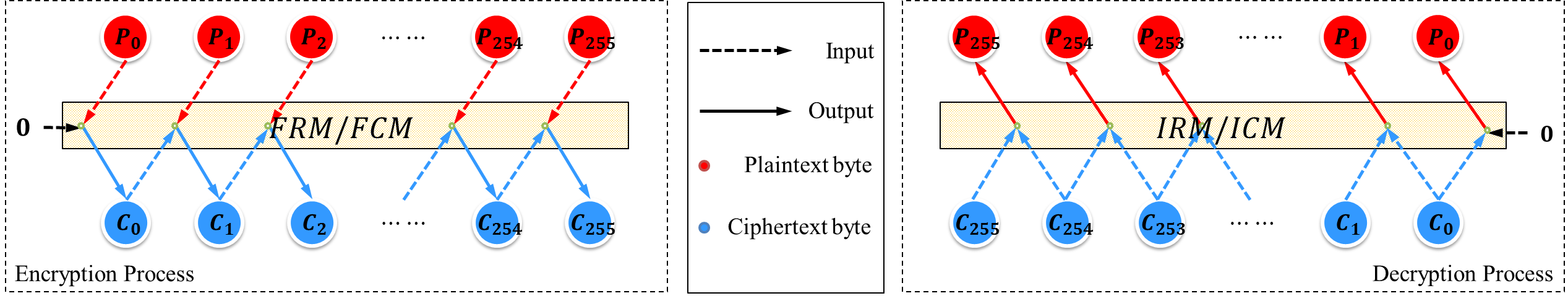}}
  \end{minipage}\hfill
  \end{minipage}\hfill
  \caption{\LSSs has an asymmetric structure for encryption and decryption}
  \label{fig:asymmetric}
  \end{figure}

Our design criteria behind the \LSSs are to achieve the following objectives,
\begin{enumerate}
  \item A ciphertext image is very sensitive to any slight change in a plaintext image.
  \item A deciphered image is insensitive to slight change in a ciphertext image.
\end{enumerate}
In practice, \LSSs is of an asymmetric structure as shown in Fig. \ref{fig:asymmetric}, in the sense that encrypting one plaintext byte requires one plaintext byte and one ciphertext byte, while decrypting one ciphertext byte requires two ciphertext byte but no plaintext byte. This process is similar to weave a thread of ciphertext bytes on the ciphertext image, which has to be done by intersecting the longitudinal threads of plaintext bytes.
As a result, changing one pixel in a plaintext image influence all pixels after it in its row in the first encryption round, further all pixels after these influenced pixels in corresponding columns in the second encryption round, and then more pixels in the third encryption round, so on and so forth until the last round. As a result, this small change in plaintext image will lead to a completely different ciphertext image. In contrast, changing one pixel in a ciphertext image only leads a two-pixel-change in the first decryption round, and at most $2^8 = 256$ pixel changes in the deciphered image. Hence the \LSSs indeed achieves both goals above and gains encryption diffusion properties against plaintext change and decryption robustness against ciphertext noise simultaneously. More results and examples about these two abilities will be discussed in future sections.

The proposed Latin square image cipher is carefully designed with respect to image data. We use the psychovisual redundancy within an image and realize the probabilistic encryption by introducing random noise in the LSB of a plaintext image block. We also notice the information redundancy within adjacent image pixels, which might lead to information leakage by estimating a pixel's intensity from its neighbors. We break these high correlated neighbor pixels by applying \LSW, which shifts pixels within a homogenous region to random-like in the sense that shifting amounts of any two highly correlated pixels within a row or a column in the homogenous region are always different. This way prevents the prediction of a pixel's intensity from its neighbors and thus decorrelates these pixels. In addition, \LSWs also allows to spatial transform a plaintext image in a random-like manner. Further our proposed \LSSs substitutes pixel bytes along rows and columns iteratively like weaving in a loom, if we consider a row of pixels is a longitudinal thread and a column of pixels is a latitudinal thread. Finally \LSPs shuffles plaintext image pixels all over the domain. In summary the SPN used in LSIC differs from the conventional SPN block cipher \cite{CryptographyBook} in:
\begin{itemize}
  \item Cipher Object: LSIC is designed for two-dimensional images while a conventional SPN block cipher works for one-dimensional bit streams.
  \item Processing Unit: LSIC processes image data byte by byte while a SPN block cipher encrypts data bit by bit.
  \item Cipher Structure: LSIC is of a loom-like processing, which includes spatial flipping and iterative substitution along rows and columns.
  \item Parametric Processes: LSIC generates dynamic \LSW, \textit{S-Box} and \textit{P-Box} while processes in a SPN block cipher are static and key independent.
\end{itemize}

\section{Simulation Results}
\subsection{Experiment Settings and Dataset}
\begin{figure}[h]
\begin{singlespace}
\BottomFloatBoxes
\begin{floatrow}
\ffigbox{%
  \begin{minipage}[b]{1\linewidth}
\begin{minipage}[b]{0.19\linewidth}
\includegraphics[width = .95\linewidth]{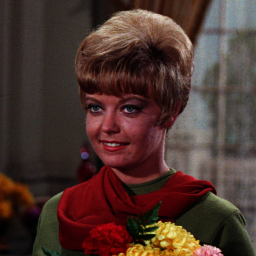}
\includegraphics[width = .95\linewidth]{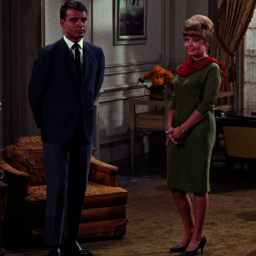}
\includegraphics[width = .95\linewidth]{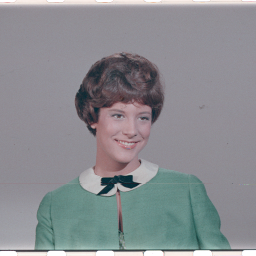}
\includegraphics[width = .95\linewidth]{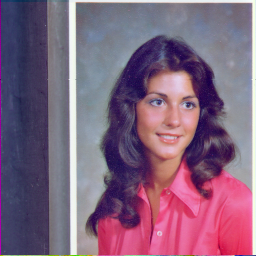}
\includegraphics[width = .95\linewidth]{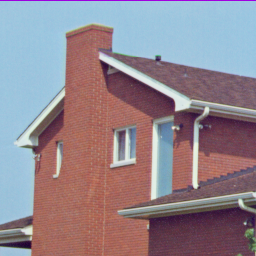}
\includegraphics[width = .95\linewidth]{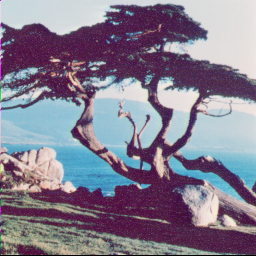}
\includegraphics[width = .95\linewidth]{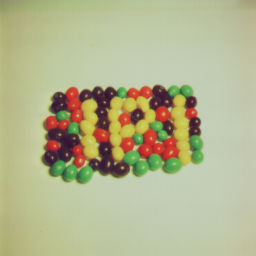}
\includegraphics[width = .95\linewidth]{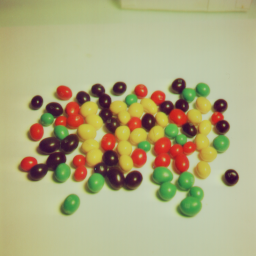}
\includegraphics[width = .95\linewidth]{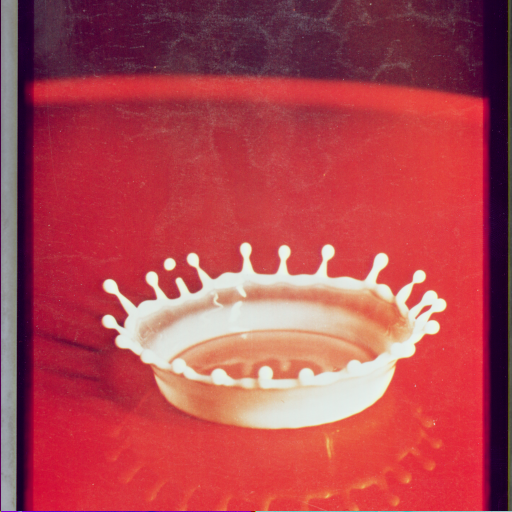}
\end{minipage}\hfill
\begin{minipage}[b]{0.19\linewidth}
\includegraphics[width = .95\linewidth]{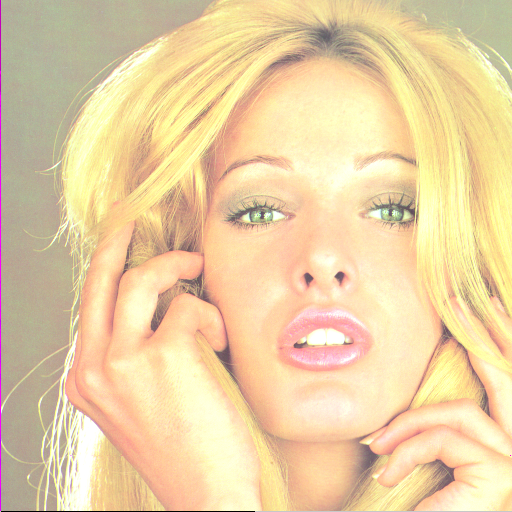}
\includegraphics[width = .95\linewidth]{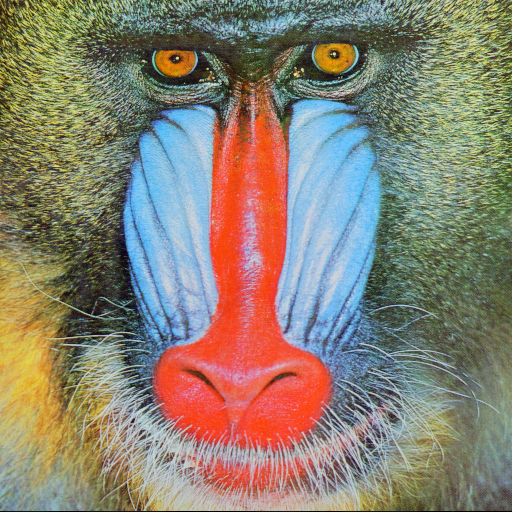}
\includegraphics[width = .95\linewidth]{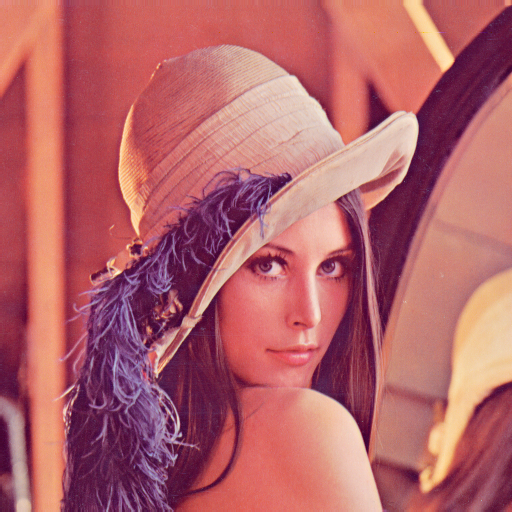}
\includegraphics[width = .95\linewidth]{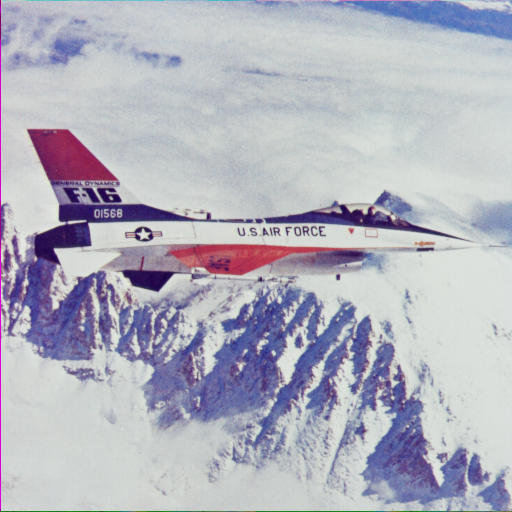}
\includegraphics[width = .95\linewidth]{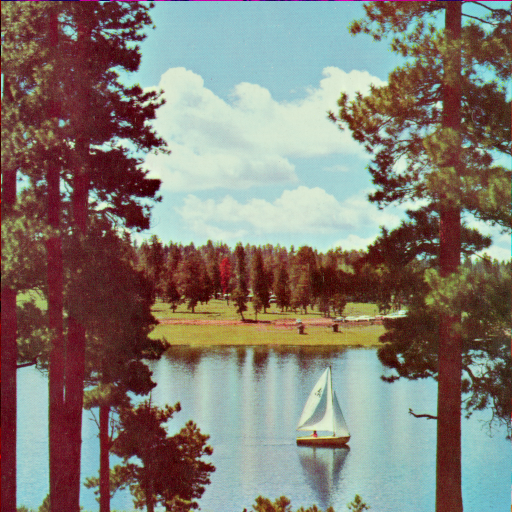}
\includegraphics[width = .95\linewidth]{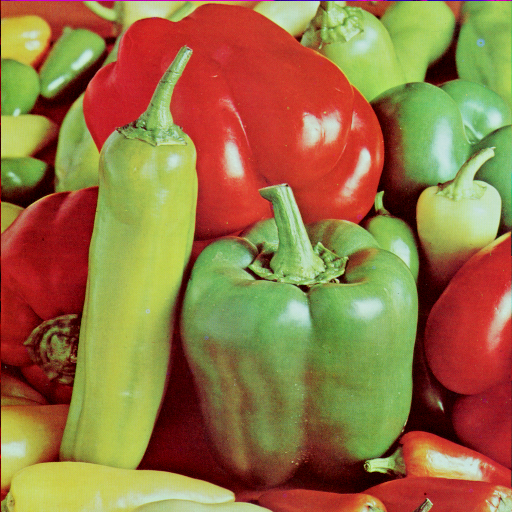}
\includegraphics[width = .95\linewidth]{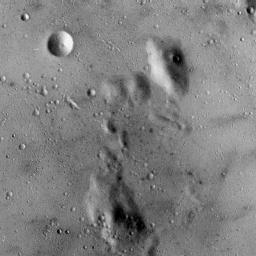}
\includegraphics[width = .95\linewidth]{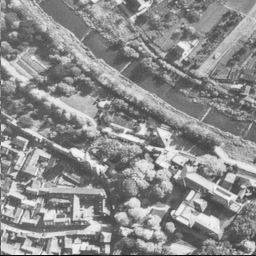}
\includegraphics[width = .95\linewidth]{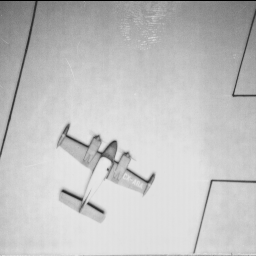}
\end{minipage}\hfill
\begin{minipage}[b]{0.19\linewidth}
\includegraphics[width = .95\linewidth]{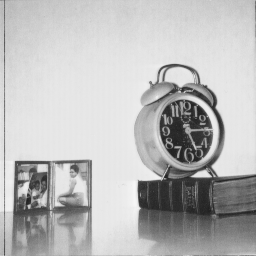}
\includegraphics[width = .95\linewidth]{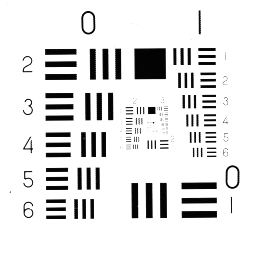}
\includegraphics[width = .95\linewidth]{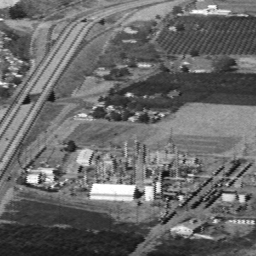}
\includegraphics[width = .95\linewidth]{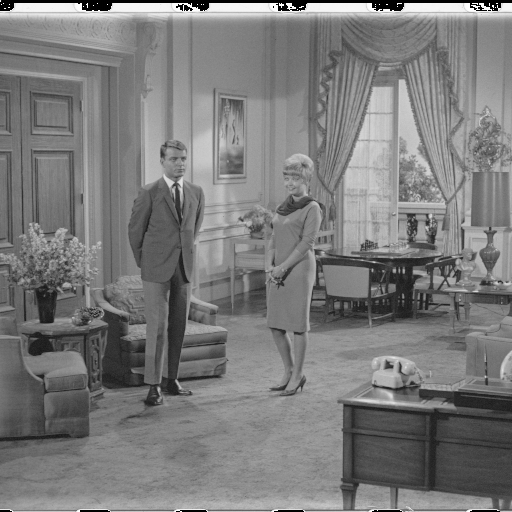}
\includegraphics[width = .95\linewidth]{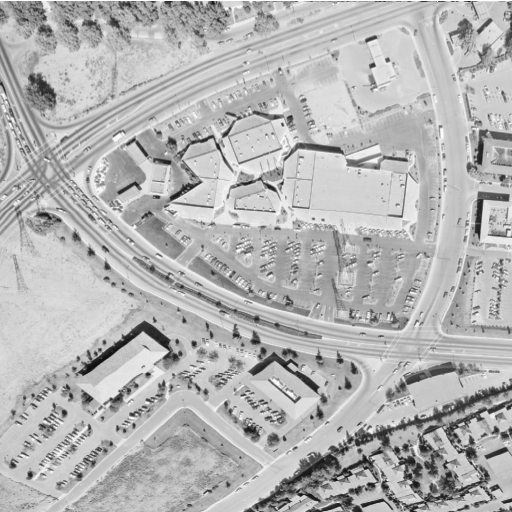}
\includegraphics[width = .95\linewidth]{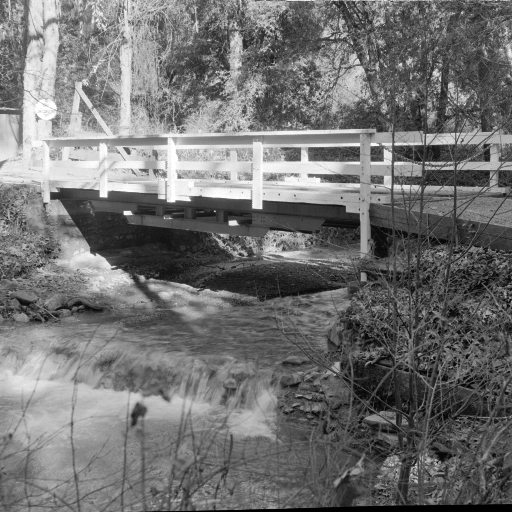}
\includegraphics[width = .95\linewidth]{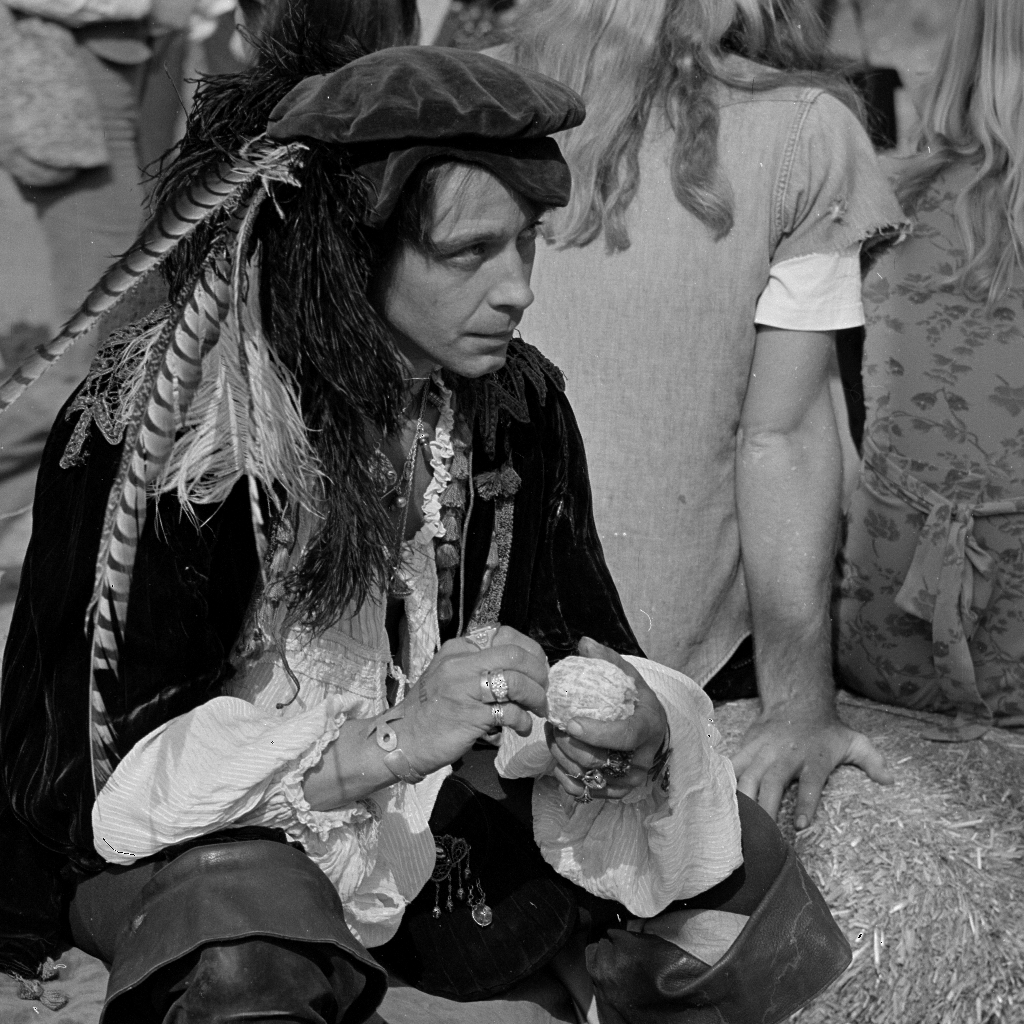}
\includegraphics[width = .95\linewidth]{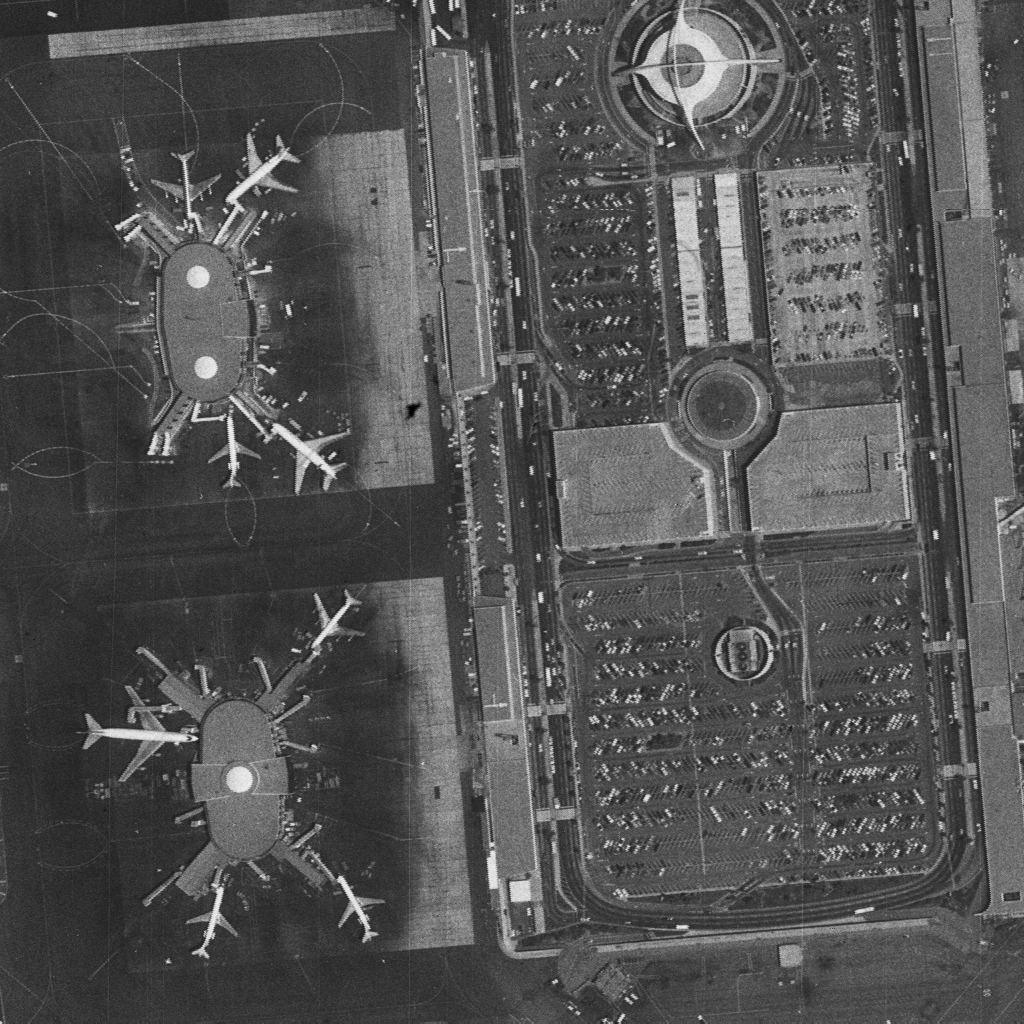}
\includegraphics[width = .95\linewidth]{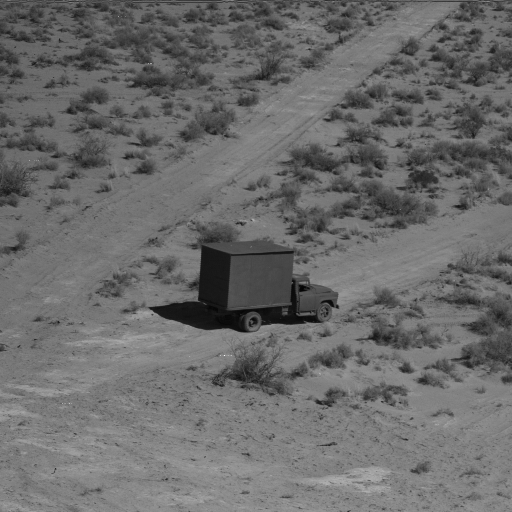}
\end{minipage}\hfill
\begin{minipage}[b]{0.19\linewidth}
\includegraphics[width = .95\linewidth]{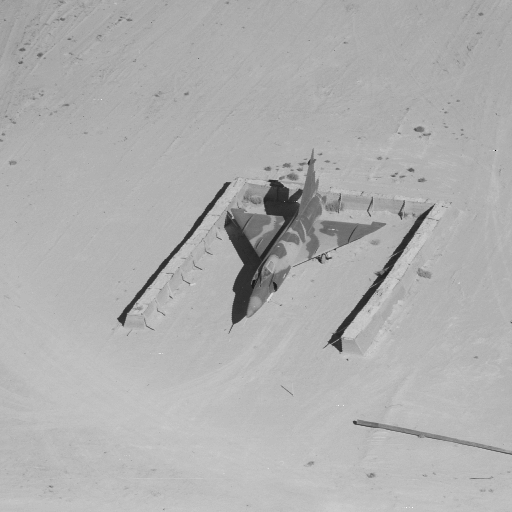}
\includegraphics[width = .95\linewidth]{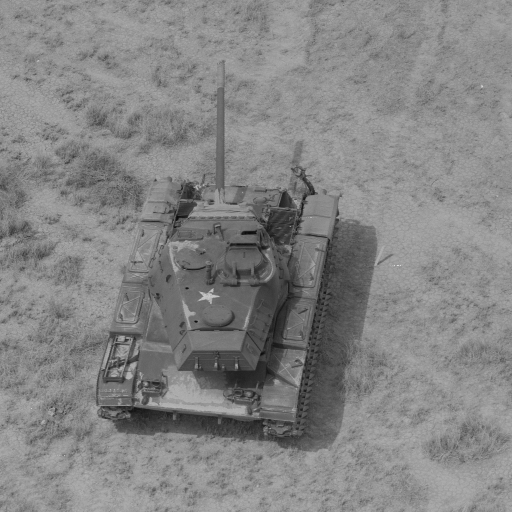}
\includegraphics[width = .95\linewidth]{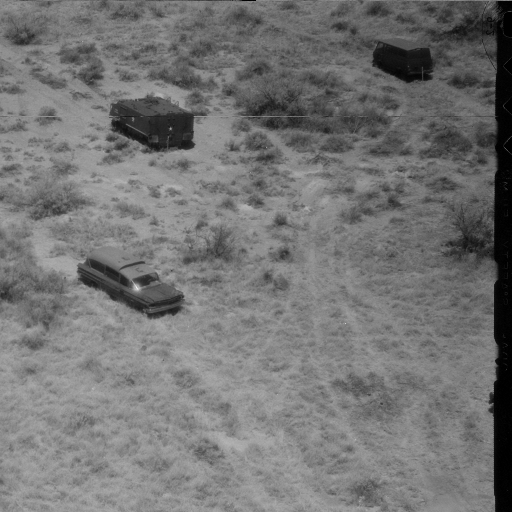}
\includegraphics[width = .95\linewidth]{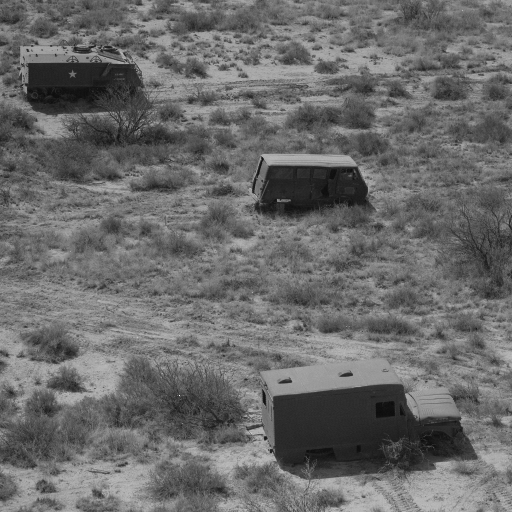}
\includegraphics[width = .95\linewidth]{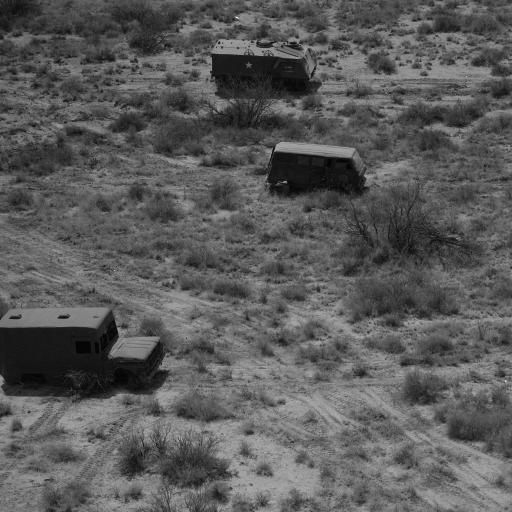}
\includegraphics[width = .95\linewidth]{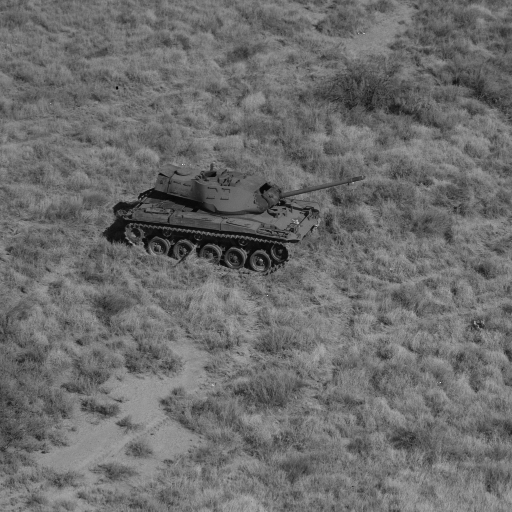}
\includegraphics[width = .95\linewidth]{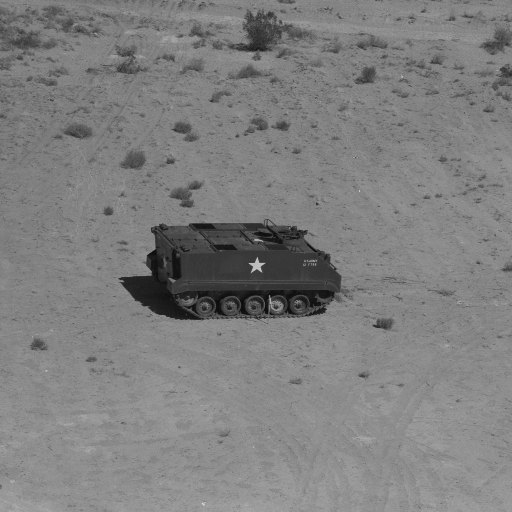}
\includegraphics[width = .95\linewidth]{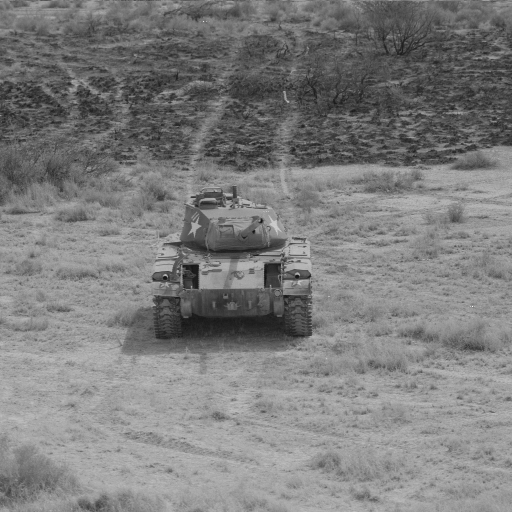}
\includegraphics[width = .95\linewidth]{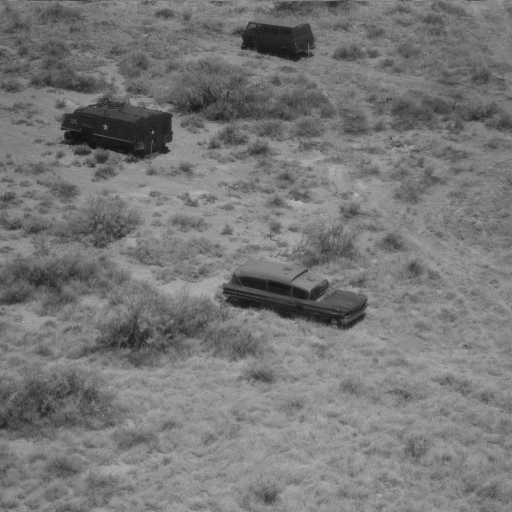}
\end{minipage}\hfill
\begin{minipage}[b]{0.19\linewidth}
\includegraphics[width = .95\linewidth]{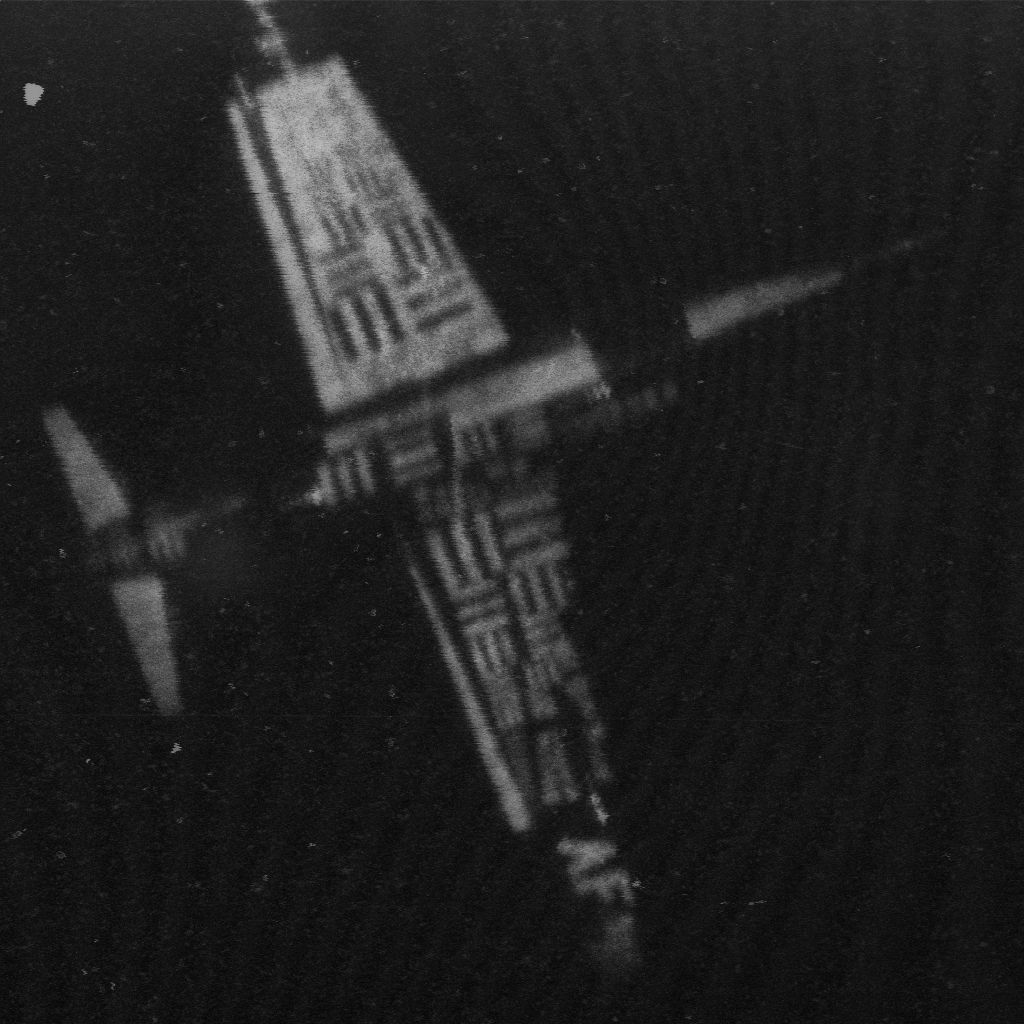}
\includegraphics[width = .95\linewidth]{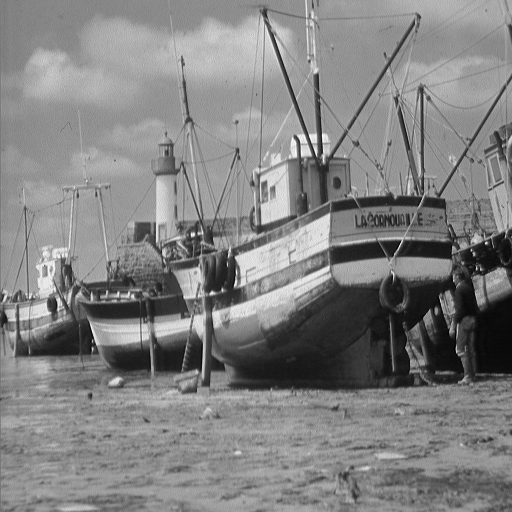}
\includegraphics[width = .95\linewidth]{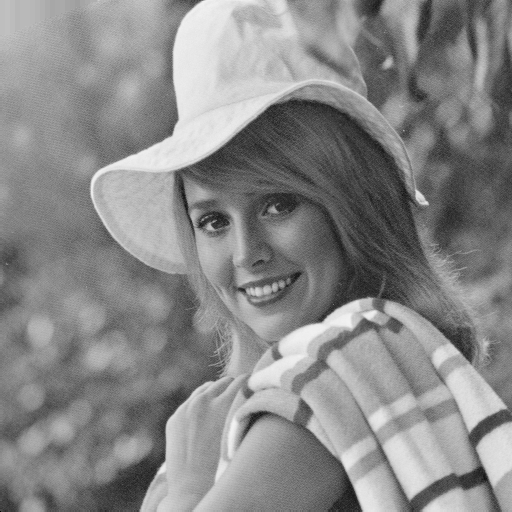}
\includegraphics[width = .95\linewidth]{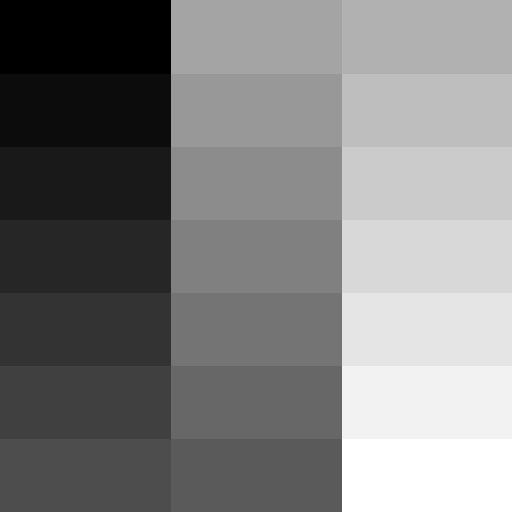}
\includegraphics[width = .95\linewidth]{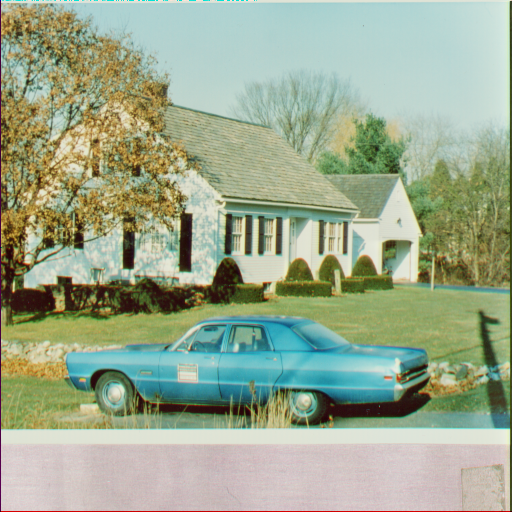}
\includegraphics[width = .95\linewidth]{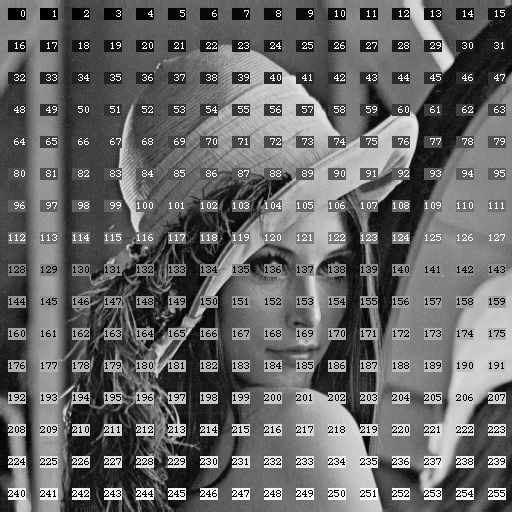}
\includegraphics[width = .95\linewidth]{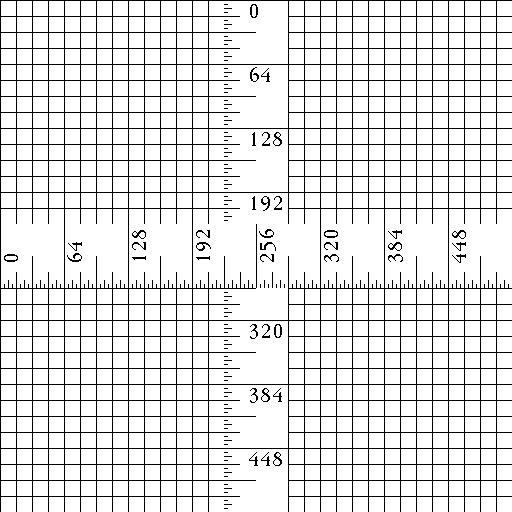}
\includegraphics[width = .95\linewidth]{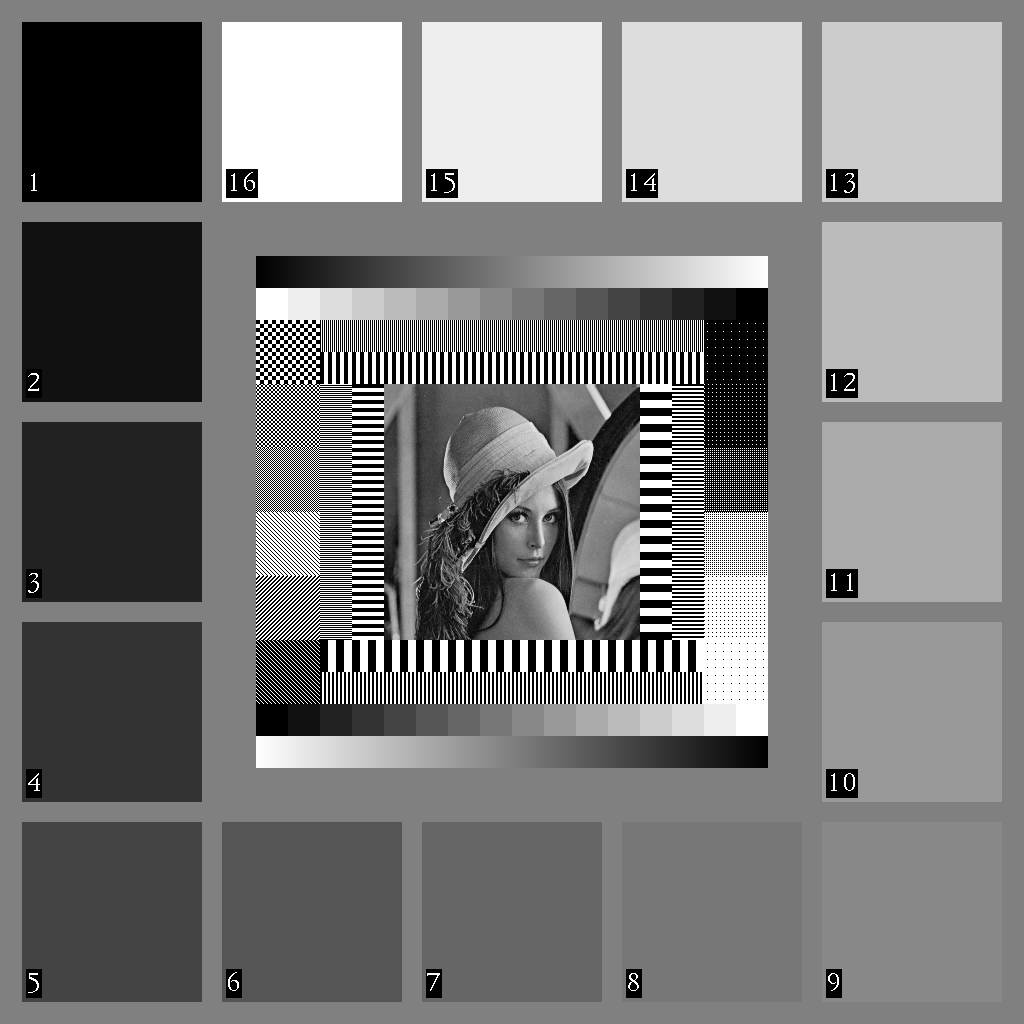}
\end{minipage}\hfill
\end{minipage}\hfill
}{%
  \caption{USC-SIPI \textit{Miscellanous} dataset}
\label{fig:USCImg}
}
\killfloatstyle\ttabbox{
\scriptsize
\centering
\begin{tabular}{llrl}
\hline\hline
\textbf{File} & \textbf{Description} & \textbf{Size} & \textbf{Type} \\\hline
4.1.01 & Girl & 256 & Color \\
4.1.02 & Couple & 256 & Color \\
4.1.03 & Girl & 256 & Color \\
4.1.04 & Girl & 256 & Color \\
4.1.05 & House & 256 & Color \\
4.1.06 & Tree & 256 & Color \\
4.1.07 & Jelly beans & 256 & Color \\
4.1.08 & Jelly beans & 256 & Color \\
4.2.01 & Splash & 512 & Color \\
4.2.02 & Girl (Tiffany) & 512 & Color \\
4.2.03 & Mandrill (a.k.a. Baboon) & 512 & Color \\
4.2.04 & Girl (Lena, or Lenna) & 512 & Color \\
4.2.05 & Airplane (F-16) & 512 & Color \\
4.2.06 & Sailboat on lake & 512 & Color \\
4.2.07 & Peppers & 512 & Color \\
5.1.09 & Moon surface & 256 & Gray \\
5.1.10 & Aerial & 256 & Gray \\
5.1.11 & Airplane & 256 & Gray \\
5.1.12 & Clock & 256 & Gray \\
5.1.13 & Resolution chart & 256 & Gray \\
5.1.14 & Chemical plant & 256 & Gray \\
5.2.08 & Couple & 512 & Gray \\
5.2.09 & Aerial & 512 & Gray \\
5.2.10 & Stream and bridge & 512 & Gray \\
5.3.01 & Man & 1024 & Gray \\
5.3.02 & Airport & 1024 & Gray \\
7.1.01 & Truck & 512 & Gray \\
7.1.02 & Airplane & 512 & Gray \\
7.1.03 & Tank & 512 & Gray \\
7.1.04 & Car and APCs & 512 & Gray \\
7.1.05 & Truck and APCs & 512 & Gray \\
7.1.06 & Truck and APCs & 512 & Gray \\
7.1.07 & Tank & 512 & Gray \\
7.1.08 & APC & 512 & Gray \\
7.1.09 & Tank & 512 & Gray \\
7.1.10 & Car and APCs & 512 & Gray \\
7.2.01 & Airplane (U-2) & 1024 & Gray \\
boat.512 & Fishing Boat & 512 & Gray \\
elaine.512 & Girl (Elaine) & 512 & Gray \\
house & House & 512 & Color \\
gray21.512 & 21 level step wedge & 512 & Gray \\
numbers.512 & 256 level test pattern & 512 & Gray \\
ruler.512 & Pixel ruler & 512 & Gray \\
testpat.1k & General test pattern & 1024 & Gray \\\hline\hline
\end{tabular}
}{
  \caption{USC-SIPI \textit{Miscellaneous} dataset information}\label{Tab:USC_SIPI}
}
\end{floatrow}
\end{singlespace}
\end{figure}

We perform the following simulations using a computer under MATLAB \textit{R2010a} environment with a Windows 7 Operating System, $6Gb$ memory and Intel Core2 $2.66GHz$ CPU. Test images are from the \textit{Miscellaneous} dataset in the USC-SIPI image database\footnotemark[5]. The detailed information about these test images are listed in Table \ref{Tab:USC_SIPI}.

\footnotetext[5]{USC-SIPI image database is a public image database held by the University of South California, available at \url{http://sipi.usc.edu/database/} \astoday.}

Although we believe the performance of an image cipher should be tested over a fairly large dataset, \eg the USC SIPI \textit{Miscellaneous} dataset, the analysis of many peer image encryption methods are merely about a small number of test images \eg \textit{Lenna} image. Therefore, we perform fair comparisons between the proposed LSIC and peer methods \cite{3DCat,Zhu2011,Liao2010,Awad2011,Kumar2011,xiang2006novel,wong2003chaotic,Sun2010,Zhang2010DNA,zhu2010chaos,wang2011new,liu2011color,Patidar2011,Fu2011,Mao2005,Pareek2006}.
 in three following aspects:
\begin{itemize}
  \item If peer methods have performance reports on \lenna image, we compare the performance of LSIC with those methods on the identical \textit{Lenna} image.
  \item If peer methods have the accessible source code (Mao \etal 2004 \cite{3DCat}) or the executive file (\eg bmpPacker\footnotemark[6]) online, we compare the performance of these methods with LSIC over the entire \textit{Miscellaneous} dataset.
  \item If peer methods have performance reports also on the entire \textit{Miscellaneous} dataset already ({Pareek \etal, 2006 \cite{Pareek2006}}), we directly cite these reports for comparisons.
\end{itemize}

\footnotetext[6]{bmpPacker is an encryption software containing many classic block ciphers written by Jens G$\ddot{o}$deke. Only its AES cipher is used in our simulations. This freeware is available on: \url{http://www.jens-goedeke.eu/tools/bmppacker/} \astoday.}

Furthermore, we turn off LSBNoiseEmbedding function in all simulations because we want to:
 \begin{itemize}
   \item prevent the possible unfair comparisons, because other compared algorithms donot have this stage
   \item separate the influences brought by these embedded random noises from Latin square encryption primitives
 \end{itemize}
However, the probabilistic encryption integrated in the LSIC does enhance the cipher security, because without this stage the two ciphertext images $C_1$ and $C_2$ by encrypting a plaintext image $P$ twice with a key $K$ are always identical,\ie $C_1 = C_2$, which provides an adversary plaintext information from ciphertext information that he/she should never know. In contrast, LSIC with \textit{LSB Noise Embedding} is able to encrypt a plaintext image to distinctive ciphertext images because of the embedded noise from time to time are not necessarily the same and thus completely prevent the risk of information leakage in the deterministic encryption.

\subsection{Simulation Results}
\begin{figure}[!h]
\scriptsize
\centering
\begin{minipage}[c]{\subw\linewidth}\centering
{\includegraphics[width = \subhf\linewidth]{4_1_01}}
{\includegraphics[width = \subh\linewidth]{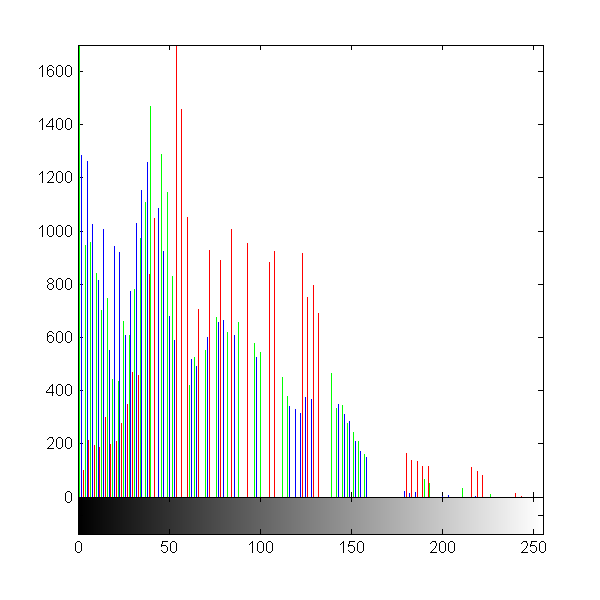}}
{\includegraphics[width = \subhf\linewidth]{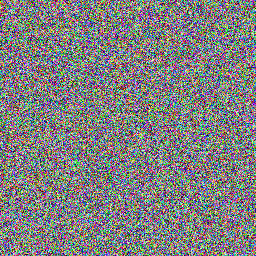}}
{\includegraphics[width = \subh\linewidth]{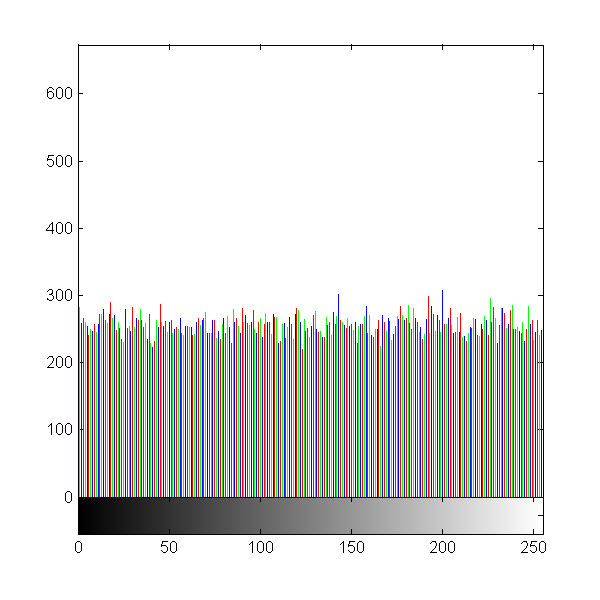}}
{\includegraphics[width = \subhf\linewidth]{4_1_01}}
\end{minipage}\hfill
\begin{minipage}[c]{\subw\linewidth}\centering
{\includegraphics[width = \subhf\linewidth]{4_2_03}}
{\includegraphics[width = \subh\linewidth]{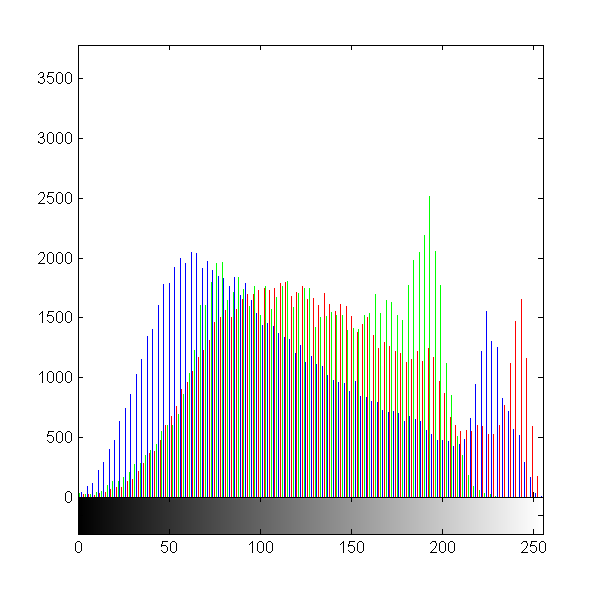}}
{\includegraphics[width = \subhf\linewidth]{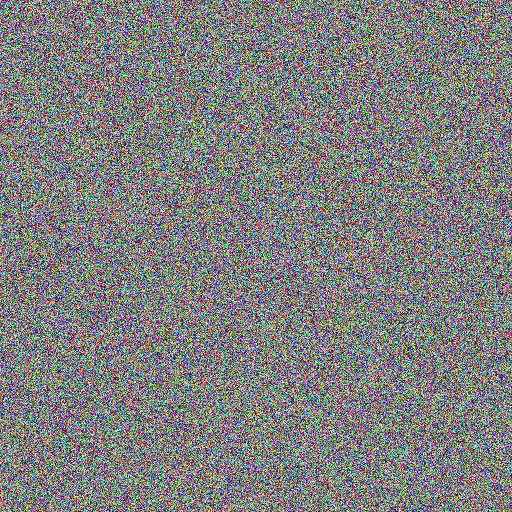}}
{\includegraphics[width = \subh\linewidth]{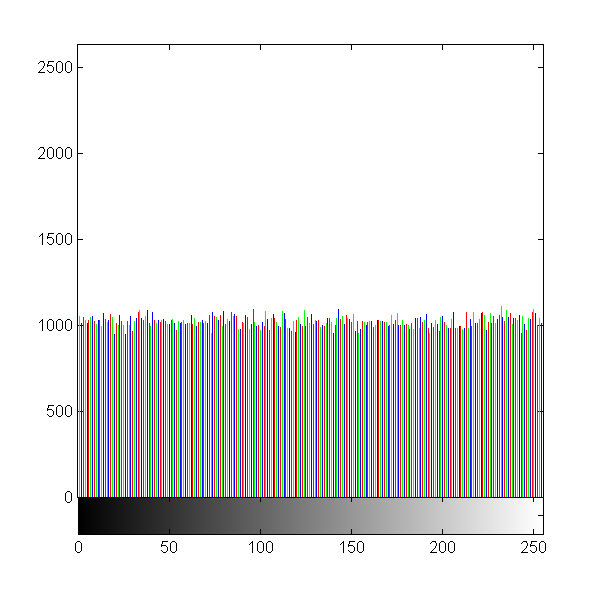}}
{\includegraphics[width = \subhf\linewidth]{4_2_03}}
\end{minipage}\hfill
\begin{minipage}[c]{\subw\linewidth}\centering
{\includegraphics[width = \subhf\linewidth]{4_2_04}}
{\includegraphics[width = \subh\linewidth]{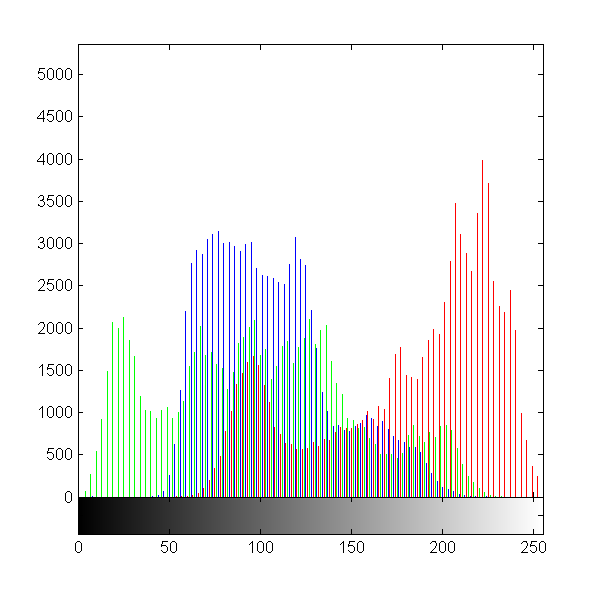}}
{\includegraphics[width = \subhf\linewidth]{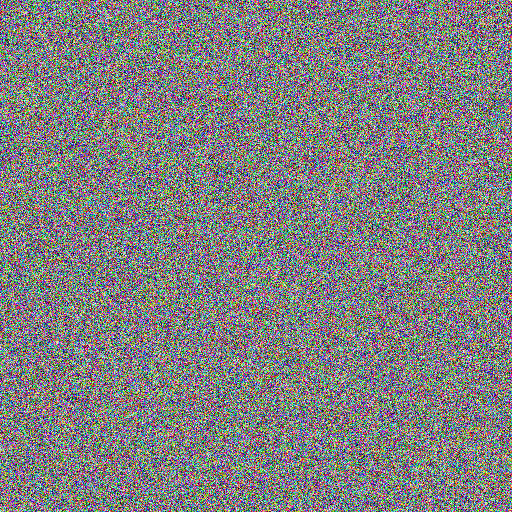}}
{\includegraphics[width = \subh\linewidth]{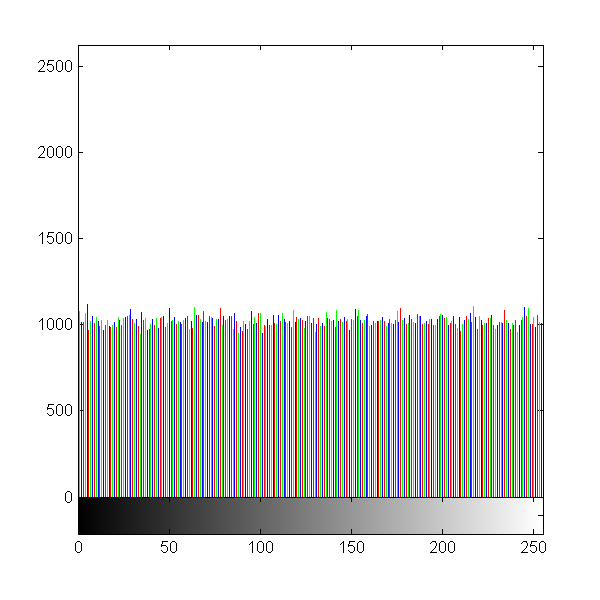}}
{\includegraphics[width = \subhf\linewidth]{4_2_04}}
\end{minipage}\hfill
\begin{minipage}[c]{\subw\linewidth}\centering
{\includegraphics[width = \subhf\linewidth]{4_2_05}}
{\includegraphics[width = \subh\linewidth]{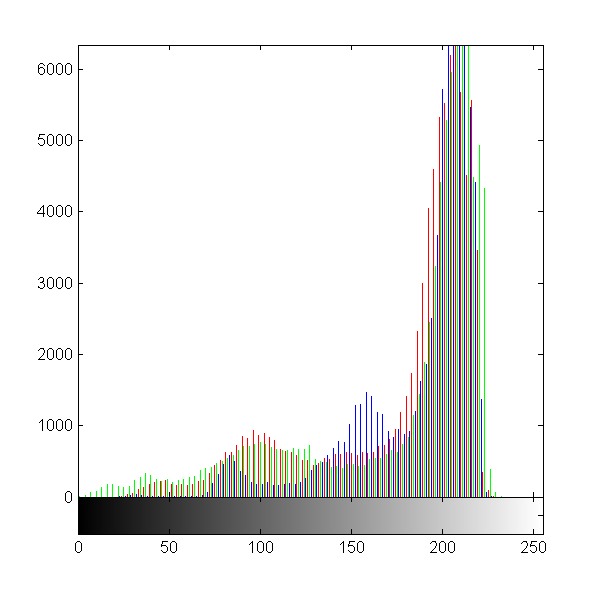}}
{\includegraphics[width = \subhf\linewidth]{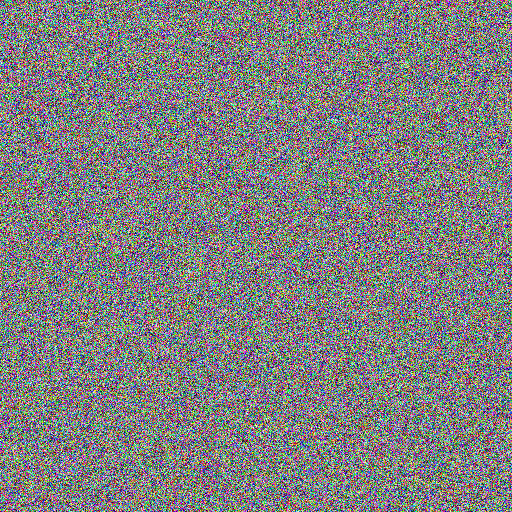}}
{\includegraphics[width = \subh\linewidth]{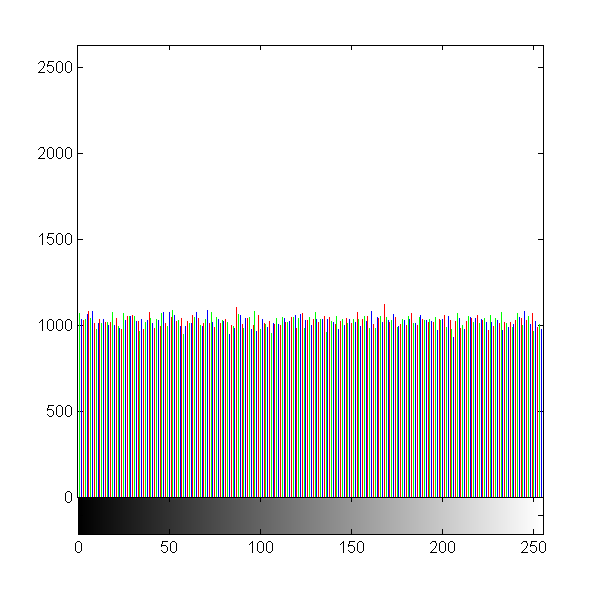}}
{\includegraphics[width = \subhf\linewidth]{4_2_05}}
\end{minipage}\hfill
\begin{minipage}[c]{\subw\linewidth}\centering
{\includegraphics[width = \subhf\linewidth]{4_2_07}}
{\includegraphics[width = \subh\linewidth]{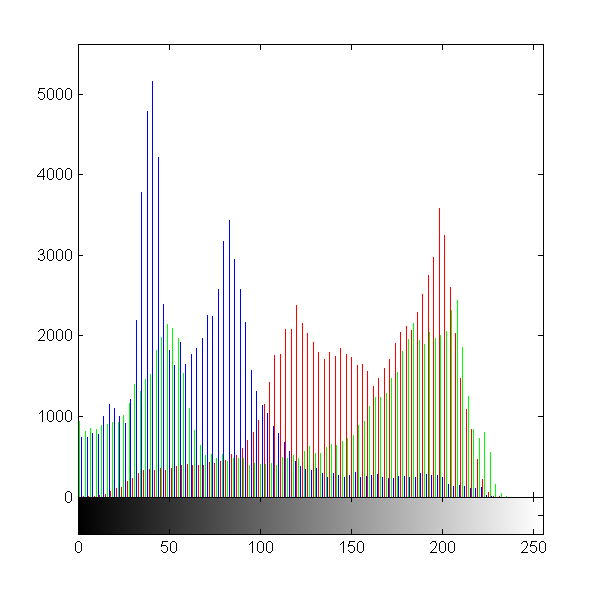}}
{\includegraphics[width = \subhf\linewidth]{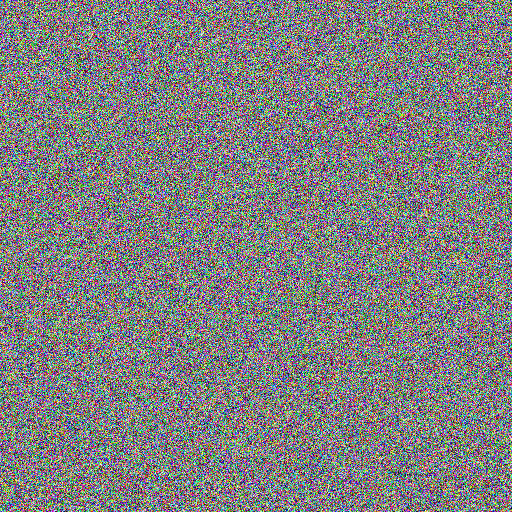}}
{\includegraphics[width = \subh\linewidth]{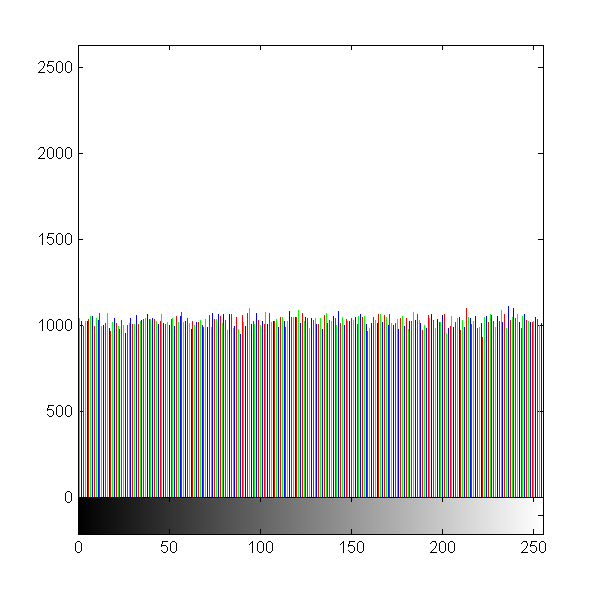}}
{\includegraphics[width = \subhf\linewidth]{4_2_07}}
\end{minipage}\hfill
\begin{minipage}[c]{\subw\linewidth}\centering
{\includegraphics[width = \subhf\linewidth]{5_1_13}}
{\includegraphics[width = \subh\linewidth]{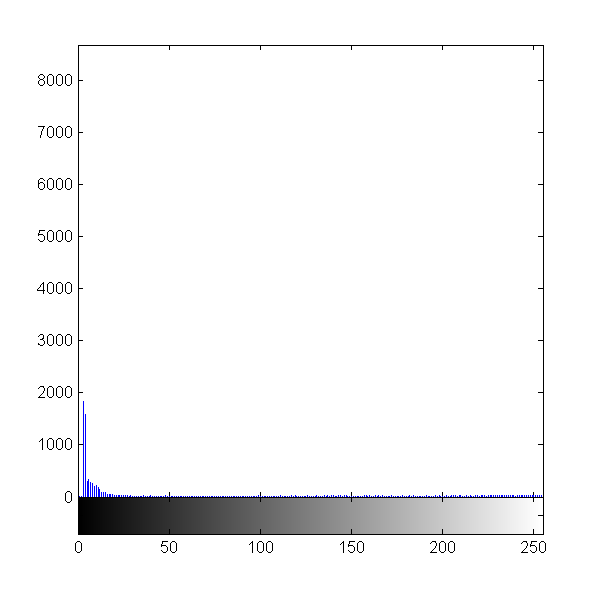}}
{\includegraphics[width = \subhf\linewidth]{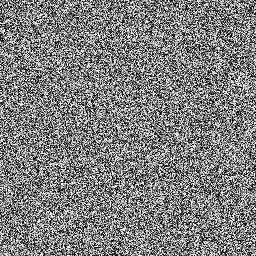}}
{\includegraphics[width = \subh\linewidth]{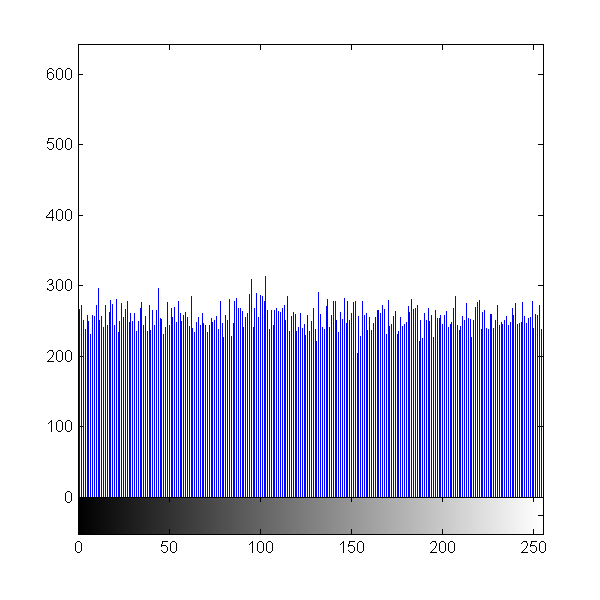}}
{\includegraphics[width = \subhf\linewidth]{5_1_13}}
\end{minipage}\hfill
\begin{minipage}[c]{\subw\linewidth}\centering
{\includegraphics[width = \subhf\linewidth]{boat_512}}
{\includegraphics[width = \subh\linewidth]{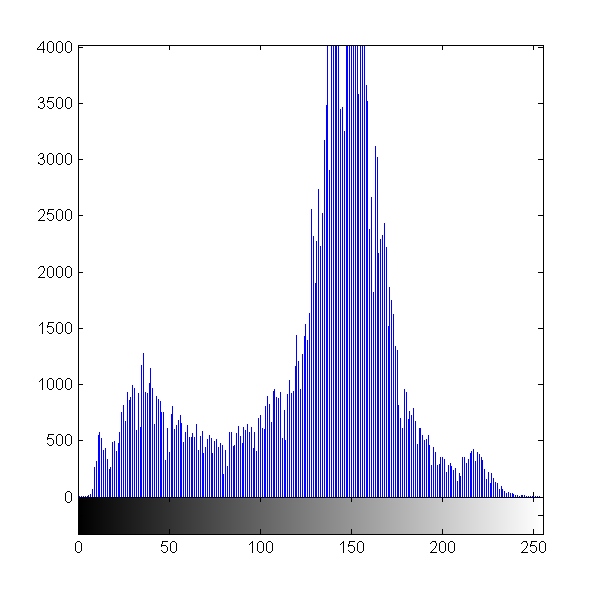}}
{\includegraphics[width = \subhf\linewidth]{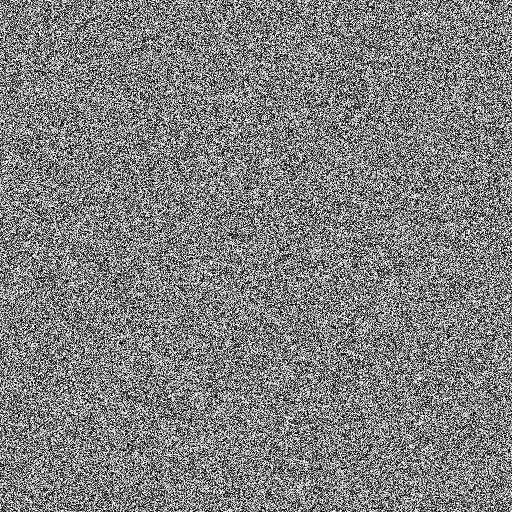}}
{\includegraphics[width = \subh\linewidth]{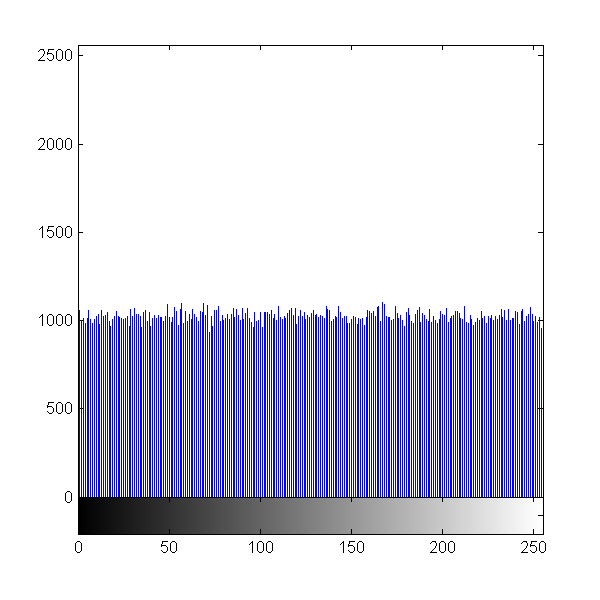}}
{\includegraphics[width = \subhf\linewidth]{boat_512}}
\end{minipage}\hfill
\begin{minipage}[c]{\subw\linewidth}\centering
{\includegraphics[width = \subhf\linewidth]{elaine_512}}
{\includegraphics[width = \subh\linewidth]{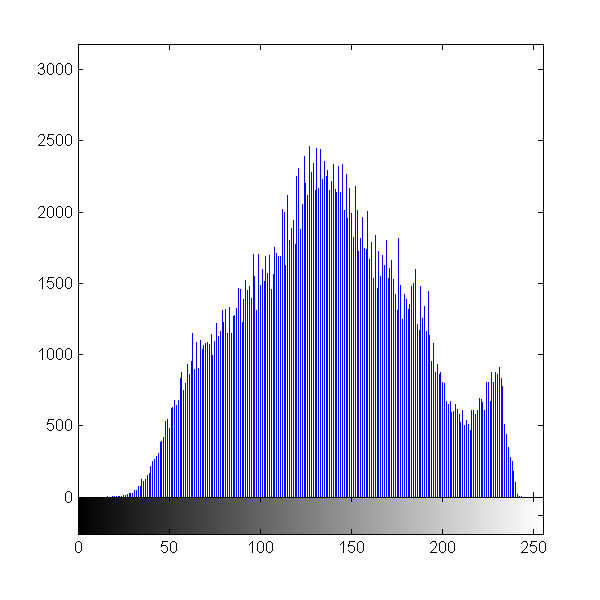}}
{\includegraphics[width = \subhf\linewidth]{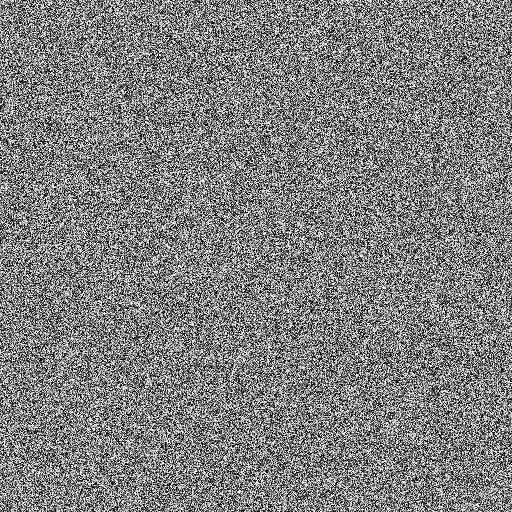}}
{\includegraphics[width = \subh\linewidth]{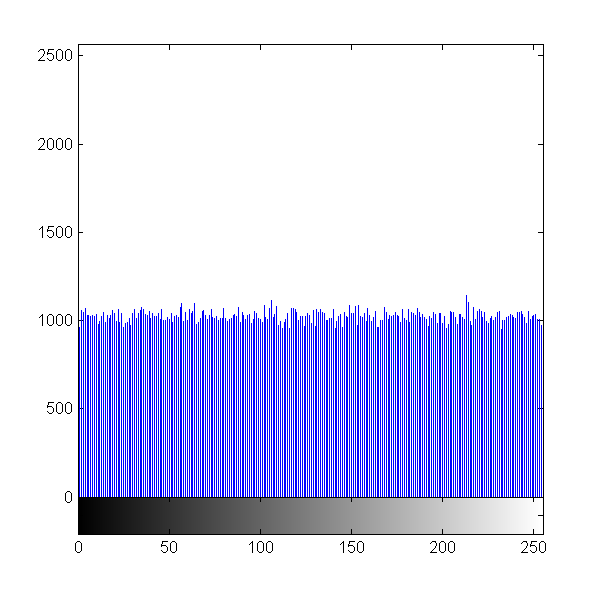}}
{\includegraphics[width = \subhf\linewidth]{elaine_512}}
\end{minipage}\hfill
\begin{minipage}[c]{\subw\linewidth}\centering
{\includegraphics[width = \subhf\linewidth]{ruler_512}}
{\includegraphics[width = \subh\linewidth]{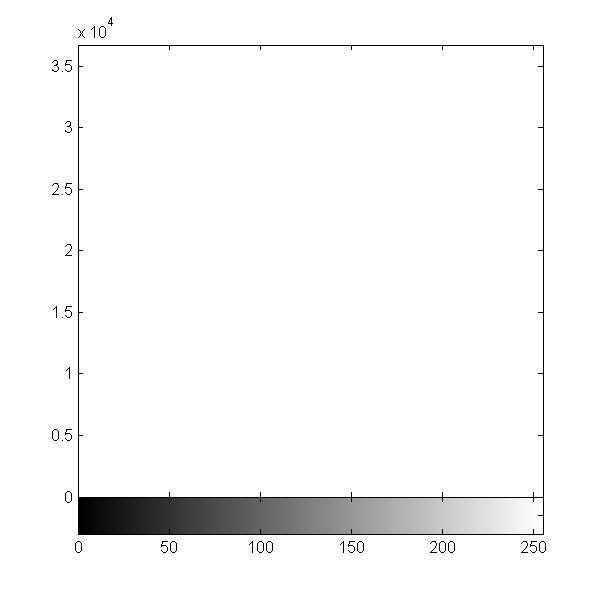}}
{\includegraphics[width = \subhf\linewidth]{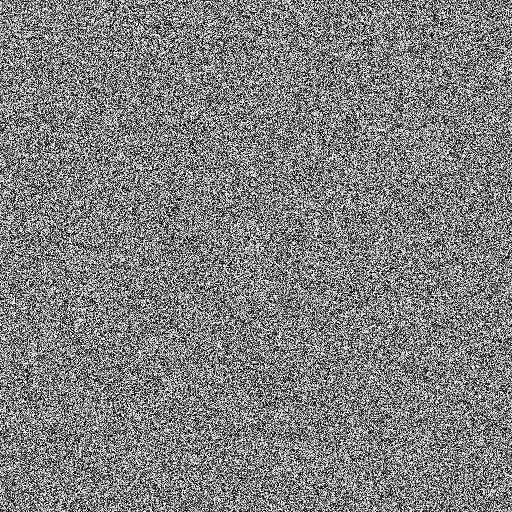}}
{\includegraphics[width = \subh\linewidth]{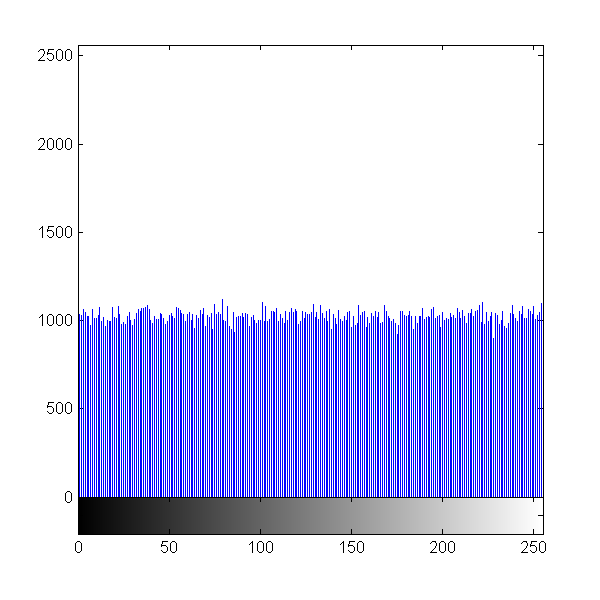}}
{\includegraphics[width = \subhf\linewidth]{ruler_512}}
\end{minipage}\hfill
\begin{minipage}[c]{\subw\linewidth}\centering
{\includegraphics[width = \subhf\linewidth]{testpat_1k}}
{\includegraphics[width = \subh\linewidth]{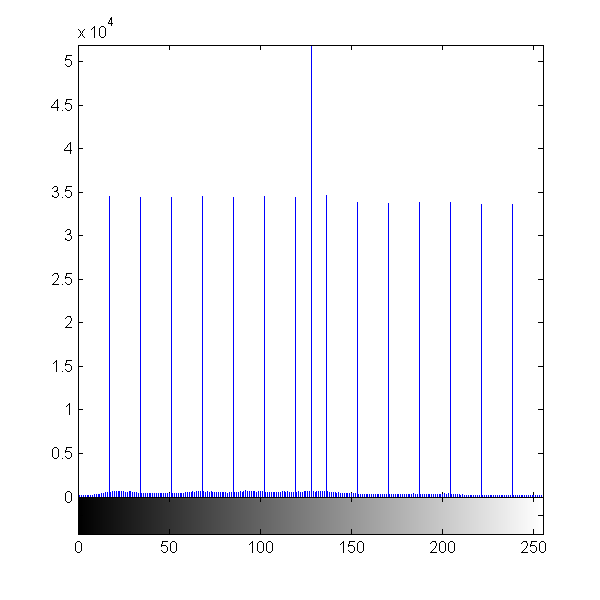}}
{\includegraphics[width = \subhf\linewidth]{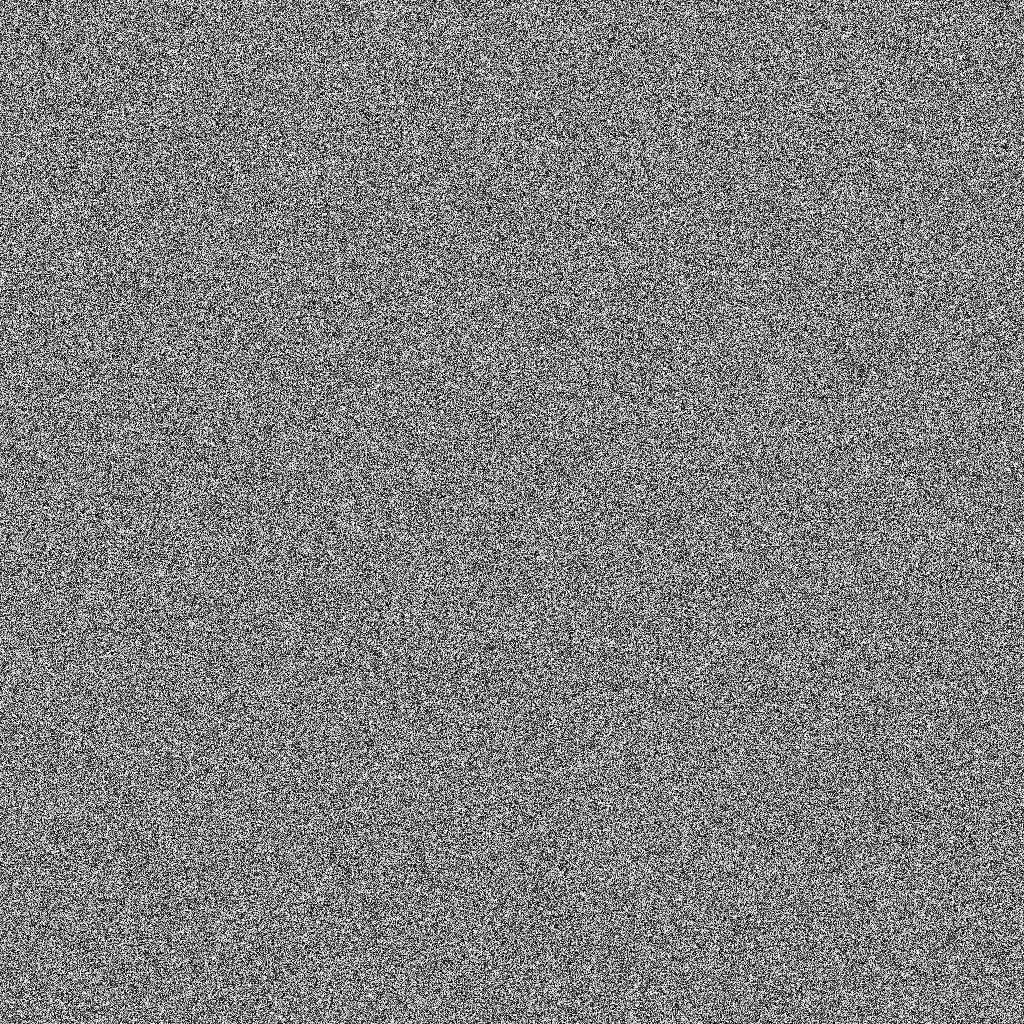}}
{\includegraphics[width = \subh\linewidth]{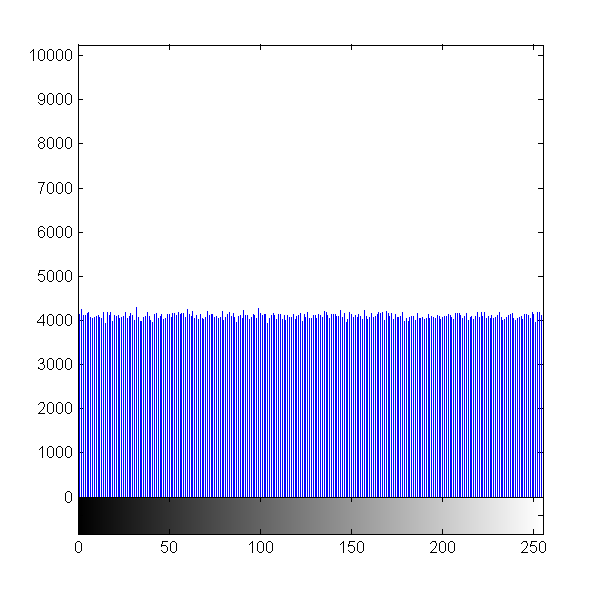}}
{\includegraphics[width = \subhf\linewidth]{testpat_1k}}
\end{minipage}\hfill
\caption{Sample results of encrypting and decrypting images using the Latin square image cipher (1st column: plaintext images; 2nd column: histograms of plaintext images; 3rd column: ciphertext images; 4th column: histograms of ciphertext images; and 5th column: decipertedtext images.)}
\label{fig:Results}
\end{figure}

Image encryption results using the proposed Latin square image cipher are shown in Fig. \ref{fig:Results}. As can be seen, the proposed LSIC is able to
\begin{itemize}
  \item generate unintelligible ciphertext images no matter what contents or types of the plaintext images;
  \item make the statistics of ciphertext images to be uniform-like;
  \item perfectly reconstruct the plaintext images from the corresponding ciphertext images.
\end{itemize}
It is also worthwhile to note that the test images \textit{5.1.13, ruler.512} and \textit{testpat.1k} are very difficult cases for image encryption because they have large size homogeneous regions, \ie they have extremely tilted histograms. However, the proposed Latin square image cipher encrypts those two images successfully.

In regard to encryption speed, the proposed Latin square image cipher is able to encrypt each $256\times 256$ image block (pixel in $8$-bit byte) with $1.2331\pm 0.10089$ seconds for all test images pulled from the USC-SIPI \textit{Miscellaneous} dataset under the MATLAB implementation. Therefore, roughly speaking, the encryption/decryption speed of the proposed Latin square image cipher about $0.415 Mb/s$ (Megabits per second) or equivalently $3.32 Mb/s$ per round under MATLAB. In regard to decryption speed, the proposed Latin square image cipher is able to decrypt each $256\times 256$ image block (pixel in $8$-bit byte) with $0.7137\pm 0.06222$ seconds (equivalently $5.74 Mb/s$ per round) for all test encrypted images from the USC-SIPI \textit{Miscellaneous} dataset. The decryption speed is faster than encryption because \textit{Latin Square S-boxes} defined in Eqs. \eqref{eqn:LatinRow} and \eqref{eqn:LatinCol} are asymmetric for encryption and decryption (see Fig. \ref{fig:asymmetric}).

\section{Security Analysis}
A good image encryption algorithm/cipher should be able to resist all known types of attacks and cryptanalysis. Its performance should be independent of a used encryption key and a plaintext image. First of all, this cipher should be theoretically secure against known attacks. Secondly, a good image encryption algorithm/cipher should have both \textit{confusion property} and \textit{diffusion property}, which are proposed in \cite{ShannonCipher} as criteria for secure ciphers. The \textit{confusion property} implies the cipher ability to encrypt an arbitrary plaintext image into random-like ciphertext, so that an adversary cannot extract any information about the plaintext image from ciphertext images. The \textit{diffusion property} implies the cipher has capability to diffuse even a slight change in a plaintext image over the entire ciphertext image. Furthermore, a good image encryption algorithm/cipher should also be secure with respect to attacks against encryption keys \cite{CryptographyBook} and robust under noisy environment \cite{Zhou2012PFibonacci}. In this section, we discuss the security of the proposed LSIC under all above analysis and deeply study the LSIC performance with extensive simulation results and comparisons.
\subsection{Theoretical Analysis}
\subsubsection{Brute-Force Attack}
In cryptography, a brute-force attack, or exhaustive key search, is a strategy that is able to crack any encrypted data by searching all possible keys in the key space until the right key is found. It is well known that the key length, \ie the size of the key space used in an encryption system, determines the practical feasibility of performing a brute-force attack: longer keys means more difficulties than shorter ones in the terms of the time complexity.

The proposed Latin square image cipher has a key space of $256$ bits, which is larger than or equal to the mainstream cipher standards such as DES \cite{DES} and AES \cite{AES}, where the later one $192$-bit key is known to be large enough to resist brute-force attacks. Therefore, the proposed Latin square image cipher also has a sufficient large key space against brute-force attack.

It is worthwhile to note the theoretical key space of the proposed Latin square image cipher has a even larger key space than $256$-bit. As stated in Section III, the proposed cipher depends on eight $256\times 256$ Latin squares. Regardless of all other methods for generating these Latin square, consider the Latin square generator in Algorithm 1 where a Latin square is uniquely determined by two length $256$ permutation sequence, then it is not difficult to see that the number of Latin squares can be generated by Algorithm 1 is $256!\times 256! \approx 2^{3368}$. Consequently, the theoretical key size allowed in the cipher is about $(2^{3368})^8 = 2^{26944}$, \ie about $26944$ bits. This implies that the adopted key length in the Latin square image cipher can be easily increased to against brute-force attacks in future, when the current $256$-bit key length is no longer long enough. Therefore, the proposed Latin square image cipher has a extremely large key space to against brute-force attacks.

\subsubsection{Ciphertext and Plaintext Attacks}
In cryptography, ciphertext and plaintext are commonly used in attack models to analyze the security of an encryption system. Specifically speaking, we consider a ciphertext-only attack assuming that an adversary is only able to access a set of ciphertext, a known plaintext attack assuming that an adversary is able to access a set of plaintexts and corresponding ciphertexts, and a chose-plaintext attack assuming that an adversary is able to access arbitrary plaintexts to be encrypted and obtains the corresponding ciphertexts. As can be seen from attack assumptions, the chosen-plaintext attack awards an adversary the most information about plaintexts and ciphertexts among the three attack models. Therefore, if a cipher is able to resist chosen-plaintext attacks, then it is also immune to ciphertext-only and known-plaintext attacks.

The proposed Latin square image cipher has good resistance to chosen-plaintext attacks because of three reasons listed below:
\begin{itemize}
  \item Nonlinear Key Translation: instead of directly applying the key the cipher for encryption in a conventional block cipher, \eg key whitening, an encryption key used in the proposed Latin square image cipher is first translated to eight $256\times 256$ Latin squares and then used for contructing key dependent encryption primitives for image encryption. Therefore, the relation between the key and ciphertext is more involved and complicated than the conventional key usage.
  \item Probabilistic Encryption: the proposed Latin square image cipher has a pre-process realizing probabilistic encryption by introducing a small amount of noise to the LSB of a plaintext image. As a result, the cipher is able to generate significantly different ciphertext images even when the key and the plaintext image are both unchanged, but to leave deciphered images visually unchanged. This way helps the Latin square image cipher achieves semantically secure \cite{646128}.
  \item Dynamic S-boxes and P-boxes: Since all S-boxes and P-boxes used in the Latin square image cipher are dependent on Latin squares used in each round, they are dynamic instead of fixed. This fact implies that even if an adversary is able to crack an encryption key, all other keys are still safe because P-boxes and S-boxes associated with these keys are unbroken. As a result, ciphertext images encrypted by other keys are still safe.
\end{itemize}
Therefore, the proposed Latin square image cipher has good resistance to these ciphertext and pliantext attacks.

\subsection{Confusion Property Analysis}
The confusion property \cite{ShannonCipher} of a secure cipher emphasizes the ability of making the relationship between the key and the ciphertext as complex and involved as possible. If a cipher has good confusion property, it is very difficult to find the encryption key, even if a large number of plaintext-ciphertext pairs produced by the same encryption key are available for an adversary. In other words, a cipher with good confusion property should have good resistances to known-plaintext attacks \cite{CryptographyBook} and ciphertext-only attacks \cite{CryptographyBook}. To demonstrate the excellent confusion property of the proposed Latin square image cipher, we perform two types of statistical analysis: entropy and correlation.
\subsubsection{Entropy Analysis}
Information entropy is a quantitative measurement of how random a signal is. Because a digital image is a type of digital signal. The entropy analysis can be used to measure the randomness of a test image. The information entropy of an image can be defined as Eq. \eqref{EqEntropy}, where $X$ denotes the test image, $x_i$ denotes the ${i^{th}}$ possible value in $X$, and ${\textrm{Pr}}(x_i)$ is the probability of $X=x_i$, \ie the probability of pulling a random pixel in $X$ and its value is $x_i$. The maximum of $H(X)$ is achieved when $X$ is uniformly distributed as shown in Eq. \eqref{EqUniform}, \ie $X$ has a complete flat histogram and the symbol $F$ denotes the number of allowed intensity scales associated with the image format. For example, $F = 256$ is the number of intensity scales of a grayscale image or a RGB color image.
\begin{equation}
    \label{EqEntropy}
    H(X) = -\sum_{i=1}^n{{\textrm{Pr}(x_i)}\log_2{\textrm{Pr}(x_i)}}
\end{equation}
\begin{equation}
    \label{EqUniform}
    {\textrm{Pr}(X=x_i)}=1/F
\end{equation}

For the test images in Table \ref{Tab:USC_SIPI}, $F = 256$ and therefore the upper-bound of information entropy is $8$. The comparison results of information entropy between the proposed Latin square image cipher and peer algorithms are listed in Table \ref{tab:entropyLenna}, where the method with the highest entropy score is shaded for each test image. These results show that the proposed Latin square image cipher outperforms listed peer image encryption algorithms/methods.
\begin{table*}[htbp]
\scriptsize
\centering
\begin{minipage}[b]{.35\linewidth}
\centering
\begin{tabular}{rp{2.5cm}|l}
  \hline\hline
  \textbf{$P$}&\textbf{Method} & \textbf{Entropy}\\\hline
  &Chen \etal, 2004 \cite{3DCat} & 7.9938 \\
  \textit{L}& Xiang \etal,2006 \cite{xiang2006novel} & 7.9950 \\
  \textit{e}&Wong \etal, 2003 \cite{wong2003chaotic} (reported in \cite{Sun2010}) & 7.9690 \\
  \textit{n}&Zhang \etal, 2010 \cite{Zhang2010DNA} & 7.9980 \\
  \textit{n}&Zhu \etal, 2010 \cite{zhu2010chaos} & 7.9901 \\
  \textit{a}&Sun \etal,2010 \cite{Sun2010} & 7.9965 \\
  & Ours & 7.997161 \\
  \hline\hline
\end{tabular}
\end{minipage}\hfill
\begin{minipage}[b]{.65\linewidth}
\caption{Algorithm comparison with information entropy} 
\label{tab:entropyLenna}
\centering
\begin{tabular}{cc|p{1.2cm}p{1cm}p{1.5cm}p{1.2cm}}
  \hline\hline
  && \multicolumn{4}{c}{\textbf{Entropy}}\\
   \textbf{$P$}&\textbf{Channel$\backslash$Method} & Wang \etal 2011 \cite{wang2011new} & Liu \etal 2011 \cite{liu2011color} &Patidar \etal 2011 \cite{Patidar2011} & Ours \\\hline
  &{Red} & 7.999324 & 7.9871 & 7.9957 & 7.999351\\
  4.2.04&{Green} & 7.999371 & 7.9802 & 7.9963 & 7.999416\\
  (\textit{ColorLenna})&{Blue} & 7.999292 & 7.9878 & 7.9951 & 7.999314\\\hline
  &{Red} &  N/a&7.9877 & 7.9952 & 7.999309\\
  4.2.07&{Green} & N/a &7.9881 & 7.9959 & 7.999285 \\
  (\textit{ColorPepper})&{Blue} & N/a &7.9877 & 7.9954 & 7.999258\\
  \hline\hline
\end{tabular}
\end{minipage}\hfill
\end{table*}
\begin{table*}[htbp]
\caption{Results of information entropy analysis}
\label{tab:entropy}
\scriptsize
\centering
\begin{minipage}[b]{.95\linewidth}
\centering
\begin{tabular}{l|c|ccc}
\hline\hline
 & &\multicolumn{3}{c}{\textbf{Ciphertext}} \\
 \textbf{File} & \textbf{Plaintext}& \textbf{Ours} & \textbf{bmpPacker-AES\footnotemark[2]} &\textbf{Chen \etal, 2004\cite{3DCat} }\\\hline
4.1.01 & 6.898139 & \cellcolor[gray]{0.9}{7.998963} & 7.988561 & 7.998430\\
4.1.02 & 6.294498 & 7.999027 & 7.979298 & \cellcolor[gray]{0.9}{7.999061}\\
4.1.03 & 5.970916 & \cellcolor[gray]{0.9}{7.999122} & 7.990187 & 7.999113\\
4.1.04 & 7.426957 & 7.998948 & 7.990089 & \cellcolor[gray]{0.9}{7.999032}\\
4.1.05 & 7.068625 & \cellcolor[gray]{0.9}{7.999140} & 7.981035 & 7.998891\\
4.1.06 & 7.537089 & \cellcolor[gray]{0.9}{7.998936} & 7.989820 & 7.998902\\
4.1.07 & 6.583485 & \cellcolor[gray]{0.9}{7.999241} & 7.990487 & 7.999033\\
4.1.08 & 6.852716 & \cellcolor[gray]{0.9}{7.999199} & 7.997651 & 7.999182\\
4.2.01 & 7.242831 & \cellcolor[gray]{0.9}{7.999808} & 7.998747 & 7.999760\\
4.2.02 & 6.416494 & 7.999756 & 7.997469 & \cellcolor[gray]{0.9}{7.999763}\\
4.2.03 & 7.762436 & \cellcolor[gray]{0.9}{7.999765} & 7.997581 & 7.999752\\
4.2.04 & 7.750197 & \cellcolor[gray]{0.9}{7.999799} & 7.997733 & 7.999772\\
4.2.05 & 6.663908 & \cellcolor[gray]{0.9}{7.999756} & 7.997463 & 7.999746\\
4.2.06 & 7.762170 & \cellcolor[gray]{0.9}{7.999789} & 7.997550 & 7.999747\\
4.2.07 & 7.669826 & 7.999766 & 7.908851 & \cellcolor[gray]{0.9}{7.999781}\\
5.1.09 & 6.709312 & \cellcolor[gray]{0.9}{7.997266} & 7.906739 & 7.997011\\
5.1.10 & 7.311807 & \cellcolor[gray]{0.9}{7.997667} & 7.941872 & 7.997144\\
5.1.11 & 6.452275 & 7.996775 & 7.929405 & \cellcolor[gray]{0.9}{7.99652}\\
5.1.12 & 6.705667 & \cellcolor[gray]{0.9}{7.996730} & 7.363568 & 7.996729\\
5.1.13 & 1.548314 & \cellcolor[gray]{0.9}{7.997299} & 7.903557 & 7.997143\\
5.1.14 & 7.342433 & 7.997022 & 7.992544 & \cellcolor[gray]{0.9}{7.997271}\\
5.2.08 & 7.201008 & \cellcolor[gray]{0.9}{7.999359} & 7.987446 & 7.999257\\
5.2.09 & 6.993994 & \cellcolor[gray]{0.9}{7.999240} & 7.984631 & 7.999199\\
5.2.10 & 5.705560 & 7.999250 & 7.998720 & \cellcolor[gray]{0.9}{7.999303}\\
5.3.01 & 7.523737 & \cellcolor[gray]{0.9}{7.999820} & 7.998599 & 7.999801\\
5.3.02 & 6.830330 & 7.999819 & 7.990318 & \cellcolor[gray]{0.9}{7.999831}\\
7.1.01 & 6.027415 & \cellcolor[gray]{0.9}{7.999353} & 7.989542 & 7.999321\\
7.1.02 & 4.004499 & 7.999284 & 7.983057 & \cellcolor[gray]{0.9}{7.999302}\\
7.1.03 & 5.495740 & \cellcolor[gray]{0.9}{7.999325} & 7.993621 & 7.999192\\
7.1.04 & 6.107418 & \cellcolor[gray]{0.9}{7.999315} & 7.983570 & 7.999313\\
7.1.05 & 6.563196 & \cellcolor[gray]{0.9}{7.999302} & 7.985406 & 7.999296\\
7.1.06 & 6.695283 & \cellcolor[gray]{0.9}{7.999279} &  7.985253 & 7.999268\\
7.1.07 & 5.991599 & \cellcolor[gray]{0.9}{7.999437} & 7.990304 & 7.999221\\
7.1.08 & 5.053448 & \cellcolor[gray]{0.9}{7.999371} & 7.983746 &  7.999236\\
7.1.09 & 6.189814 & \cellcolor[gray]{0.9}{7.999412} & 7.985088 & 7.999271\\
7.1.10 & 5.908790 & \cellcolor[gray]{0.9}{7.999391} & 7.998478 & 7.999235\\
7.2.01 & 5.641454 & 7.999811 & 7.985051 & \cellcolor[gray]{0.9}{7.999829}\\
boat.512 & 7.191370 & 7.999353 & 7.989566 & \cellcolor[gray]{0.9}{7.999401}\\
elaine.512 & 7.505984 & \cellcolor[gray]{0.9}{7.999252} & 6.492821 & 7.999231\\
gray21.512 & 4.392295 & 7.999364 & 7.979908 & \cellcolor[gray]{0.9}{7.999772}\\
house & 7.485787 & \cellcolor[gray]{0.9}{7.999786} & 7.997553 & 7.999781\\
numbers.512 & 7.729247 & \cellcolor[gray]{0.9}{7.999263} &6.893247 & 7.999201\\
ruler.512 & 0.500033 & 7.999258 & 7.903134 & \cellcolor[gray]{0.9}{7.999298}\\
testpat.1k & 4.407726 & \cellcolor[gray]{0.9}{7.999809} & 7.903132 & 7.999800\\\hline
& \textbf{\# Best} & 31 & 0 & 13\\
\textbf{Statistics} & \textbf{Mean} & 7.999105&	 7.905055&7.9990486\\
& \textbf{Std} & 0.000856 &	0.290957 &	0.000903\\
\hline\hline
\end{tabular}
\end{minipage}\hfill
\end{table*}

The results of information entropy analysis for the complete set of test images are listed in Table \ref{tab:entropy}. These results illustrate that the proposed Latin square image cipher
\begin{itemize}
   \item obtains ciphertext images whose information entropy scores much closer to the entropy upper-bound than those encrypted by other encryption methods;
   \item is effective and robust to different image contents and pixel intensity statistics;
   \item outperforms the conventional AES cipher implemented by bmpPacker and the method proposed in \cite{3DCat} in most cases.
\end{itemize}

\subsubsection{Adjacent Pixel Correlation Analysis}
Adjacent pixel correlation (APC) is a quantitative measurement for image randomness. It measures the correlations between adjacent pixels within an image. The adjacent pixel correlation can be defined as an autocorrelation function as shown in Eq. \eqref{EqCorrd}, where $X_t$ denotes an extracted pixel sequence and $X_{t+1}$ denotes a pixel sequence where each pixel is the adjacent pixel of the corresponding pixel in $X_{t}$, $\mu$ is the mean value defined by Eq. \eqref{EqMean} and $\sigma$ is the standard deviation defined by Eq. \eqref{EqStd}, the definition of mathematical expectation is given in Eq. \eqref{EqExpectation}.

\begin{equation}
    \label{EqCorrd}
    APC = {\textrm{E}}[(X_t-\mu)(X_{t+1}-\mu)]/\sigma^2
\end{equation}
\begin{equation}
    \label{EqMean}
    \mu = {\textrm{E}}[X]
\end{equation}
\begin{equation}
    \label{EqStd}
    \sigma = \sqrt{{\textrm{E}}[(X-\mu)^2]}
\end{equation}
\begin{equation}
    \label{EqExpectation}
    {\textrm{E}}[x] = \sum_{i=1}^{N}x_i/N
\end{equation}
The closer to zero this correlation coefficient is, the weaker relationship between the pixel sequence and its adjacent pixel sequence. Because the adjacent pixel sequence can be extracted from the horizontal, vertical, or diagonal direction, the adjacent pixel correlation will be analyzed with respect to these three directions.

\begin{figure}[h]
  \scriptsize
  \centering
\begin{minipage}[c]{.9\linewidth}
  \begin{minipage}[c]{.25\linewidth}
  \centering
  \centerline{\includegraphics[width=.85\linewidth]{Lenna}}
  \vspace{5pt}
  \centerline{(a)}
  \end{minipage}\hfill
  \begin{minipage}[c]{.25\linewidth}
  \centering
  \centerline{\includegraphics[width=.95\linewidth]{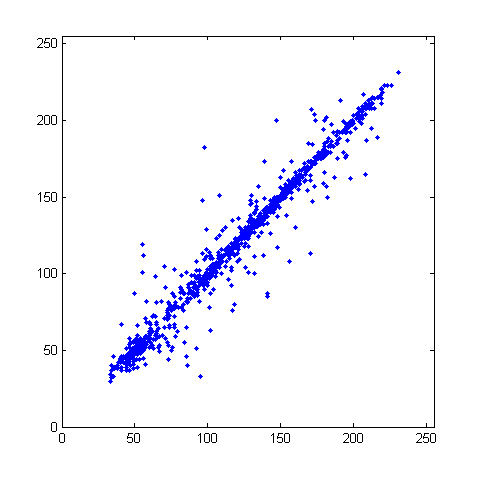}}
  \centerline{(b)}
  \end{minipage}\hfill
  \begin{minipage}[c]{.25\linewidth}
  \centering
  \centerline{\includegraphics[width=.95\linewidth]{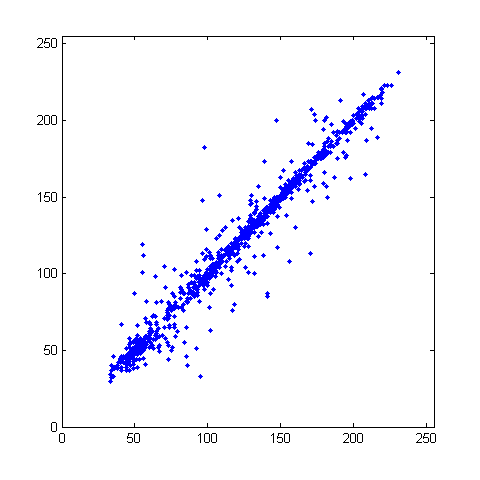}}
  \centerline{(c)}
  \end{minipage}\hfill
    \begin{minipage}[c]{.25\linewidth}
  \centering
  \centerline{\includegraphics[width=.95\linewidth]{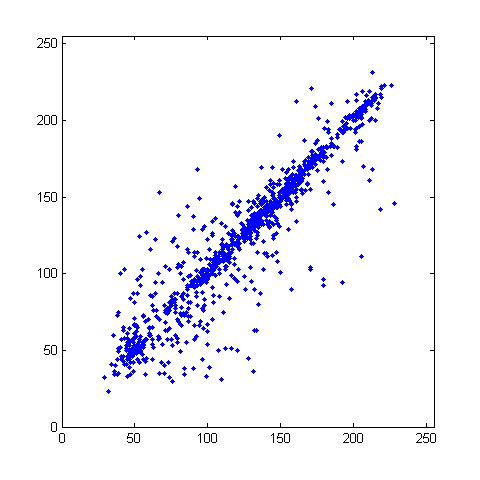}}
  \centerline{(d)}
  \end{minipage}\hfill
  \begin{minipage}[c]{.25\linewidth}
  \centerline{\includegraphics[width=.85\linewidth]{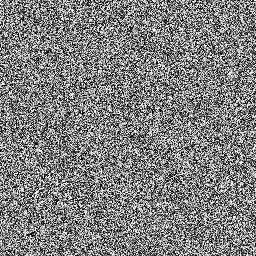}}
  \vspace{5pt}
  \centerline{(e)}
  \end{minipage}\hfill
  \begin{minipage}[c]{.25\linewidth}
  \centering
  \centerline{\includegraphics[width=.95\linewidth]{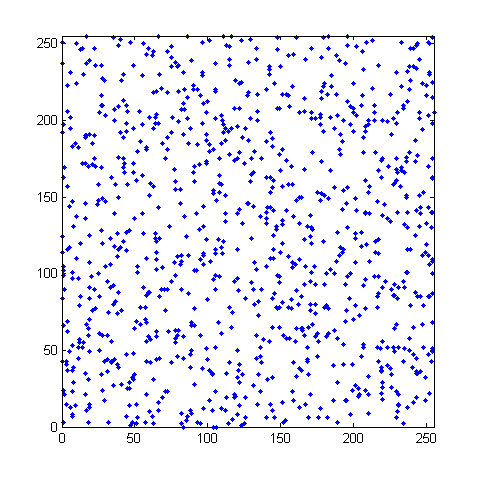}}
  \centerline{(f)}
  \end{minipage}\hfill
  \begin{minipage}[c]{.25\linewidth}
  \centering
  \centerline{\includegraphics[width=.95\linewidth]{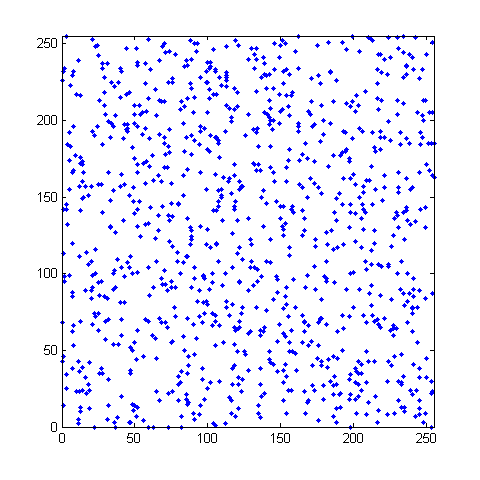}}
  \centerline{(g)}
  \end{minipage}\hfill
    \begin{minipage}[c]{.25\linewidth}
  \centering
  \centerline{\includegraphics[width=.95\linewidth]{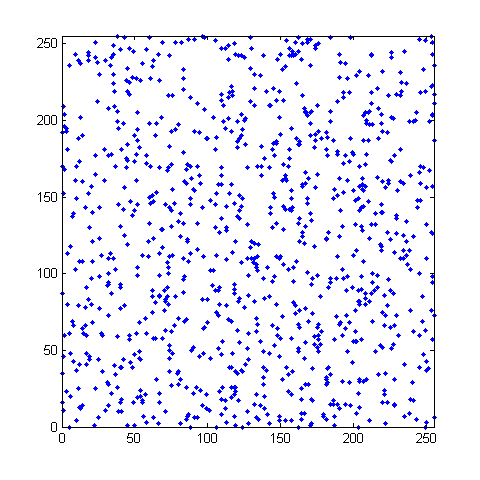}}
  \centerline{(h)}
  \end{minipage}\hfill\end{minipage}\hfill
  \caption{Adjacent pixels correlations before and after encryption - (a) plaintext \textit{Lenna} $P$, (b) horizontal adjacent pixels in $P$, (c) vertical adjacent pixels in $P$, (d) diagonal adjacent pixels in $P$, (e) ciphertext $C = \mathfrak{E}(P,K)$, (f) horizontal adjacent pixels in $C$, (g) vertical adjacent pixels in $C$, and (h) diagonal adjacent pixels in $C$. }\label{Fig:APC}
\end{figure}

\begin{table}[h]
\caption{Comparison of APC scores for the image \textit{Lenna}}
\label{Tab:Lenna_APCA}
\centering
\scriptsize
\begin{tabular}{lr@{.}lr@{.}lr@{.}l}
\hline\hline
\textbf{Encryption Method} & \multicolumn{2}{c}{\textbf{Horizontal}} & \multicolumn{2}{c}{\textbf{Vertical}} & \multicolumn{2}{c}{\textbf{Diagonal}}\\\hline
Original \textit{Lenna} & 0 & 9400 & 0 & 9709 & 0 & 9710\\
Awad, 2011 \cite{Awad2011} & 0& 0127 & -0 &0093 &-0 & 0059\\
Mao \etal, 2004 \cite{3DCat} & -0& 00024 & -0 &24251 & 0 & 23644\\
Liao \etal, 2010 \cite{Liao2010} & 0 & 0127 & -0 & 0190 & -0 & 0012\\
Fu \etal, 2011 \cite{Fu2011} & 0 & 0368 & 0 & 0392 & 0 & 0068\\
Ours & 0 & 0053365 & -0 & 0027616 & 0 & 0016621\\\hline\hline
\end{tabular}
\end{table}

Fig. \ref{Fig:APC} shows the randomly selected 1024 pairs of adjacent pixels along the horizontal, vertical and diagonal directions. The horizontal axis of the scatter plot denotes the intensity of one randomly selected pixel and the vertical axis denotes the intensity of its adjacent pixel. The APC results for the test image \textit{Lenna} are listed in Table \ref{Tab:Lenna_APCA}. The proposed Latin square image cipher decorrelates adjacent pixels in the plaintext image and outperforms the listed encryption algorithms \cite{Awad2011,3DCat,Kumar2011,Liao2010,Fu2011} due to its smaller correlation coefficients.

Table \ref{tab:ajacentPixel} shows the APCs for test plaintext images and their corresponding ciphertext images using the Latin square image cipher and peer algorithms, where the best APC score is shaded for each test image. These results further demonstrates that the Latin square image cipher outperforms peer algorithms in most trails with the smallest mean and  standard deviation of APCs.

\begin{table}[h]
\caption{Results of adjacent pixel correlation analysis}
\label{tab:ajacentPixel}
\scriptsize
\centering
\begin{minipage}[b]{.98\linewidth}
\centering
\begin{tabular}{l|p{1cm}|p{1.5cm}p{1.5cm}p{1.5cm}p{1.5cm}p{.5cm}}
\hline\hline
 \textbf{} & \multicolumn{1}{c|}{\textbf{$|$APC(Plaintext)$|$}}& \multicolumn{5}{c}{\textbf{$|$APC(Ciphertext)$|\times 10^{-3}$}}\\
\textbf{File} & \multicolumn{1}{c|}{\textbf{$\times10^{-3}$}} & {bmpPacker-AES} & {Mao \etal, 2005 \cite{Mao2005}} & {Chen \etal, 2004 \cite{3DCat}} & {Pareek \etal, 2006 \cite{Pareek2006}} & {Ours} \\\hline\hline
4.1.01 & 955.730 & 11.397 & 1.780 & \cellcolor[gray]{0.9}{1.240} & 3.961 & 1.382  \\
4.1.02 & 926.227 & 13.327 & 3.000 & 1.973 & 8.119 & \cellcolor[gray]{0.9}{1.384}  \\
4.1.03 & 922.433 & 24.657 & \cellcolor[gray]{0.9}{1.760} & 2.170 & 5.314 & 1.956  \\
4.1.04 & 959.193 & 9.840 & 1.443 & \cellcolor[gray]{0.9}{0.883} & 11.863 & 1.097  \\
4.1.05 & 953.143 & 11.130 & 0.640 & 1.437 & 9.243 & \cellcolor[gray]{0.9}{0.432}  \\
4.1.06 & 932.417 & 26.303 & 1.487 & 0.910 & 3.994 & \cellcolor[gray]{0.9}{0.208}  \\
4.1.07 & 979.317 & 10.723 & 2.050 & \cellcolor[gray]{0.9}{1.680} & 1.905 & 2.572  \\
4.1.08 & 972.013 & 11.790 & 1.393 & 2.187 & 2.922 & \cellcolor[gray]{0.9}{0.722}  \\
4.2.01 & 988.877 & 6.757 & 0.790 & \cellcolor[gray]{0.9}{0.427} & 7.686 & 0.533  \\
4.2.02 & 945.423 & 3.790 & 1.137 & \cellcolor[gray]{0.9}{1.050} & 8.184 & 1.402  \\
4.2.03 & 857.587 & 6.603 & 1.083 & \cellcolor[gray]{0.9}{0.897} & 7.076 & 1.036  \\
4.2.04 & 978.600 & 6.940 & 1.523 & 0.743 & 2.325 & \cellcolor[gray]{0.9}{0.557}  \\
4.2.05 & 943.307 & 7.493 & 1.067 & \cellcolor[gray]{0.9}{0.587} & 2.833 & 2.199  \\
4.2.06 & 959.510 & 7.117 & 1.263 & \cellcolor[gray]{0.9}{0.977} & 8.145 & 1.950  \\
4.2.07 & 974.480 & 5.347 & 1.290 & 0.990 & 0.815 & \cellcolor[gray]{0.9}{0.146}  \\
5.1.09 & 911.973 & 59.040 & 3.053 & 5.233 & \cellcolor[gray]{0.9}{0.779} & 4.829  \\
5.1.10 & 853.567 & 61.290 & 7.663 & 7.397 & 7.672 & \cellcolor[gray]{0.9}{0.188}  \\
5.1.11 & 890.580 & 33.700 & 2.580 & 4.567 & 4.110 & \cellcolor[gray]{0.9}{1.766}  \\
5.1.12 & 954.440 & 45.793 & 4.097 & 4.943 & 11.780 & \cellcolor[gray]{0.9}{3.443}  \\
5.1.13 & 831.833 & 162.693 & 4.160 & 3.697 & 17.896 & \cellcolor[gray]{0.9}{2.424}  \\
5.1.14 & 892.687 & 67.513 & 4.340 & \cellcolor[gray]{0.9}{1.407} & 8.989 & 4.538  \\
5.2.08 & 884.630 & 10.943 & 1.600 & 2.137 & 6.210 & \cellcolor[gray]{0.9}{0.615}  \\
5.2.09 & 850.460 & 18.137 & 1.817 & 1.253 & 6.024 & \cellcolor[gray]{0.9}{0.168}  \\
5.2.10 & 917.130 & 21.857 & 1.403 & \cellcolor[gray]{0.9}{1.070} & 1.512 & 2.442  \\
5.3.01 & 974.543 & 3.887 & 1.400 & 1.203 & \cellcolor[gray]{0.9}{0.297} & 0.752  \\
5.3.02 & 890.127 & 3.523 & 0.863 & 0.730 & 1.944 & \cellcolor[gray]{0.9}{0.158}  \\
7.1.01 & 926.903 & 10.853 & 1.890 & 2.217 & 6.857 & \cellcolor[gray]{0.9}{0.670}  \\
7.1.02 & 928.663 & 13.270 & 3.263 & 1.747 & 6.561 & \cellcolor[gray]{0.9}{1.017}  \\
7.1.03 & 925.900 & 22.620 & 0.957 & 1.877 & 11.244 & \cellcolor[gray]{0.9}{0.352}  \\
7.1.04 & 958.447 & 8.640 & 1.297 & \cellcolor[gray]{0.9}{0.780} & 2.139 & 2.735  \\
7.1.05 & 914.000 & 22.010 & 1.857 & \cellcolor[gray]{0.9}{1.397} & 6.582 & 1.407  \\
7.1.06 & 908.887 & 22.580 & 1.373 & 1.170 & 1.338 & \cellcolor[gray]{0.9}{0.331}  \\
7.1.07 & 866.260 & 21.397 & 0.483 & 1.577 & 2.992 & \cellcolor[gray]{0.9}{0.432}  \\
7.1.08 & 934.933 & 12.813 & 1.257 & 2.833 & 6.279 & \cellcolor[gray]{0.9}{0.045}  \\
7.1.09 & 935.977 & 24.920 & 0.633 & 1.213 & 10.011 & \cellcolor[gray]{0.9}{0.025}  \\
7.1.10 & 946.177 & 20.797 & 2.547 & \cellcolor[gray]{0.9}{1.707} & 6.344 & 2.182  \\
7.2.01 & 951.530 & 5.330 & 1.180 & 0.613 & 4.317 & \cellcolor[gray]{0.9}{0.273}  \\
boat.512 & 942.427 & 19.453 & 1.207 & \cellcolor[gray]{0.9}{0.887} & 9.823 & 2.120  \\
elaine.512 & 969.757 & 14.950 & \cellcolor[gray]{0.9}{1.607} & 2.777 & 7.839 & 4.381  \\
house & 993.220 & 116.010 & 1.213 & 2.507 & 10.950 & \cellcolor[gray]{0.9}{0.757}  \\
gray21.512 & 941.383 & 5.990 & 1.590 & \cellcolor[gray]{0.9}{1.280} & 12.766 & 1.359  \\
numbers.512 & 692.123 & 26.230 & 2.257 & \cellcolor[gray]{0.9}{1.733} & 11.103 & 1.948  \\
ruler.512 & 313.253 & 51.003 & 1.947 & \cellcolor[gray]{0.9}{0.890} & 3.126 & 1.270  \\
testpat.1k & 752.093 & 45.560 & 1.557 & 1.067 & 0.720 & \cellcolor[gray]{0.9}{0.703}  \\\hline
 & \multicolumn{1}{c|}{\textbf{\#Best}} & 0 & 2 & 17 & 2 & 23 \\
Statistics & \multicolumn{1}{c|}{\textbf{Mean}} & 25.364 & 1.882 & 1.820 & 6.195 & 1.384 \\
 & \multicolumn{1}{c|}{\textbf{Std}} & 30.331 & 1.275 & 1.406 & 3.968 & 1.222 \\ \hline\hline
\end{tabular}
\label{tab:APC}
\end{minipage}\hfill
\end{table}


\subsection{Diffusion Property Analysis}
The diffusion property describes the cipher ability of diffusing a change in a plaintext image over its corresponding ciphertext image. If a cipher is weak in diffusion, it might be vulnerable against the differential attacks \cite{CryptographyBook}. In image encryption, the number of changing pixel rate (NPCR) and the unified averaged changed intensity (UACI) are two most common quantities used for evaluating the resistance of differential attacks against an image encryption method/algorithm/cipher \cite{3DCat,Zhu2011,Liao2010}.

Mathematically, the NPCR ${\cal N}(C^1,C^2)$ and UACI ${\cal U}(C^1,C^2)$ scores between two ciphertext images $C^1$ and $C^2$, whose plaintext images are slightly different, can be defined as Eqs. \eqref{EqNPCR} and \eqref{EqUACI}, respectively. The image of their difference $Diff(i,j)$ is defined in Eq. \eqref{EqNPCRdiff} and denotes whether two pixels located at the image grid $(i,j)$ of $C^1$ and $C^2$ are equal. The symbols $T$ and $S$ denote the number of pixels in the ciphertext image and the number of allowed pixel intensity scales, respectively.

\begin{equation}
    \label{EqNPCR}
    {\cal{N}}(C^1,C^2) = \sum_{i=1}^{M}\sum_{j=1}^{N}\frac{Diff(i,j)}{T}\times 100\%
\end{equation}

\begin{equation}
    \label{EqUACI}
    {\cal{U}}(C^1,C^2) = \sum_{i=1}^{M}\sum_{j=1}^{N}\frac{|C^1(i,j)-C^2(i,j)|}{S\cdot T}\times 100\%
\end{equation}

\begin{equation}
\label{EqNPCRdiff}
    Diff(i,j) =
  \left\{
   \begin{aligned}
   0, & {\textrm{ if } C^1(i,j) = C^2(i,j)}\\
   1, & {\textrm{ if } C^1(i,j) \neq C^2(i,j)}\\
   \end{aligned}
   \right.
\end{equation}
It is noticeable that NPCR concentrates on the absolute number of pixels which changes values in differential attacks, while the UACI focuses on the averaged difference between the paired ciphertext images.

\begin{figure*}[htbp]
  \centering
     \begin{minipage}[c]{0.5\linewidth}
     \begin{minipage}[c]{0.33\linewidth}
      \centering
     \centerline{\includegraphics[width=.85\linewidth]{Lenna}}
     \centerline{\scriptsize (a)}
    \end{minipage}\hfill
     \begin{minipage}[c]{0.33\linewidth}
      \centering
     \centerline{\includegraphics[width=.85\linewidth]{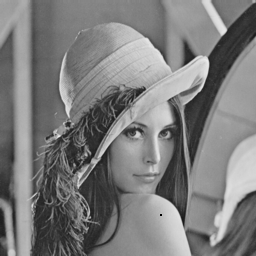}}
     \centerline{\scriptsize (b)}
    \end{minipage}\hfill
         \begin{minipage}[c]{0.33\linewidth}
      \centering
     \centerline{\includegraphics[width=.85\linewidth]{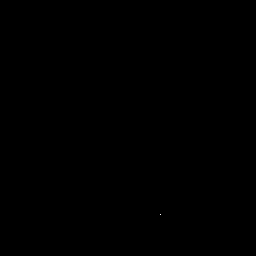}}
     \centerline{\scriptsize (c)}
    \end{minipage}\hfill
         \begin{minipage}[c]{0.33\linewidth}
      \centering
     \centerline{\includegraphics[width=.95\linewidth]{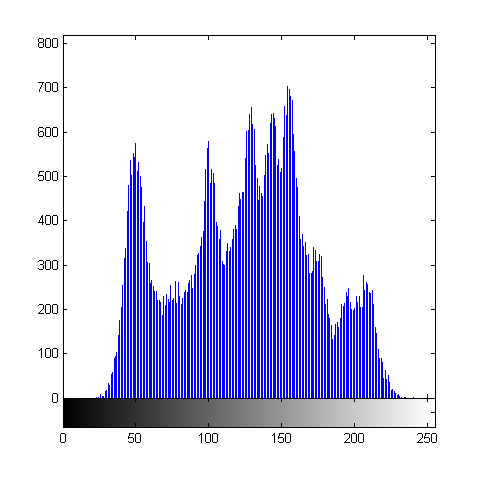}}
    \end{minipage}\hfill
     \begin{minipage}[c]{0.33\linewidth}
      \centering
     \centerline{\includegraphics[width=.95\linewidth]{Hlenna3}}
    \end{minipage}\hfill
             \begin{minipage}[c]{0.33\linewidth}
      \centering
     \centerline{\includegraphics[width=.95\linewidth]{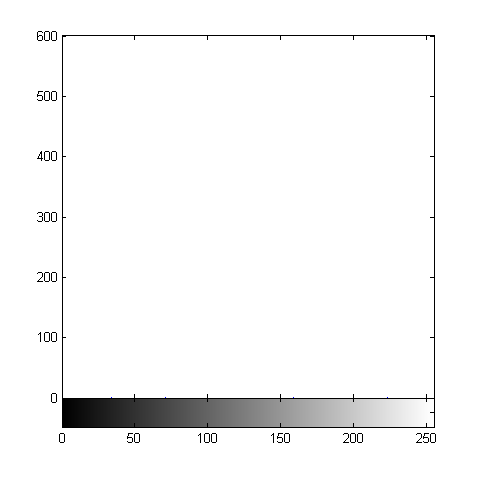}}
    \end{minipage}\hfill  \end{minipage}\hfill
      \begin{minipage}[c]{0.5\linewidth}
     \begin{minipage}[c]{0.33\linewidth}
      \centering
     \centerline{\includegraphics[width=.85\linewidth]{Elenna}}
     \centerline{\scriptsize (d)}
    \end{minipage}\hfill
     \begin{minipage}[c]{0.33\linewidth}
      \centering
     \centerline{\includegraphics[width=.85\linewidth]{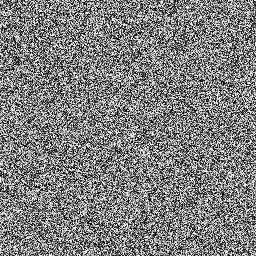}}
     \centerline{\scriptsize (e)}
    \end{minipage}\hfill
         \begin{minipage}[c]{0.33\linewidth}
      \centering
     \centerline{\includegraphics[width=.85\linewidth]{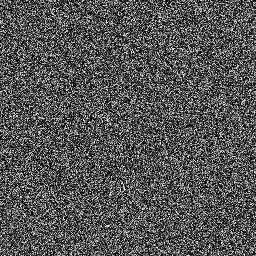}}
     \centerline{\scriptsize (f)}
    \end{minipage}\hfill
         \begin{minipage}[c]{0.33\linewidth}
      \centering
     \centerline{\includegraphics[width=.95\linewidth]{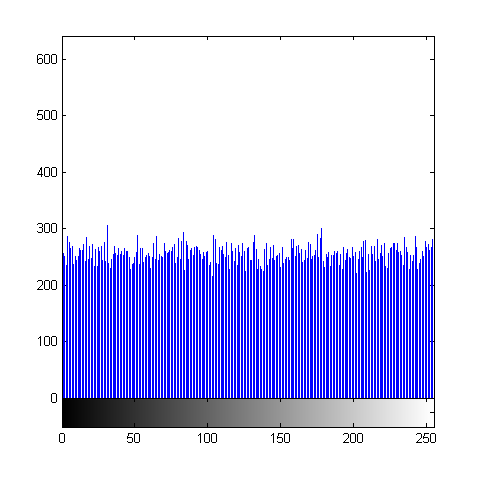}}
    \end{minipage}\hfill
     \begin{minipage}[c]{0.33\linewidth}
      \centering
     \centerline{\includegraphics[width=.95\linewidth]{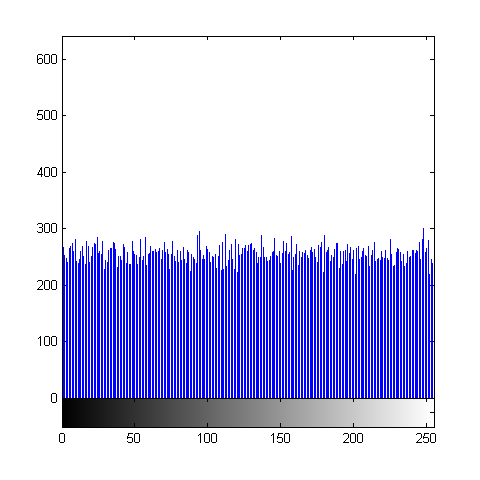}}
    \end{minipage}\hfill
             \begin{minipage}[c]{0.33\linewidth}
      \centering
     \centerline{\includegraphics[width=.95\linewidth]{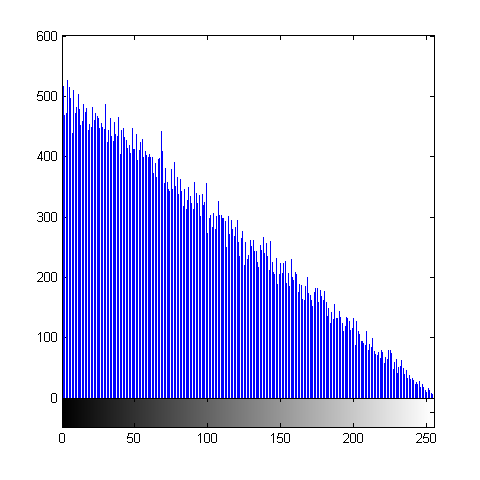}}
    \end{minipage}\hfill\end{minipage}\hfill
  \caption{Sample results of the cipher diffusion property - (a)plaintext \lenna original $P^1$, (b) plaintext \lenna modified $P^2$, (c) plaintext difference $|P^1-P^2|$, (d) ciphertext $C^1 = \mathfrak{E}(P^1,K)$, (e) ciphertext $C^2 = \mathfrak{E}(P^2,K)$, and (f) ciphertext difference $|C^1-C^2|$}
  \label{Fig:NPCRLenna}
\end{figure*}

Fig. \ref{Fig:NPCRLenna} shows the sample results of the diffusion property of the Latin square image cipher. The plaintext image $P^1$ differs from $P^2$ only one pixel on \textit{Lenna}'s shoulder. $C^1$ and $C^2$ are corresponding ciphertext images using the same encryption key. As can be seen, a single pixel difference between $P^1$ and $P^2$ diffuses to the entire cipher image and leads to a significant difference between $C^1$ and $C^2$. Table \ref{Tab:Lenna_NPCR_UACI} shows corresponding NPCR and UACI scores of $C^1$ and $C^2$ using the proposed Latin square image cipher and other encryption algorithms.

\begin{table}[h]
\caption{Comparison of the NPCR and UACI scores of the image \textit{Lenna} }
\label{Tab:Lenna_NPCR_UACI}
\centering
\scriptsize
\begin{tabular}{lll}
\hline\hline
\textbf{Encryption Method} & \multicolumn{1}{c}{\textbf{NPCR\%}} & \multicolumn{1}{c}{\textbf{UACI\%}} \\\hline
Zhu \etal, 2011 \cite{Zhu2011} & 99.63 & 33.48 \\
Chen \etal, 2004 \cite{3DCat} & 99.25 & 33.14 \\
Liao \etal, 2010 \cite{Liao2010} & 99.65 & 33.48 \\
Ours & 99.6689 & 33.4936 \\\hline\hline
\end{tabular}
\end{table}

\begin{table}[h]
\caption{Results of Diffusion Properties}
\label{tab:diffusion}
\scriptsize
\centering
\begin{tabular}{l|c|c}
\hline\hline
 \textbf{File} & \textbf{UACI\%}& \textbf{NPCR\%}\\\hline
4.1.01 & 33.4402 & 99.5885 \\
4.1.02 & 33.3715 & 99.6272 \\
4.1.03 & 33.5130 & 99.6297 \\
4.1.04 & 33.5303 & 99.6023 \\
4.1.05 & 33.4018 & 99.6028 \\
4.1.06 & 33.5235 & 99.5956 \\
4.1.07 & 33.5855 & 99.6302 \\
4.1.08 & 33.4874 & 99.6089 \\
4.2.01 & 33.4763 & 99.6058 \\
4.2.02 & 33.4431 & 99.6179 \\
4.2.03 & 33.5031 & 99.6120 \\
4.2.04 & 33.5113 & 99.6146 \\
4.2.05 & 33.4475 & 99.6231 \\
4.2.06 & 33.4145 & 99.6113 \\
4.2.07 & 33.4708 & 99.6031 \\
5.1.09 & 33.6739 & 99.6170 \\
5.1.10 & 33.5039 & 99.6170 \\
5.1.11 & 33.5079 & 99.5529 \\
5.1.12 & 33.4536 & 99.6033 \\
5.1.13 & 33.3681 & 99.6033 \\
5.1.14 & 33.6094 & 99.5636 \\
5.2.08 & 33.5379 & 99.6067 \\
5.2.09 & 33.3704 & 99.6151 \\
5.2.10 & 33.5169 & 99.5987 \\
5.3.01 & 33.5039 & 99.6115 \\
5.3.02 & 33.4856 & 99.6077 \\
7.1.01 & 33.5084 & 99.5918 \\
7.1.02 & 33.4485 & 99.6124 \\
7.1.03 & 33.5236 & 99.6346 \\
7.1.04 & 33.5454 & 99.6029 \\
7.1.05 & 33.3958 & 99.6063 \\
7.1.06 & 33.4756 & 99.6262 \\
7.1.07 & 33.4469 & 99.5953 \\
7.1.08 & 33.5569 & 99.6128 \\
7.1.09 & 33.3964 & 99.5953 \\
7.1.10 & 33.3942 & 99.5880 \\
7.2.01 & 33.4813 & 99.6142 \\
boat.512 & 33.3792 & 99.5941 \\
elaine.512 & 33.4430 & 99.6185 \\
gray21.512 & 33.4651 & 99.6174 \\
house & 33.4570 & 99.6063 \\
numbers.512 & 33.4484 & 99.6258 \\
ruler.512 & 33.4471 & 99.6002 \\
testpat.1k & 33.4748 & 99.6018 \\\hline\hline
\end{tabular}
\end{table}

\subsection{Key Sensitivity Analysis}
\begin{figure*}[htbp]
\scriptsize
\centering
\begin{minipage}[b]{.9\linewidth}
  \begin{minipage}[b]{.25\linewidth}
    \centerline{\includegraphics[width=.85\linewidth]{Lenna}}
    \centerline{(a)}
  \end{minipage}\hfill
  \begin{minipage}[b]{.25\linewidth}
    \centerline{\includegraphics[width=.85\linewidth]{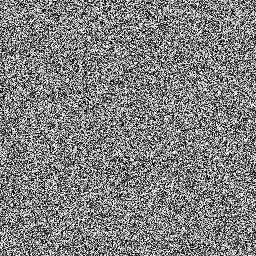}}
    \centerline{(b)}
  \end{minipage}\hfill
  \begin{minipage}[b]{.25\linewidth}
    \centerline{\includegraphics[width=.85\linewidth]{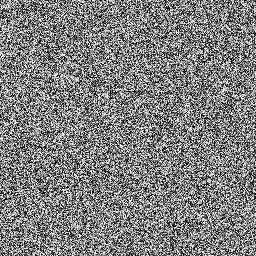}}
    \centerline{(c)}
  \end{minipage}\hfill
  \begin{minipage}[b]{.25\linewidth}
    \centerline{\includegraphics[width=.85\linewidth]{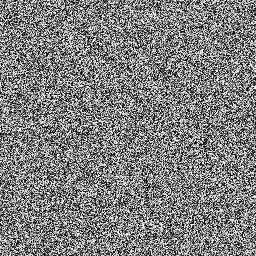}}
    \centerline{(d)}
  \end{minipage}\hfill
\end{minipage}\hfill
\begin{minipage}[b]{.682\linewidth}
    \begin{minipage}[b]{.33\linewidth}
    \centerline{\includegraphics[width=.85\linewidth]{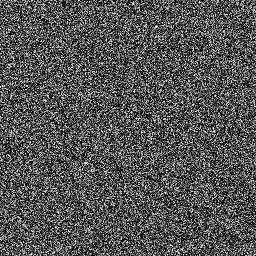}}
    \centerline{(e)}
  \end{minipage}\hfill
  \begin{minipage}[b]{.33\linewidth}
    \centerline{\includegraphics[width=.85\linewidth]{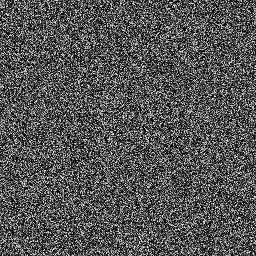}}
    \centerline{(f)}
  \end{minipage}\hfill
  \begin{minipage}[b]{.33\linewidth}
    \centerline{\includegraphics[width=.85\linewidth]{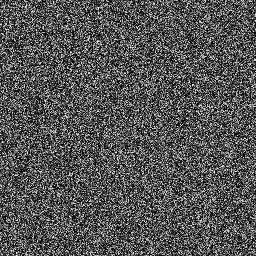}}
    \centerline{(g)}
  \end{minipage}\hfill
\end{minipage}\hfill
    \caption{Key sensitivity analysis for the encryption stage - (a) plaintext $P$ \textit{Lenna}, (b) ciphertext $C^1 = \mathfrak{E}(P,K^1)$, (c) ciphertext $C^2 = \mathfrak{E}(P,K^2)$, (d) ciphertext $C^3 = \mathfrak{E}(P,K^3)$, (e) ciphertext difference $|C^1-C^2|$, (f) ciphertext difference  $|C^2-C^3|$, and (g) ciphertext difference $|C^3-C^1|$}\label{fig:KeySensitivityEnc}
\end{figure*}

\begin{figure}[h]
\scriptsize
\centering
\begin{minipage}[b]{.8\linewidth}
  \begin{minipage}[b]{.25\linewidth}
    \centerline{\includegraphics[width=.85\linewidth]{E1_lenna}}
    \centerline{(a)}
  \end{minipage}\hfill
  \begin{minipage}[b]{.25\linewidth}
    \centerline{\includegraphics[width=.85\linewidth]{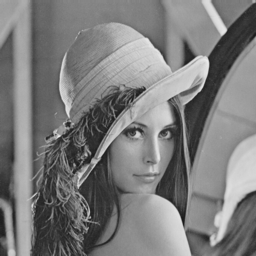}}
    \centerline{(b)}
  \end{minipage}\hfill
  \begin{minipage}[b]{.25\linewidth}
    \centerline{\includegraphics[width=.85\linewidth]{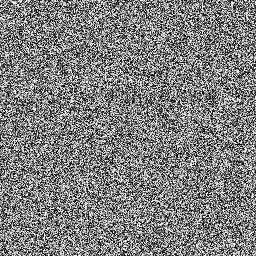}}
    \centerline{(c)}
  \end{minipage}\hfill
  \begin{minipage}[b]{.25\linewidth}
    \centerline{\includegraphics[width=.85\linewidth]{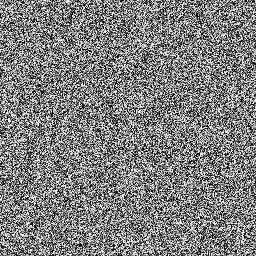}}
        \centerline{(d)}
  \end{minipage}\hfill
\end{minipage}\hfill
    \caption{Key sensitivity analysis for the decryption stage - (a) ciphertext $C^1$, (b) decipheredtext $D^1 = \mathfrak{D}(C^1,K^1)$, (c) decipheredtext $D^2= \mathfrak{D}(C^1,K^2)$, and (d) decipheredtext $D^3 = \mathfrak{D}(C^1,K^3)$}\label{fig:KeySensitivityDec}
\end{figure}

A secure cipher should have high key sensitivities in both encryption and decryption processes. Simulation results with respect to encryption and decryption stages are shown in Fig. \ref{fig:KeySensitivityEnc} and \ref{fig:KeySensitivityDec}. The encryption keys in our simulations are listed below in the HEX format:
\begin{eqnarray*}
  K^1 &=& 'B9B5ED7585C8B15D7454ED271AA3A3A3A07B00321C11759D0FDE340234384BC9' \\
  K^2 &=& 'B9B5ED7585C8B15D7454ED271AA3A3A3A07B00321C11759D0FDE340234384BC8' \\
  K^3 &=& '39B5ED7585C8B15D7454ED271AA3A3A3A07B00321C11759D0FDE340234384BC8'
\end{eqnarray*}
It is noticeable that $K^1$ differs $K^2$ only for the last bit; $K^2$ differs $K^3$ only for the first bit; and $K^1$ differs $K^3$ only for the first and the last bit. Although the hamming distances between $K^1,K^2$ and $K^3$ are very small, \ie they are very similar to each other, their corresponding ciphertext images $C^1,C^2$ and $C^3$ have significant differences. These can be verified by the fact that their differences \ie $|C^1-C^2|$, $|C^2-C^3|$ and $|C^3-C^1|$ are random-like images as shown in Fig. \ref{fig:KeySensitivityEnc}. The example shows the proposed Latin square image cipher is very sensitive to the encryption key in the encryption stage.

Similar results can also be obtained in the decryption stage as shown in Fig. \ref{fig:KeySensitivityDec}. We decrypt the same ciphertext image $C^1$ using the encryption keys $K^1,K^2$ and $K^3$, respectively. As can be seen, using the correct encryption key $K^1$, decrypted image $D^1$ perfectly reconstructs the original plaintext image. However, decrypted images $D^2$ and $D^3$ using key $K^2$ and $K^3$ are random-like ones which do not contain any information related to the original plaintext image.

\subsection{Noise Robustness Analysis}
\begin{figure*}[htbp]
  \centering
     \begin{minipage}[c]{0.5\linewidth}
     \begin{minipage}[c]{0.33\linewidth}
      \centering
     \centerline{\includegraphics[width=.85\linewidth]{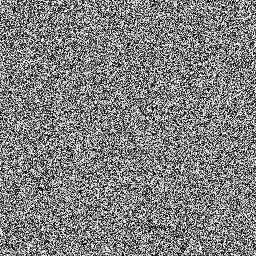}}
     \centerline{\scriptsize (a)}
    \end{minipage}\hfill
     \begin{minipage}[c]{0.33\linewidth}
      \centering
     \centerline{\includegraphics[width=.85\linewidth]{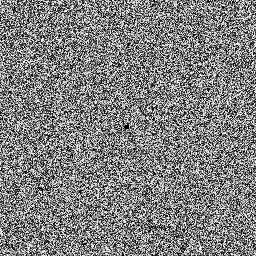}}
     \centerline{\scriptsize (b)}
    \end{minipage}\hfill
         \begin{minipage}[c]{0.33\linewidth}
      \centering
     \centerline{\includegraphics[width=.85\linewidth]{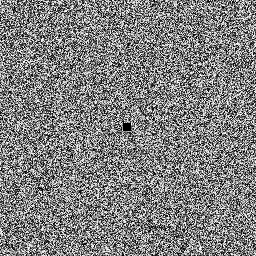}}
     \centerline{\scriptsize (c)}
    \end{minipage}\hfill
         \begin{minipage}[c]{0.33\linewidth}
      \centering
     \centerline{\includegraphics[width=.85\linewidth]{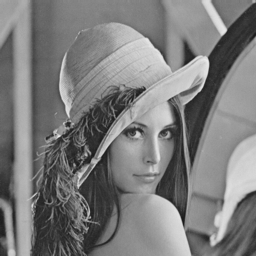}}
    \end{minipage}\hfill
     \begin{minipage}[c]{0.33\linewidth}
      \centering
     \centerline{\includegraphics[width=.85\linewidth]{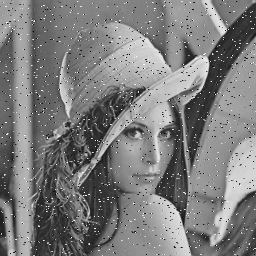}}
    \end{minipage}\hfill
             \begin{minipage}[c]{0.33\linewidth}
      \centering
     \centerline{\includegraphics[width=.85\linewidth]{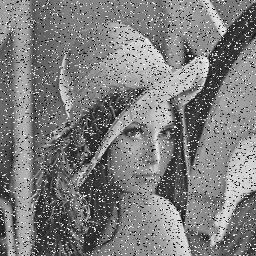}}
    \end{minipage}\hfill  \end{minipage}\hfill
      \begin{minipage}[c]{0.5\linewidth}
     \begin{minipage}[c]{0.33\linewidth}
      \centering
     \centerline{\includegraphics[width=.85\linewidth]{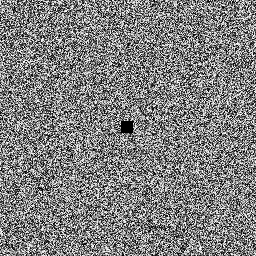}}
     \centerline{\scriptsize (d)}
    \end{minipage}\hfill
     \begin{minipage}[c]{0.33\linewidth}
      \centering
     \centerline{\includegraphics[width=.85\linewidth]{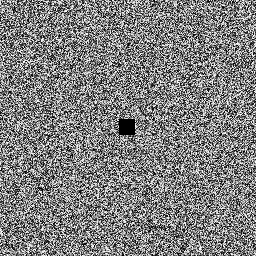}}
     \centerline{\scriptsize (e)}
    \end{minipage}\hfill
         \begin{minipage}[c]{0.33\linewidth}
      \centering
     \centerline{\includegraphics[width=.85\linewidth]{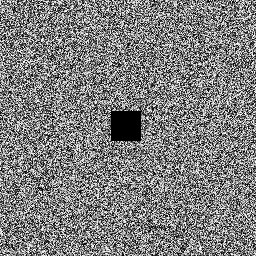}}
     \centerline{\scriptsize (f)}
    \end{minipage}\hfill
         \begin{minipage}[c]{0.33\linewidth}
      \centering
     \centerline{\includegraphics[width=.85\linewidth]{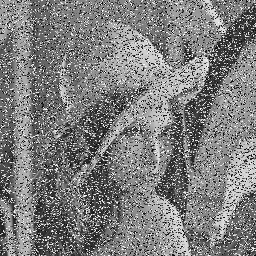}}
    \end{minipage}\hfill
     \begin{minipage}[c]{0.33\linewidth}
      \centering
     \centerline{\includegraphics[width=.85\linewidth]{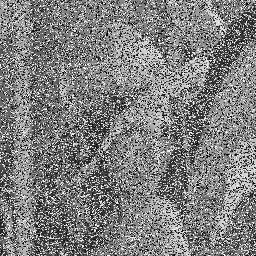}}
    \end{minipage}\hfill
             \begin{minipage}[c]{0.33\linewidth}
      \centering
     \centerline{\includegraphics[width=.85\linewidth]{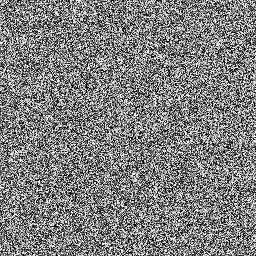}}
    \end{minipage}\hfill\end{minipage}\hfill
  \caption{Sample results of noise robustness in decryption - (a) ciphertext $C$ and its decryptedtext $D = \mathfrak{D}(C,K)$, (b) $C^1$ with $0.025\%$ noise and $D^1= \mathfrak{D}(C^1,K)$, (c) $C^2$ with $0.1\%$ noise and $D^2 = \mathfrak{D}(C^2,K)$, (d) $C^3$ with $0.22\%$ noise and $D^3 = \mathfrak{D}(C^3,K)$, (e) $C^4$ with $0.39\%$ noise and $D^4 = \mathfrak{D}(C^4,K)$, and (f) $C^5$ with $1.56\%$ noise and $D^5 = \mathfrak{D}(C^5,K)$}
  \label{Fig:NoiseLenna}
\end{figure*}

A good cipher should also be able to tolerate a certain amount of noise, \eg noise in a channel or decoding errors. As discussed previously, the proposed Latin square image cipher adopts an asymmetric structure for encryption and decryption, and one noisy pixel in ciphertext image will only propagate in a factor of two in each round. Fig. \ref{Fig:NoiseLenna} shows the results of the decryption robustness of the Latin square image cipher against various noise ratio in ciphertext images. After decryption, noise concentrating in the center square of a ciphertext image now distributed almost evenly over the deciphered image. Due to the psychovisual redundancy within an image, human vision system is still able to recognize the deciphered image contents as long as it is not fully unintelligible.

\section{Conclusion}
In this paper, we introduce a symmetric-key Latin square image cipher with probabilistic encryption for grayscale and color images. This new image cipher has distinctive characteristics: 1) LSIC is purely defined on integers, and thus it can be easily implemented in software and hardware without causing finite precision or descritization problems; 2) LSIC  constructs all encryption primitives based on one keyed Latin square, including whitening, substitution and permutation, and thus it attains high sensitivities to any key change; 3) LSIC encrypts image pixels in the unit of byte instead of bit and processes a $256\times 256$ plaintext image at one time, and thus it is efficient for image data; 4) LSIC arranges all these encryption primitives in the framework of substitution-permutation network (SPN) and thus it attains good confusion and diffusion properties \cite{ShannonCipher}; 5) LSIC also integrates probabilistic encryption in a pre-processing stage and thus it allows to encrypt a plaintext image into different ciphertext images when the same encryption key is used; and 6) LSIC's decryption stage is robust against a certain level of noise and thus it is suitable to transmit cipher data over a corrupted channel. The effectiveness and robustness of LSIC have been demonstrated by extensive simulation results using the complete USC-SIPI \textit{Miscellaneous} image dataset. Theoretical security analysis shows that LSIC has good resistances to brute-force attacks, ciphertext-only attacks, known-plaintext attacks and chosen-plaintext attacks. Experimental security analysis with comparisons to peer algorithms indicate that LSIC outperforms or reaches state of the art. All these analysis and results demonstrate that the LSIC is very suitable for digital image encryption. Finally, we open source the LSIC MATLAB code under webpage \url{https://sites.google.com/site/tuftsyuewu/source-code}.

Although in our simulations we have not found any '\textit{weak}' Latin squares, it is still an open question whether such '\textit{weak}' Latin squares exist. Meanwhile, it is natural to ask the next question that if such '\textit{weak}' Latin squares exist, how to recognize them in the very beginning and thus we are able to avoid using these Latin squares for encryption. Eventually, we need some means to evaluate the suitability of a Latin square for image encryption. However, Neither question is answered in this paper.

\bibliographystyle{IEEEtran}
\bibliography{report}







\end{document}